\documentclass[a4paper,fleqn,usenatbib]{mnras}

\usepackage{newtxtext,newtxmath}

\usepackage[T1]{fontenc}
\usepackage{ae,aecompl}

\usepackage{graphicx}	
\usepackage{amsmath}	
\usepackage{amssymb}	

\usepackage{tablefootnote}
\usepackage{multirow}

\usepackage{textcase}

\title[Instability of low--frequency modes in rotating stars]{Domains of pulsational instability of low--frequency
modes in rotating  upper main--sequence stars}

\author[Szewczuk \& Daszy\'nska--Daszkiewicz]{
Wojciech Szewczuk\thanks{E-mail: szewczuk@astro.uni.wroc.pl (WS)},
Jadwiga Daszy\'nska--Daszkiewicz
\\
Astronomical Institute of the Wroc{\l}aw Univeristy, Kopernika 11, 51--622 Wroc{\l}aw, Poland}

\date{Accepted XXX. Received YYY; in original form ZZZ}

\pubyear{2015}

\begin{document}
\label{firstpage}
\pagerange{\pageref{firstpage}--\pageref{lastpage}}
\maketitle

\begin{abstract}
We determine instability domains on the Hertzsprung--Russel diagram
for rotating main sequence stars with masses 2--20 $\mathrm M_{\sun}$.
The effects of the Coriolis force are treated in the framework of the traditional approximation.
High--order g--modes with the harmonic  degrees, $\ell$, up to 4 and mixed gravity--Rossby modes with $|m|$ up to 4 are considered.
Including the effects of rotation results in wider instability strips for a given $\ell$ comparing to the non--rotating case
and in an extension of the pulsational instability to hotter and more massive models.
We present results for the fixed value of the initial rotation velocity
as well as for the fixed ratio of the angular rotation frequency
to its critical value.
Moreover, we check how the initial hydrogen abundance, metallicity, overshooting from the convective core and
the opacity data affect
the pulsational instability domains. The effect of rotation on the period spacing is also discussed.
\end{abstract}

\begin{keywords}
stars: early--type -- stars: oscillations  -- stars: rotation
\end{keywords}



\section{Introduction}

Two main classes of pulsating variables are usually distinguished amongst main sequence stars with masses
$M \gtrsim 2.5\,\mathrm M_{\sun}$:
$\beta$ Cephei \citep{1902ApJ....15..340F, 2005ApJS..158..193S}
and Slowly Pulsating B--type stars \citep[SPB,][]{Waelkens1991}.
The first class consists of early B--type stars with masses of 8--16 $\mathrm M_{\sun}$
and pulsate mainly in frequencies corresponding to low radial--order
pressure and gravity (p and g) modes.
The second class consists of stars of middle B spectral types pulsating with frequencies associated with high radial--order g modes.
Pulsations of both types are driven in the metal (Z) opacity bump at $\log T\approx 5.3$
\citep{1992A&A...256L...5M,1992ApJ...393..272C,WD_PM_AP1993,Gautschy1993}.
However,
this simple division into low--order p/g mode and high--order g mode pulsators is no longer valid.
Recently, high--order g modes were detected in early B--type stars and low--order p/g modes,
in mid B--type stars. First such hybrid pulsators were discovered from the ground
\citep[e.g.][]{2005MNRAS.360..619J,2006MNRAS.365..327H,2006A&A...448..697C}.
Subsequently, hybrid pulsators with far more rich oscillation spectra
than those seen from the ground were discovered from space by projects such as MOST, CoRoT, Kepler and BRITE
\citep{Degroote2010N, Balona2011, McNamara2012, Papics2012, Papics2014, Ppics2015, 2015MNRAS.451.1445B, 2016A&A...588A..55P}. 
From the theoretical point of view, the important fact is that computations for non--rotating models showed that g mode pulsations
can be excited in very hot and massive stars \citep{Pamyatnykh1999}. This result was obtained with both
OPAL \citep{Iglesias1996} and OP opacity tables \citep{Seaton1996}.

The theoretical instability strip for g--mode pulsators has been constantly recomputed when the updated opacity data were released \citep[e.\,g.,][]{Seaton2005}
or the new solar mixture was determined \citep{2005ASPC..336...25A, Asplund2009}. For example, \citet{Miglio2007} and \citet{Zdravkov2008} investigated the influence
of  updated opacity data and chemical mixture on the SPB instability strip for non--rotating models and pulsational modes with $\ell=1$, 2.
\citet{2007CoAst.151...48M} extended the computations to $\ell=3$.

\citet{2012MNRAS.422.3460S} tested the effects of increasing the opacity in the region of the iron--group bump
and of changing the chemical mixture on the excitation of pulsation modes in B--type stars in the Magellanic Clouds.
Another example of this type of research
are pulsational studies for massive stars by \citet{2014A&A...565A..76C} who identified a new opacity bump at $\log T=5.06$ in the Kurucz stellar atmosphere models \citep{2003IAUS..210P.A20C}.
Recently, \citet{PW2015} found that calculations with the new Los Alamos opacity data, OPLIB \citep{2013HEDP....9..369C,2015HEDP...14...33C},
result in a wider SPB instability domain  than those with OP or OPAL tables. Similarly, a wider instability strip was obtained by
\citet{Moravveji2016} who artificially enhanced the iron and nickel contribution to the Rosseland mean opacity by 75 \%.
This was motivated by the work of \citet{2015Natur.517...56B} who reported
higher than predicted laboratory measurement of iron opacity at solar interior temperatures.
Laboratory measurements and theoretical computations of iron and nickel opacities
for envelopes of massive stars were also performed by \citet{2013HEDP....9..473T} who concluded that
a significant increase in comparison with the OP data is predicted for the nickel opacity.

The instability domains
of gravity modes with the effects of rotation on pulsations taken into account
have been computed by \citet{Townsend2005, 2005MNRAS.364..573T}.
His  calculations were carried out in the so--called traditional approximation
and all effects of rotation on the evolutionary models were omitted; the chemical mixture of  \citet[][GN93]{1993oee..conf...15G} was used.
This author considered $\ell\le2$ g modes  for models
with masses smaller than 13\,$\mathrm M_{\sun}$ \citep{Townsend2005}
and mixed gravity--Rossby modes
for models with masses smaller than 8\,$\mathrm M_{\sun}$ \citep{2005MNRAS.364..573T}.

Here we extend Townsend's computations to the harmonic degree up to $\ell=4$ and $|m|\le\ell$
as well as mixed gravity--Rossby modes, also known as $r$ modes, with $|m|\le 4$
\citep{2006MNRAS.365..677L,JDD_WD_AP2007} for models with masses 2 -- 20\,$\mathrm M_{\sun}$.
Including higher $\ell$ and $r$ modes was motivated by the high precision space--based photometry.
Furthermore, the most recent chemical mixture of \citet{Asplund2009} was used in our calculations.
In the present paper by SPB modes we mean high radial--order g modes.

In Section \ref{sec_l_1_4}, we present instability strips for our reference rotating models
with  masses in the
range 2 -- 20\,$\mathrm M_{\sun}$.
In Section \ref{par_effects}, the effects of the initial hydrogen abundance, metallicity, core overshooting and
the opacity data
on the SPB instability strip are discussed.
Section \ref{fixed_omega_section} is devoted to checking the impact on the instability strip
of  using a
fixed ratio of the angular rotation frequency
to its critical value instead of a fixed equatorial velocity.
The influence of rotation on the period spacing for high radial--order g modes
is examined in Section \ref{DP}. We end with a summary in Section \ref{conclusions}.

\section{Low--frequency modes in rotating models of upper main sequence stars}
\label{sec_l_1_4}
In the present paper we will consider only low--frequency (slow) modes,
i.\,e., high--order gravity as well as mixed gravity--Rossby modes.
The latter become propagative in the radiative envelope only if the rotation is fast enough
\citep[e.\,g.,][]{2005A&A...443..557S, 2005MNRAS.364..573T}.
Low radial--order p and g modes which can be also excited in B--type stars and are responsible
for $\beta$ Cephei phenomenon are omitted here.

Since the periods of the slow modes are often of the order of the rotation  period,
the effects of rotation cannot be regarded as a small perturbation to the pulsations another treatment is needed.
In particular, the effects of the Coriolis force have to be taken into account.
As far as the centrifugal force is concerned,
\citet{Ballot2012, 2013LNP...865...91B} showed that its effects on g modes
can be safely neglected if the rotation rate is well below the critical value.
In the first paper these authors used the condition $\Omega <0.7 \Omega_\mathrm{crit}$
where $\Omega$ and $\Omega_\mathrm{crit}$ are the angular frequency of rotation and its critical value, respectively,
but they did not justify their choice. Moreover, in the second paper \citep{2013LNP...865...91B} they mentioned
that an impact of the centrifugal force is negligible below a lower threshold $\Omega <0.4 \Omega_\mathrm{crit}$
and it affects mostly prograde sectoral modes ($\ell=m$).
In the present paper, following \citet{Townsend2005}, \citet{WD_JDD_AP2007} and \citet{2015MNRAS.446.1438D},
as an upper limit of the applicability of the traditional approximation, we used a less rigorous condition, i.e.,
$\Omega\lesssim 0.7\Omega_\mathrm{crit}$ corresponding to the value of \citet{Ballot2012}.

While including the effects of Coriolis force one often adopts the traditional approximation
\citep[e.\,g.,][]{Lee_Saio1997,Townsend2003b,Townsend2003a,Townsend2005,2005MNRAS.364..573T,
WD_JDD_AP2007,JDD_WD_AP2007} or a truncated expansion in the associated Legendre polynomials for the eigenfunctions
\citep[e.\,g.,][]{Lee1989,Lee2001}. In the present paper we will use the first approach
in which the horizontal component of the Coriolis force related to the radial motion and
the radial component of the Coriolis force related to the horizontal motion are neglected
\citep[e.\,g.,][]{Townsend2003b}.

Here, we computed evolutionary models with the Warsaw--New Jersey code \citep[e.\,g.][]{Pamyatnykh1998}.
In this code, a very simple approach of including rotation is used, i.e., the effect of the averaged centrifugal force is taken into account
in the equation of hydrostatic equilibrium.
The rotation is assumed uniform and the global angular momentum is conserved during the evolution.

\subsection{Instability domains of high--order g modes}
\label{sub_sec_g}

We constructed a grid of evolutionary models
with masses ranging from 2 to 20\,$\mathrm M_{\sun}$ and with steps:
0.05\,$\mathrm M_{\sun}$ for $M\le 3.5\,\mathrm M_{\sun}$,
0.1$\,\mathrm M_{\sun}$ for $3.5\,\mathrm M_{\sun}<M\le 12.2\,\mathrm M_{\sun}$,
0.2$\,\mathrm M_{\sun}$ for $12.2\,\mathrm M_{\sun}<M\le 14\,\mathrm M_{\sun}$,
and 0.5 $\,\mathrm M_{\sun}$ for $M>14\,\mathrm M_{\sun}$.
As a reference set of parameters, we chose the initial hydrogen abundance,
$X_0=0.7$, metallicity, $Z=0.015$, and no overshooting from the convective core,
$\alpha_\mathrm{ov}=0.0$. We used the OP opacity tables \citep{Seaton2005}
and the AGSS09 chemical mixture \citep{Asplund2009}.
At lower temperatures, $\log T < 3.95$, the opacities
were supplemented with the \citet{FergusonAlexander2005} data.
We adopted OPAL equation of state \citep{2002ApJ...576.1064R} and
the nuclear reaction rates from \citet{1995RvMP...67..781B}.
In the envelope, convection was treated in the framework of the standard mixing-length theory (MLT)
with the parameter $\alpha_\mathrm{MLT}=1.0$. The value of $\alpha_\mathrm{MLT}$ does not affect the pulsational computations
for masses considered in this paper.
Three values of the rotation velocity on the Zero Age Main Sequence (ZAMS)
were assumed, $V_\mathrm{rot}=0,\,100$ and 200 $\mathrm{km\,s^{-1}}$.
The last value was chosen because it is more or less the upper limit of the applicability
of the traditional approximation as well as because most B--type stars rotate with equatorial velocity
below or around 200 $\mathrm{km\,s^{-1}}$ \citep[see e.g.][]{2006ApJ...648..580H}.
We have to add that at $V_\mathrm{rot}=200 \,\mathrm{km\,s^{-1}}$
a small number of low mass models close to Terminal Age Main Sequence (TAMS) has $\Omega$
slightly exceeding $0.7\Omega_\mathrm{crit}$ (see Fig.\,\ref{figA1} in Appendix \ref{appA}),
but this fact does not spoil the overall picture of our results.
Nevertheless, the computed mode frequencies and instabilities
have to be treated with some caution, especially for the prograde sectoral modes which are mostly
affected by the centrifugal force.
It should also be kept in mind that fast rotation can stabilize retrograde g modes
if a truncated expansion of the Legendre functions for the eigenfunctions is used
\citep{2008CoAst.157..203L,2011MNRAS.412.2265A}.
The ranges of $\Omega/\Omega_\mathrm{crit}$, corresponding to fixed values of equatorial velocities
for models in a given instability strip are listed in the last columns of Tables\,\ref{reference_ranges},
\ref{r_tab} and \ref{rest_ranges}.


Linear nonadiabatic pulsational calculations were done with the Dziembowski code in its version which includes
the effects of rotation in the traditional approximation \citep{WD_JDD_AP2007}.
This code uses the same approach as that adopted by \citet{Townsend2005}. Moreover,
the freezing approximation is assumed for the convective
flux; this is fully justified for the range of masses we consider.

For all models from ZAMS to TAMS, the pulsations were computed for the rotation velocities fixed to the ZAMS values, i.\,e.,
for $V_\mathrm{rot, MS}=V_\mathrm{rot, ZAMS}=0,~100$ and 200 $\mathrm{km\,s^{-1}}$.
We considered high--order g modes with the spherical harmonic degrees, $\ell\le 4$, and azimuthal orders, $-\ell\le m \le \ell$,
with the convention $m>0$ for prograde modes.

\begin{figure*}
	\includegraphics[width=2\columnwidth]{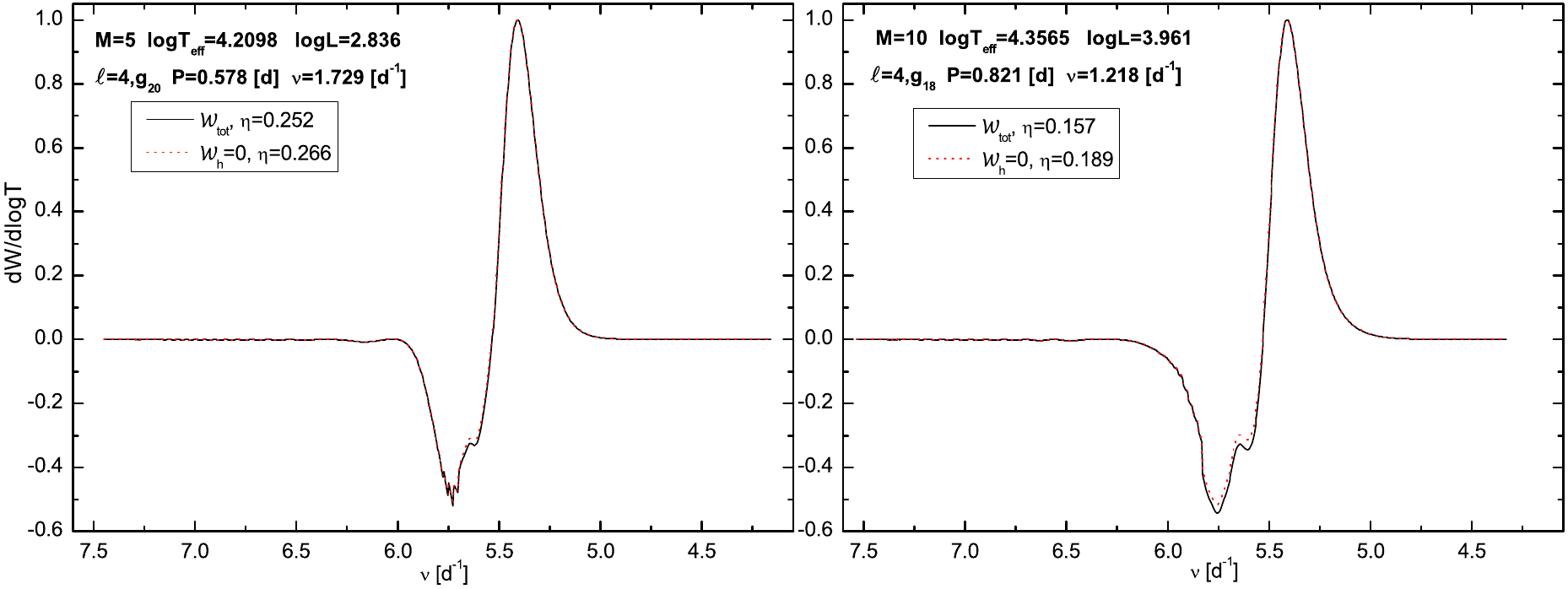}
    \caption{The run of the differential work integral, $\mathcal{W}=dW/d\log T$, inside the models of
    5\,$\mathrm M_{\sun}$ (the left panel) and 10\,$\mathrm M_{\sun}$ (the right panel)
             for unstable high--order g modes with the degree $\ell=4$.
             Models were computed with the OP opacities for $X_0 = 0.7$, $Z = 0.015$,
            $\alpha_\mathrm{ov} = 0$ and the AGSS09 chemical mixture.
            The dotted lines indicate the work integrals without the horizontal losses.
            The values of the normalized instability parameter, $\eta$, are given in the insets. Both modes are unstable.}
    \label{work_int}
\end{figure*}

One of the important assumptions in the traditional approximation is
that of a small contribution of the horizontal heat losses to the overall work integral.
This is because replacing the eigenvalue  $\ell(\ell+1)$ by $\lambda$ does not correctly include these losses which are
proportional to the mode degree \citep[cf.][]{Townsend2005, WD_JDD_AP2007}. Here we consider modes up to $\ell=4$, thus, it is interesting to check whether
neglecting the horizontal heat losses is valid for such high degree modes.
In Fig.\,\ref{work_int}, we plot the differential work integral for high--order g modes with $\ell=4$ considering two models with masses typical for SPB and $\beta$ Cephei stars: 5 and 10\,$\mathrm M_{\sun}$, respectively. The corresponding effective temperatures of these models are $\log T=4.2098$ and $\log T=4.3565$, respectively.
The work integral was computed with the zero-rotation approximation. In addition, in Fig.\,\ref{work_int}
we marked the instability parameter $\eta$ for both modes.
The parameter $\eta$ is the normalized total work integral and modes which are pulsational unstable have positive values of $\eta$.
As one can see, the horizontal heat losses are negligible and have the largest contribution below the Z--bump ($\log T<5.3$).
One can also see that the effect of the horizontal heat losses slightly increases with mass.

\begin{figure*}

	\includegraphics[width=2.6\columnwidth, angle=270]{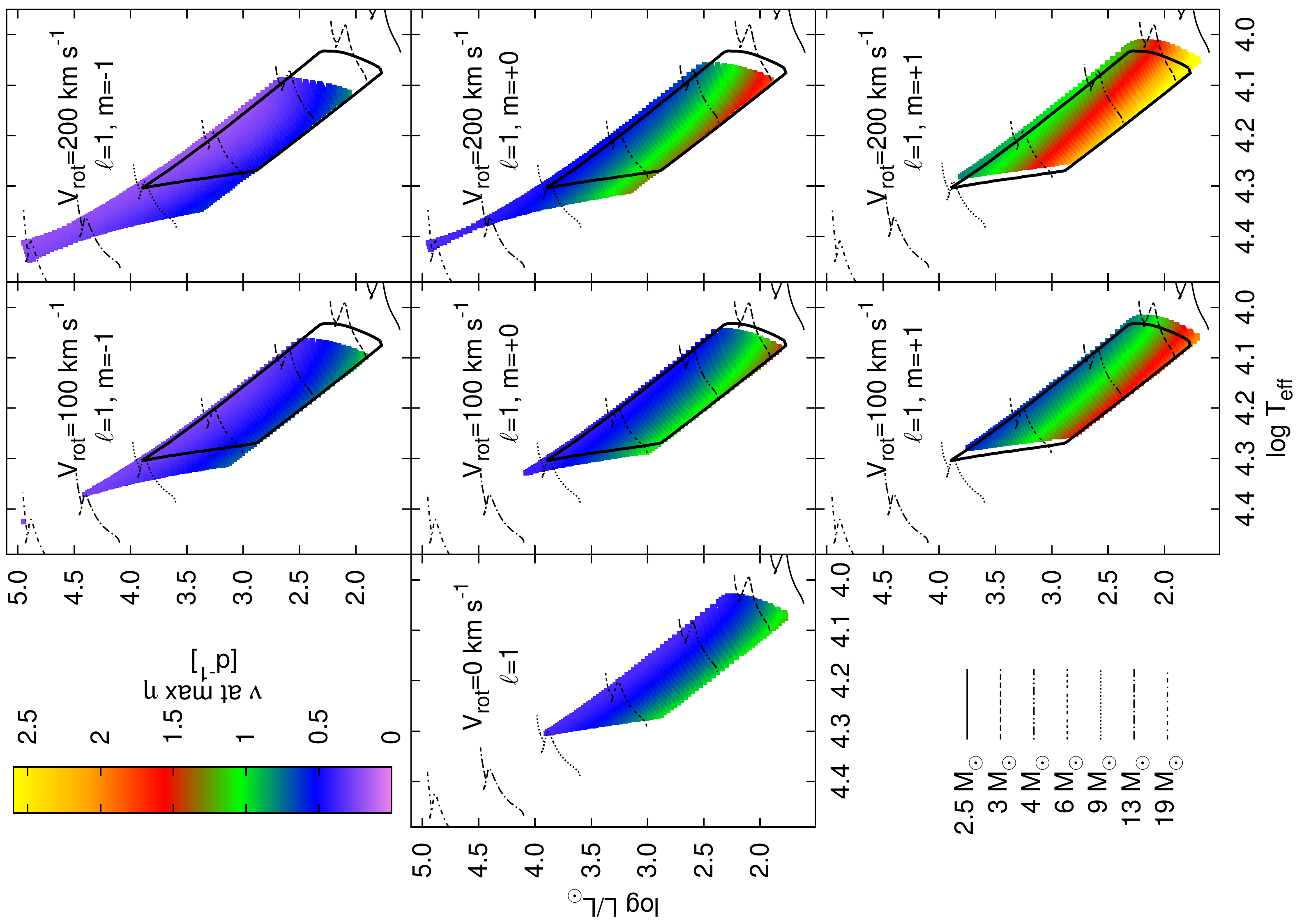}
    \caption{The SPB instability strip on the H--R diagram.
    There are shown the dipole modes with all possible
            azimuthal orders. Models were computed with the OP opacity tables assuming $X_0 = 0.7$, $Z = 0.015$,
            $\alpha_\mathrm{ov} = 0$ and three values of the rotation velocity, $V_\mathrm{rot}=0$, 100 and 200 $\mathrm{km\,s^{-1}}$.
            Colours code the values of the mode frequency in the inertial frame at which the instability parameter, $\eta$, reaches maximum.
             There are also shown the evolutionary tracks for 2.5, 3, 4, 6, 9, 13 and 19$\,\mathrm M_{\sun}$.
             For a better comparison, the instability domain for the non--rotating case is additionally
             marked on the panels with the rotating models with thick black lines.
            (Colour figure only in the electronic edition of the journal.)
            }
    \label{l1_OP_X0_7Z0_015_ov0_0}
\end{figure*}

\begin{figure*}
	\includegraphics[width=2.7\columnwidth, angle=270]{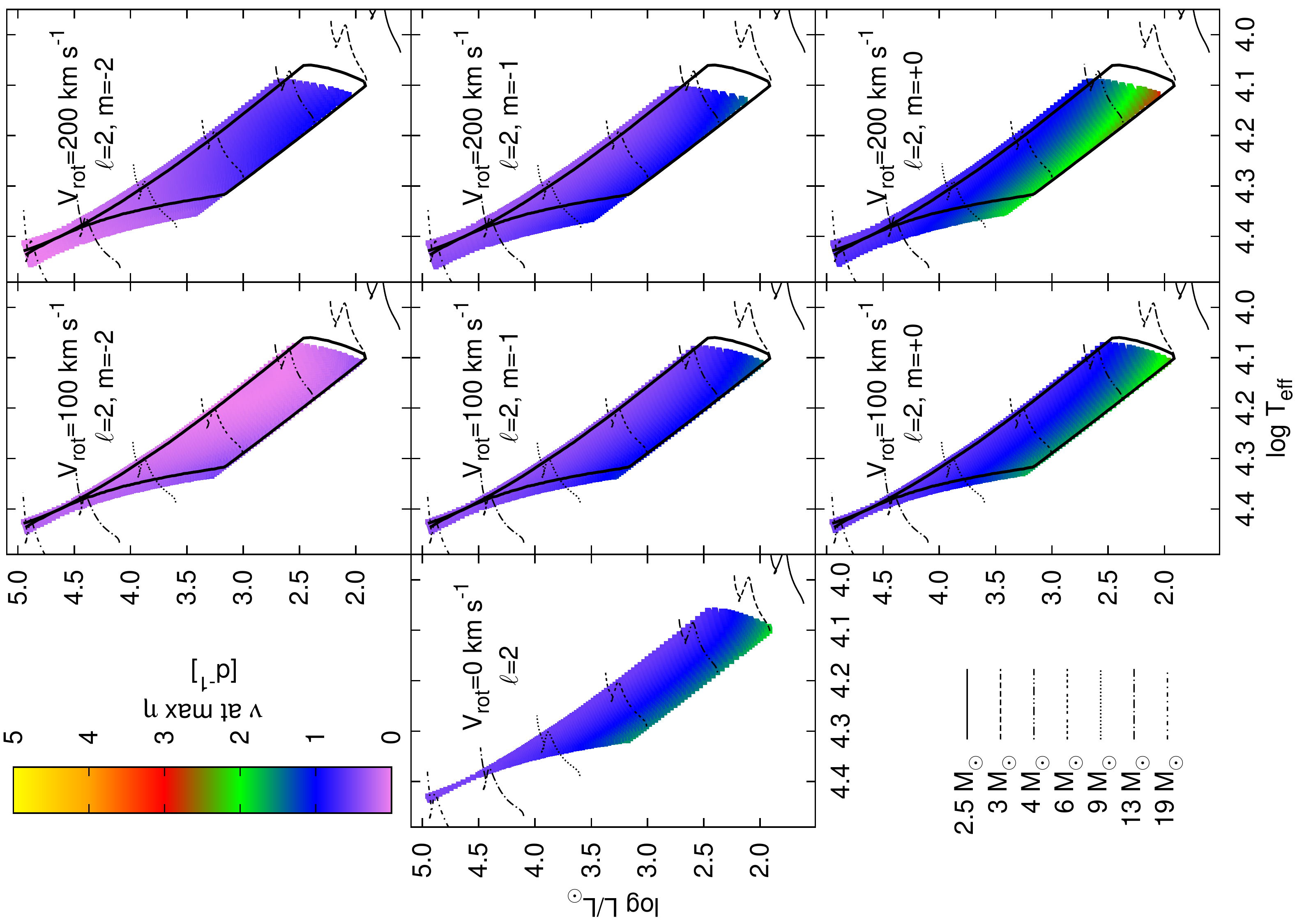}
    \caption{The same as in Fig.\,\ref{l1_OP_X0_7Z0_015_ov0_0} but for modes with $\ell=2$
             and $m\le 0$.
            }
    \label{l2a_OP_X0_7Z0_015_ov0_0}
\end{figure*}

\begin{figure*}
	\includegraphics[width=2.7\columnwidth, angle=270]{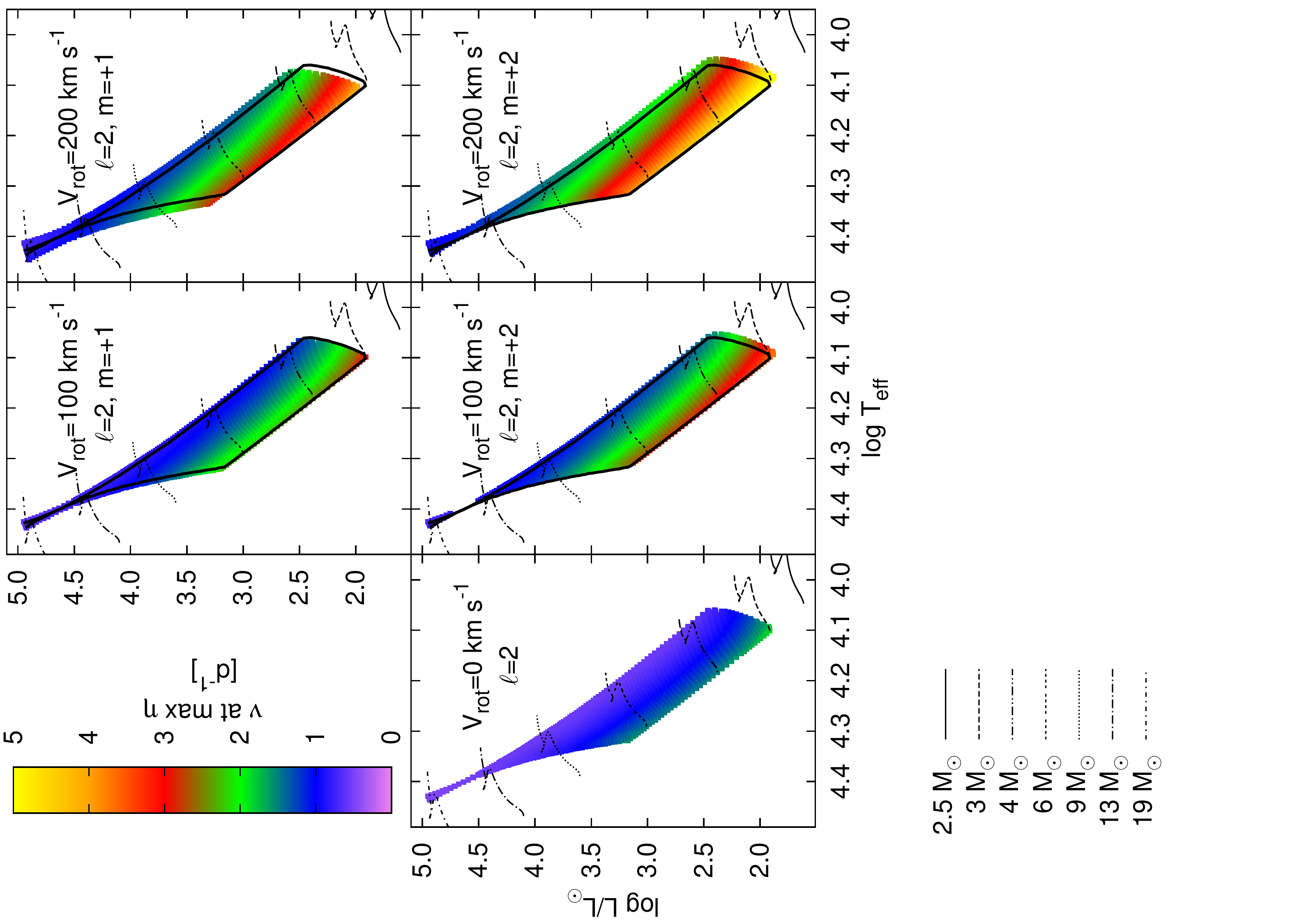}
    \caption{The same as in Fig.\,\ref{l1_OP_X0_7Z0_015_ov0_0} but for modes with $\ell=2$
             and $m\ge+1$.
            }
    \label{l2b_OP_X0_7Z0_015_ov0_0}
\end{figure*}

\begin{table*}
	\caption{The ranges of mass, effective temperature, luminosity and surface gravity for the instability domains of the reference
	models with $V_\mathrm{rot}=0$, 100,  200 km\,s$^{-1}$.
        Results for all pulsational modes with $\ell=1-4$ are shown. Note that in the non--rotating case,
        modes are degenerated in the azimuthal order $m$.
        The asterisk  indicates the limit of the model grid rather than the limit of the instability domain.
        In the penultimate column there is given the range of frequencies
        of all unstable modes (in the observer frame). Frequencies of ,,reflected'' modes, i.e. those which formally have the negative values of frequencies, are given in parentheses.
        In the last column the ranges of the ratio of angular rotation rate to its critical value for models within the given instability domain are listed.
	}
	\label{reference_ranges}
	\centering
	\begin{tabular}{lcccccccc} 
		\hline
		Grid & $V_\mathrm{rot}$                   & mode                      & $M$    & $\log T_\mathrm{eff}$ & $\log L/\mathrm L_{\sun}$ & $\log g$ & $\nu$ &$\Omega/\Omega_\mathrm{crit}$ \\
		     & $\left[\mathrm{km\,s^{-1}}\right]$ & $\left( \ell,\,m \right)$ & $\left[\mathrm M_{\sun}\right]$   &                       &                           &  & $\left[\mathrm d^{-1}\right]$ &\\
		\hline
		\multirow{53}{*}{\rotatebox[origin=c]{90}{OP~~~$X_0=0.70~~~Z=0.015~~~\alpha_\mathrm{ov}=0.0$}} & \multirow{4}{*}{0}  & (1)  & 2.75 -- 9.10 & 4.032 -- 4.305 & 1.77 -- 3.90 & 3.67 -- 4.36 & 0.2162 -- 1.6026 & 0.00 \\
		                    &                     &  (2)  & 3.00 -- 20.00$^*$ & 4.060 -- 4.437$^*$ & 1.91 -- 4.95$^*$ & 3.45$^*$ -- 4.36 & 0.3206 -- 2.6318 & 0.00  \\
		                    &                     &  (3)  & 3.15 -- 20.00$^*$ & 4.080 -- 4.456$^*$ & 1.99 -- 4.95$^*$ & 3.45$^*$ -- 4.36 & 0.4101 -- 3.5454 & 0.00  \\
		                    &                     &  (4)  & 3.30 -- 20.00$^*$ & 4.095 -- 4.469$^*$ & 2.07 -- 4.95$^*$ & 3.45$^*$ -- 4.36 & 0.4929 -- 4.3470 & 0.00  \\
		                    \cline{2-9}
 &\multirow{25}{*}{100}&  $\left(1,\,-1\right)$ & 3.00 -- 13.00 & 4.066 -- 4.371 & 1.91 -- 4.41 & 3.57 -- 4.34 & 0.0991 -- 1.3095 & 0.28 -- 0.44\\
 &                     &  $\left(1,\,-1\right)$ & 20.00$^*$ -- 20.00$^*$ & 4.425$^*$ -- 4.425$^*$ & 4.95$^*$ -- 4.95$^*$ & 3.43$^*$ -- 3.43$^*$ & 0.2178 -- 0.2327 & 0.35 -- 0.35\\
 &                     &  $\left(1,\,+0\right)$ & 2.85 -- 10.30 & 4.046 -- 4.328 & 1.83 -- 4.08 & 3.63 -- 4.34 & 0.3174 -- 1.7661 & 0.29 -- 0.45\\
 &                     &  $\left(1,\,+1\right)$ & 2.65 -- 8.20 & 4.019 -- 4.280 & 1.71 -- 3.74 & 3.67 -- 4.34 & 0.5429 -- 2.3164 & 0.29 -- 0.46\\
 &                     &  $\left(2,\,-2\right)$ & 3.10 -- 20.00$^*$ & 4.076 -- 4.445$^*$ & 1.97 -- 4.95$^*$ & 3.43$^*$ -- 4.34 & (0.3607) -- 1.2483 & 0.27 -- 0.44\\
 &                     &  $\left(2,\,-1\right)$ & 3.15 -- 20.00$^*$ & 4.081 -- 4.445$^*$ & 1.99 -- 4.95$^*$ & 3.43$^*$ -- 4.34 & 0.2094 -- 2.0576 & 0.27 -- 0.44\\
 &                     &  $\left(2,\,+0\right)$ & 3.10 -- 20.00$^*$ & 4.074 -- 4.442$^*$ & 1.97 -- 4.95$^*$ & 3.43$^*$ -- 4.34 & 0.4036 -- 2.7899 & 0.28 -- 0.44\\
 &                     &  $\left(2,\,+1\right)$ & 3.05 -- 20.00$^*$ & 4.065 -- 4.438$^*$ & 1.94 -- 4.95$^*$ & 3.43$^*$ -- 4.34 & 0.5296 -- 3.5429 & 0.28 -- 0.45\\
 &                     &  $\left(2,\,+2\right)$ & 2.95 -- 14.00 & 4.053 -- 4.383 & 1.88 -- 4.50 & 3.56 -- 4.34 & 0.8523 -- 4.2689 & 0.28 -- 0.45\\
 &                     &  $\left(2,\,+2\right)$ & 17.00 -- 20.00$^*$ & 4.409 -- 4.433$^*$ & 4.75 -- 4.95$^*$ & 3.43$^*$ -- 3.50 & 0.6533 -- 0.7864 & 0.34 -- 0.35\\

 &                     &  $\left(3,\,-3\right)$ & 3.25 -- 20.00$^*$ & 4.089 -- 4.460$^*$ & 2.04 -- 4.95$^*$ & 3.43$^*$ -- 4.34 & (1.0289) - 1.3043 & 0.27 -- 0.44\\
 &                     &  $\left(3,\,-2\right)$ & 3.30 -- 20.00$^*$ & 4.092 -- 4.460$^*$ & 2.07 -- 4.95$^*$ & 3.43$^*$ -- 4.34 & (0.1497) - 2.0969 & 0.27 -- 0.44\\
 &                     &  $\left(3,\,-1\right)$ & 3.30 -- 20.00$^*$ & 4.093 -- 4.460$^*$ & 2.07 -- 4.95$^*$ & 3.43$^*$ -- 4.34 & 0.2817 - 2.8573 & 0.27 -- 0.44\\
 &                     &  $\left(3,\,+0\right)$ & 3.25 -- 20.00$^*$ & 4.090 -- 4.459$^*$ & 2.04 -- 4.95$^*$ & 3.43$^*$ -- 4.34 & 0.4778 - 3.6251 & 0.27 -- 0.44\\
 &                     &  $\left(3,\,+1\right)$ & 3.25 -- 20.00$^*$ & 4.086 -- 4.458$^*$ & 2.04 -- 4.95$^*$ & 3.43$^*$ -- 4.34 & 0.6087 - 4.4622 & 0.27 -- 0.44\\
 &                     &  $\left(3,\,+2\right)$ & 3.20 -- 20.00$^*$ & 4.082 -- 4.456$^*$ & 2.02 -- 4.95$^*$ & 3.43$^*$ -- 4.34 & 0.7315 - 5.2454 & 0.27 -- 0.44\\
 &                     &  $\left(3,\,+3\right)$ & 3.15 -- 20.00$^*$ & 4.075 -- 4.455$^*$ & 1.99 -- 4.95$^*$ & 3.43$^*$ -- 4.34 & 0.8657 - 6.0119 & 0.27 -- 0.44\\

 &                     &  $\left(4,\,-4\right)$ & 3.40 -- 20.00$^*$ & 4.102 -- 4.471$^*$ & 2.12 -- 4.95$^*$ & 3.43$^*$ -- 4.34 & (1.6677) -- 1.3370 & 0.27 -- 0.43\\
 &                     &  $\left(4,\,-3\right)$ & 3.40 -- 20.00$^*$ & 4.104 -- 4.472$^*$ & 2.12 -- 4.95$^*$ & 3.43$^*$ -- 4.34 & (0.7022) -- 2.0928 & 0.27 -- 0.43\\
 &                     &  $\left(4,\,-2\right)$ & 3.40 -- 20.00$^*$ & 4.105 -- 4.471$^*$ & 2.12 -- 4.95$^*$ & 3.43$^*$ -- 4.34 & (0.0529) -- 2.8552 & 0.27 -- 0.43\\
 &                     &  $\left(4,\,-1\right)$ & 3.40 -- 20.00$^*$ & 4.105 -- 4.471$^*$ & 2.12 -- 4.95$^*$ & 3.43$^*$ -- 4.34 & 0.3346 -- 3.6039 & 0.27 -- 0.43\\
 &                     &  $\left(4,\,+0\right)$ & 3.40 -- 20.00$^*$ & 4.103 -- 4.471$^*$ & 2.12 -- 4.95$^*$ & 3.43$^*$ -- 4.34 & 0.5493 -- 4.4406 & 0.27 -- 0.43\\
 &                     &  $\left(4,\,+1\right)$ & 3.40 -- 20.00$^*$ & 4.102 -- 4.471$^*$ & 2.12 -- 4.95$^*$ & 3.43$^*$ -- 4.34 & 0.6787 -- 5.2348 & 0.27 -- 0.43\\
 &                     &  $\left(4,\,+2\right)$ & 3.35 -- 20.00$^*$ & 4.099 -- 4.469$^*$ & 2.09 -- 4.95$^*$ & 3.43$^*$ -- 4.34 & 0.8146 -- 6.0251 & 0.27 -- 0.43\\
 &                     &  $\left(4,\,+3\right)$ & 3.35 -- 20.00$^*$ & 4.096 -- 4.469$^*$ & 2.09 -- 4.95$^*$ & 3.43$^*$ -- 4.34 & 0.9455 -- 6.8330 & 0.27 -- 0.44\\
 &                     &  $\left(4,\,+4\right)$ & 3.30 -- 20.00$^*$ & 4.091 -- 4.468$^*$ & 2.07 -- 4.95$^*$ & 3.43$^*$ -- 4.34 & 1.0731 -- 7.6288 & 0.27 -- 0.44\\
		                    \cline{2-9}
 &\multirow{24}{*}{200}&  $\left(1,\,-1\right)$ & 3.30 -- 20.00$^*$ & 4.089 -- 4.447$^*$ & 2.06 -- 4.94$^*$ & 3.35$^*$ -- 4.28 & (0.0919) -- 1.2089 & 0.53 -- 0.83\\
 &                     &  $\left(1,\,+0\right)$ & 3.00 -- 20.00$^*$ & 4.059 -- 4.425$^*$ & 1.90 -- 4.94$^*$ & 3.35$^*$ -- 4.28 & 0.2997 -- 2.0555 & 0.55 -- 0.85\\
 &                     &  $\left(1,\,+1\right)$ & 2.65 -- 8.60 & 4.014 -- 4.282 & 1.70 -- 3.81 & 3.60 -- 4.28 & 0.7865 -- 3.1563 & 0.57 -- 0.88\\

 &                     &  $\left(2,\,-2\right)$ & 3.35 -- 20.00$^*$ & 4.093 -- 4.456$^*$ & 2.08 -- 4.94$^*$ & 3.35$^*$ -- 4.28 & (1.5798) -- 0.1472 & 0.53 -- 0.83\\
 &                     &  $\left(2,\,-1\right)$ & 3.45 -- 20.00$^*$ & 4.107 -- 4.460$^*$ & 2.13 -- 4.94$^*$ & 3.35$^*$ -- 4.28 & 0.0700 -- 1.8633 & 0.52 -- 0.83\\
 &                     &  $\left(2,\,+0\right)$ & 3.30 -- 20.00$^*$ & 4.092 -- 4.455$^*$ & 2.06 -- 4.94$^*$ & 3.35$^*$ -- 4.28 & 0.4238 -- 3.1703 & 0.53 -- 0.83\\
 &                     &  $\left(2,\,+1\right)$ & 3.15 -- 20.00$^*$ & 4.073 -- 4.446$^*$ & 1.98 -- 4.94$^*$ & 3.35$^*$ -- 4.28 & 0.6247 -- 4.5348 & 0.54 -- 0.84\\
 &                     &  $\left(2,\,+2\right)$ & 2.95 -- 20.00$^*$ & 4.049 -- 4.435$^*$ & 1.88 -- 4.94$^*$ & 3.35$^*$ -- 4.28 & 0.8393 -- 5.9309 & 0.55 -- 0.86\\

 &                     &  $\left(3,\,-3\right)$ & 3.40 -- 20.00$^*$ & 4.100 -- 4.467$^*$ & 2.11 -- 4.94$^*$ & 3.35$^*$ -- 4.28 & (3.1684) -- (0.0175) & 0.52 -- 0.83\\
 &                     &  $\left(3,\,-2\right)$ & 3.50 -- 20.00$^*$ & 4.111 -- 4.468$^*$ & 2.16 -- 4.94$^*$ & 3.35$^*$ -- 4.28 & (1.1147) -- 0.9181 & 0.52 -- 0.82\\
 &                     &  $\left(3,\,-1\right)$ & 3.60 -- 20.00$^*$ & 4.119 -- 4.469$^*$ & 2.20 -- 4.94$^*$ & 3.35$^*$ -- 4.28 & 0.1859 -- 2.4609 & 0.52 -- 0.82\\
 &                     &  $\left(3,\,+0\right)$ & 3.50 -- 20.00$^*$ & 4.109 -- 4.467$^*$ & 2.16 -- 4.94$^*$ & 3.35$^*$ -- 4.28 & 0.5110 -- 3.9127 & 0.52 -- 0.82\\
 &                     &  $\left(3,\,+1\right)$ & 3.40 -- 20.00$^*$ & 4.099 -- 4.464$^*$ & 2.11 -- 4.94$^*$ & 3.35$^*$ -- 4.28 & 0.7335 -- 5.4101 & 0.52 -- 0.83\\
 &                     &  $\left(3,\,+2\right)$ & 3.30 -- 20.00$^*$ & 4.086 -- 4.461$^*$ & 2.06 -- 4.94$^*$ & 3.35$^*$ -- 4.28 & 0.9571 -- 6.9289 & 0.53 -- 0.84\\
 &                     &  $\left(3,\,+3\right)$ & 3.15 -- 20.00$^*$ & 4.071 -- 4.456$^*$ & 1.98 -- 4.94$^*$ & 3.35$^*$ -- 4.28 & 1.1772 -- 8.5067 & 0.53 -- 0.85\\

 &                     &  $\left(4,\,-4\right)$ & 3.60 -- 20.00$^*$ & 4.109 -- 4.475$^*$ & 2.20 -- 4.94$^*$ & 3.35$^*$ -- 4.28 & (4.6932) -- (0.1415) & 0.52 -- 0.82\\
 &                     &  $\left(4,\,-3\right)$ & 3.60 -- 20.00$^*$ & 4.117 -- 4.476$^*$ & 2.20 -- 4.94$^*$ & 3.35$^*$ -- 4.28 & (2.6540) -- 0.1937 & 0.51 -- 0.82\\
 &                     &  $\left(4,\,-2\right)$ & 3.70 -- 20.00$^*$ & 4.123 -- 4.476$^*$ & 2.24 -- 4.94$^*$ & 3.35$^*$ -- 4.28 & (0.7842) -- 1.6143 & 0.51 -- 0.82\\
 &                     &  $\left(4,\,-1\right)$ & 3.70 -- 20.00$^*$ & 4.127 -- 4.476$^*$ & 2.24 -- 4.94$^*$ & 3.35$^*$ -- 4.28 & 0.2784 -- 3.0516 & 0.51 -- 0.81\\
 &                     &  $\left(4,\,+0\right)$ & 3.70 -- 20.00$^*$ & 4.122 -- 4.476$^*$ & 2.24 -- 4.94$^*$ & 3.35$^*$ -- 4.28 & 0.6091 -- 4.5461 & 0.51 -- 0.82\\
 &                     &  $\left(4,\,+1\right)$ & 3.60 -- 20.00$^*$ & 4.116 -- 4.475$^*$ & 2.20 -- 4.94$^*$ & 3.35$^*$ -- 4.28 & 0.8347 -- 6.1004 & 0.52 -- 0.82\\
 &                     &  $\left(4,\,+2\right)$ & 3.50 -- 20.00$^*$ & 4.109 -- 4.474$^*$ & 2.16 -- 4.94$^*$ & 3.35$^*$ -- 4.28 & 1.0480 -- 7.6309 & 0.52 -- 0.82\\
 &                     &  $\left(4,\,+3\right)$ & 3.45 -- 20.00$^*$ & 4.099 -- 4.472$^*$ & 2.13 -- 4.94$^*$ & 3.35$^*$ -- 4.28 & 1.2823 -- 9.2755 & 0.52 -- 0.83\\
 &                     &  $\left(4,\,+4\right)$ & 3.35 -- 20.00$^*$ & 4.089 -- 4.469$^*$ & 2.08 -- 4.94$^*$ & 3.35$^*$ -- 4.28 & 1.5098 -- 10.8095 & 0.52 -- 0.83\\
		
		\hline
\end{tabular}

\end{table*}



\begin{table*}
	\caption{The instability ranges for the g modes in the models near ZAMS ($X_\mathrm{c}=0.65$), in the middle of the main--sequence evolution
	($X_\mathrm{c}=0.35$)
	and near TAMS ($X_\mathrm{c}=0.10$) for two stellar masses, $M=4$ and 9 $\mathrm M_{\sun}$. Three values of the rotation velocity
	are considered, $V_\mathrm{rot}=0$, 100 and 200 km\,s$^{-1}$. For each $\left(\ell,\,m\right)$ pair, the range of radial orders,
	$n$,
	and frequencies in the observer frame, $\nu$, spanned  by the unstable modes is listed (formally negative values for reflected modes
	are given in parentheses).
	}
	\label{ranges_nu_selected_models}
	\centering
	\begin{tabular}{ccccccccc} 
\hline
$M\,\left[\mathrm M_{\sun}\right]$ & $V_\mathrm{rot}$\,[km\,s$^{-1}$]& $\left( \ell,\,m \right)$ & \multicolumn{2}{c}{$X_\mathrm{c}=0.65$} &  \multicolumn{2}{c}{$X_\mathrm{c}=0.35$} & \multicolumn{2}{c}{$X_\mathrm{c}=0.10$} \\
                              &                                 &                           & $n$ & $\nu\,\left[\mathrm d^{-1}\right]$ & $n$ & $\nu\,\left[\mathrm d^{-1}\right]$ & $n$ & $\nu\,\left[\mathrm d^{-1}\right]$  \\

\hline
\multirow{18}{*}{4} &\multirow{2}{*}{0}   & $\left(1\right)$     & 21 -- 9 & 0.6005 -- 1.3448   & 39 -- 14 & 0.3586 -- 0.9588   & 71 -- 28 & 0.2397 -- 0.5967  \\
                    &                     & $\left(2\right)$     & 24 -- 9 & 0.9176 -- 2.3130   & 44 -- 15 & 0.5508 -- 1.5505   & 72 -- 30 & 0.4091 -- 0.9649   \\
\cline{2-9}

                    & \multirow{8}{*}{100} &$\left(1,\,-1\right)$& 26 -- 8 & 0.3846 -- 1.1399   & 44 -- 14 & 0.2396 -- 0.7616   & 59 -- 29 & 0.2339 -- 0.4731\\
                    &                      &$\left(1,\,+0\right)$& 23 -- 8 & 0.8161 -- 1.5202   & 43 -- 14 & 0.5336 -- 1.0354   & 71 -- 28 & 0.4002 -- 0.6927\\
                    &                      &$\left(1,\,+1\right)$& 18 -- 8 & 1.3222 -- 1.9226   & 35 -- 14 & 0.8865 -- 1.3267   & 67 -- 27 & 0.6375 -- 0.9293\\

                    &                      &$\left(2,\,-2\right)$& 26 -- 8 &(0.3388) -- 1.0396  & 44 -- 15 &(0.3024) -- 0.5381  & 55 -- 32 &(0.1240) -- 0.2000\\
                    &                      &$\left(2,\,-1\right)$& 27 -- 8 & 0.6657 -- 1.8619   & 42 -- 15 & 0.4728 -- 1.1538   & 49 -- 33 & 0.4714 -- 0.6710\\
                    &                      &$\left(2,\,+0\right)$& 26 -- 8 & 1.2692 -- 2.5410   & 44 -- 15 & 0.8732 -- 1.6416   & 56 -- 31 & 0.7498 -- 1.0735\\
                    &                      &$\left(2,\,+1\right)$& 25 -- 8 & 1.8392 -- 3.1646   & 44 -- 15 & 1.2665 -- 2.0865   & 65 -- 30 & 0.9934 -- 1.4278\\
                    &                      &$\left(2,\,+2\right)$& 23 -- 8 & 2.4286 -- 3.7593   & 42 -- 15 & 1.6725 -- 2.5109   & 73 -- 29 & 1.2426 -- 1.7716\\
\cline{2-9}
                    & \multirow{8}{*}{200} &$\left(1,\,-1\right)$& 27 -- 9 & 0.1034 -- 1.0280   & 38 -- 17 & 0.1203 -- 0.6279   &    --    &   --\\
                    &                      &$\left(1,\,+0\right)$& 27 -- 8 & 1.0325 -- 1.8549   & 46 -- 15 & 0.7053 -- 1.2313   & 60 -- 32 & 0.5949 -- 0.8200\\
                    &                      &$\left(1,\,+1\right)$& 19 -- 8 & 2.0549 -- 2.6284   & 38 -- 14 & 1.4118 -- 1.8385   & 72 -- 28 & 1.0379 -- 1.3044\\

                    &                      &$\left(2,\,-2\right)$& 27 -- 9 &(1.4572) -- (0.2283)& 37 -- 18 &(0.9846) -- (0.4108)&    --    &   --\\
                    &                      &$\left(2,\,-1\right)$& 25 -- 10& 0.6510 -- 1.5472   & 30 -- 19 & 0.6447 -- 0.9755   &    --    &   --\\
                    &                      &$\left(2,\,+0\right)$& 27 -- 9 & 1.7391 -- 2.8637   & 37 -- 18 & 1.3211 -- 1.8475   &    --    &   --\\
                    &                      &$\left(2,\,+1\right)$& 28 -- 9 & 2.7910 -- 3.9632   & 44 -- 16 & 2.0001 -- 2.7406   & 50 -- 36 & 1.6660 -- 1.8523\\
                    &                      &$\left(2,\,+2\right)$& 25 -- 9 & 3.9040 -- 5.0272   & 45 -- 15 & 2.7425 -- 3.5717   & 74 -- 31 & 2.0663 -- 2.5213\\
\hline

\multirow{18}{*}{9} &\multirow{2}{*}{0}    & $\left(1\right)$    &  --    & --                 &  --      &    --              &   --     &   -- \\
                    &                      & $\left(2\right)$    &  --    & --                 &  --      &    --              & 35 -- 19 & 0.4460 -- 0.7921 \\
\cline{2-9}

                    & \multirow{8}{*}{100} &$\left(1,\,-1\right)$&  --    & --                 &  --      &    --              & 39 -- 18 & 0.1934 -- 0.3899 \\
                    &                      &$\left(1,\,+0\right)$&  --    & --                 &  --      &    --              & 29 -- 20 & 0.3794 -- 0.4833\\
                    &                      &$\left(1,\,+1\right)$&  --    & --                 &  --      &    --              &    --    &        --\\

                    &                      &$\left(2,\,-2\right)$&  --    & --                 & 16 -- 11 & 0.1729 -- 0.4235   & 42 -- 18 & (0.0421) -- 0.3556\\
                    &                      &$\left(2,\,-1\right)$&  --    & --                 & 17 -- 11 & 0.5485 -- 0.8059   & 45 -- 18 & 0.2775 -- 0.6381 \\
                    &                      &$\left(2,\,+0\right)$&  --    & --                 & 15 -- 11 & 0.9182 -- 1.1168   & 41 -- 18 & 0.5092 -- 0.8710\\
                    &                      &$\left(2,\,+1\right)$&  --    & --                 &    --    &        --          & 37 -- 18 & 0.7352 -- 1.0847\\
                    &                      &$\left(2,\,+2\right)$&  --    & --                 &    --    &        --          & 33 -- 18 & 0.9672 -- 1.2884\\
\cline{2-9}
                    & \multirow{8}{*}{200} &$\left(1,\,-1\right)$&  --    & --                 & 22 -- 11 & 0.2191 -- 0.4948   & 54 -- 19 & 0.0508 -- 0.3731 \\
                    &                      &$\left(1,\,+0\right)$&  --    & --                 &    --    &        --          & 41 -- 19 & 0.4171 -- 0.6173\\
                    &                      &$\left(1,\,+1\right)$&  --    & --                 &    --    &        --          &    --    &    -- \\

                    &                      &$\left(2,\,-2\right)$&  --    & --                 & 23 -- 11 &(0.4953) -- (0.0688)& 54 -- 19 &(0.4811) -- (0.0397)\\
                    &                      &$\left(2,\,-1\right)$&  --    & --                 & 27 -- 10 & 0.3795 -- 0.8596   & 61 -- 19 & 0.1776 -- 0.6018\\
                    &                      &$\left(2,\,+0\right)$&  --    & --                 & 23 -- 11 & 0.9679 -- 1.3527   & 54 -- 19 & 0.6109 -- 1.0132\\
                    &                      &$\left(2,\,+1\right)$&  --    & --                 & 18 -- 11 & 1.5665 -- 1.8618   & 46 -- 19 & 1.0212 -- 1.3907\\
                    &                      &$\left(2,\,+2\right)$&  --    & --                 &    --    &        --          & 35 -- 19 & 1.4595 -- 1.7554\\
\hline
\end{tabular}

\end{table*}



The instability strip of the dipole modes on the Hertzsprung--Russel (H--R) diagram
is shown in Fig.\,\ref{l1_OP_X0_7Z0_015_ov0_0}.
In the case of retrograde and axisymetric modes, rotation shifts the whole instability domains
towards higher masses and effective temperatures and makes the instability domains wider.
For prograde sectoral modes  the shift caused by rotation is reversed.
Moreover, for $V_\mathrm{rot} = 200~\mathrm{km\,s^{-1}}$ and dipole modes with $m=0,-1$,
the instability extends beyond the mass range considered in our grid.
For retrograde modes ($\ell=1,\,m=-1$) and $V_\mathrm{rot} = 100~\mathrm{km\,s^{-1}}$,
the instability ends at $M=13\,\mathrm M_{\sun}$ and appears again at the edge of our grid
for $M=20\,\mathrm M_{\sun}$.

The boundary values of masses, $M$, effective temperatures, $\log T_\mathrm{eff}$,
luminosities, $\log L/\log \mathrm{L}_{\sun}$, and gravities, $\log g$,
for instability domains for our references models with $\ell$ up to 4,
$\left|m\right|\le\ell$ and $V_\mathrm{rot}=0$, 100
and 200 km\,s$^{-1}$ are summarized in Table\,\ref{reference_ranges}.

In the non--rotating models, increasing $\ell$
shifts the instability strip to  higher masses and higher effective temperatures.
This behaviour can be seen easily when we compare Fig.\,\ref{l1_OP_X0_7Z0_015_ov0_0} with Fig.\,\ref{l2a_OP_X0_7Z0_015_ov0_0} or
\ref{l2b_OP_X0_7Z0_015_ov0_0} where the instability strips for dipole and quadrupole modes are shown, respectively.
The instability strips for modes with $\ell=3$ and 4 are presented in Appendix \ref{appB}
(Figs.\,\ref{l3a_OP_X0_7Z0_015_ov0_0}--\ref{l4c_OP_X0_7Z0_015_ov0_0}).
For modes with $\ell\ge 2$ we did not find  upper boundaries of the mass
and effective temperature within the considered grid of models.
An exception are the modes (2,\,+2) at $V_\mathrm{rot} = 100~\mathrm{km\,s^{-1}}$.
In this case there is a gap in the instability strip, i.e. models with masses
between 14 and 17\,$\mathrm M_{\sun}$ are pulsationally stable during their whole
main--sequence evolution.

It should be mentioned that the location of TAMS and consequently our instability borders are sensitive
to the amount of overshooting from the convective core. Including overshooting 
prolongs the main sequence stage and extends instability to higher luminosities.
This effect is investigated in Section\,\ref{par_effects}.

When the effects of rotation are taken into account, increasing of
$V_\mathrm{rot}$ acts in the same direction as
increasing of the mode degree $\ell$. The exception
are the prograde sectoral modes, $\ell=m$, for which the instability strip is
shifted toward lower masses and effective temperatures, as discussed for dipole $m=1$
modes (see Fig.\,\ref{l1_OP_X0_7Z0_015_ov0_0}). This behaviour is connected with the
eigenvalue $\lambda$ which is a counterpart of the eigenvalue $\ell \left(\ell+1\right)$
in the non--rotating case \citep[e.\,g.\,][]{WD_JDD_AP2007,JDD_WD_AP2008}.
Let us remind the reader that in the traditional approximation the latitudinal relationships are
described by the Hough functions $\Theta_\lambda^m\left(\theta\right)$
which are the solutions of the tidal Laplace equation for a given eigenvalue $\lambda$.
For all modes but the prograde sectoral ones, the value of $\lambda$ is a monotonically increasing function
of the spin parameter, $s$, defined  as the doubled ratio of the rotation angular frequency to the pulsation frequency in the co--rotating
frame, $s={2\Omega}/{\omega}$. In the case of the prograde sectoral modes, $\lambda$
slowly decreases with increasing $s$, or with the rotation rate at a fixed value of the pulsation frequency
and tends to an asymptotic value.
The behaviour of $\lambda$ has an impact on the behaviour of eigenfrequencies
and has implications
for the instability of different modes. Therefore prograde sectoral modes are
less sensitive to the rotation velocity than
other pulsational modes.
An exact explanation of the instability properties
of prograde sectoral and other modes was given by
\citet{Townsend2005}.

\subsection{Impact of rotation on the range of unstable frequencies}
Now, let us discuss the values of the frequencies of excited modes.
In Figs.\,\ref{l1_OP_X0_7Z0_015_ov0_0}--\ref{l2b_OP_X0_7Z0_015_ov0_0}
and Figs.\,B1--B6, we coded in colours
the observer's frame frequencies of the most unstable modes, i.e.,
the modes for which the instability parameter $\eta$ reaches maximum.
If not stated otherwise, the frequencies in the inertial
 frame are used throughout the paper.

The mode frequencies at the maximum of $\eta$
can be treated as a kind of a mean frequency of unstable modes in a given model
for specified angular indices ($\ell,\,m$), i.e. these modes are more or less in the middle of
frequency range of unstable modes.
Sometimes, especially on the border of the instability
domains, there is only one unstable mode. However, more often there are many unstable modes.
Examples of the evolution of frequencies and their ranges for dipole modes excited in models
with masses 4, 6, and $9\,\mathrm{M_{\sun}}$ are presented in Appendix \ref{appC},
Figs.\,\ref{f_ranges_4}--\ref{f_ranges_9}. In Fig.\,\ref{f_ranges_l2-4} the evolution of
the frequencies of $\ell=2-4$ modes
for non-rotating, and $\ell=2$
for rotating $4\,\mathrm{M_{\sun}}$ models
is shown.

As was mentioned above, for a given model, there are many unstable modes
with the same angular indices ($\ell,\,m$). Therefore, in the penultimate column of Table\,\ref{reference_ranges},
there were given frequency ranges of all unstable modes. Of course, these ranges are much wider than the ranges
of frequencies at the maximum $\eta$ presented in the figures.

The range of the excited frequencies is a function of many variables. Here,
we discuss some of them.
Firstly, the frequencies of the excited modes depend on the model parameters. Generally,
the less evolved the model the higher frequencies are excited.
Moreover, for the close-to-ZAMS models the frequency values increase with decreasing mass.
This behaviour is the consequence of the sensitivity of the frequency of excited modes
to the location of the driving region. However, we would like to stress that
there are exceptions to this trend:
the most important is that the reflected modes obey it in the corotating frame only (see below).
Secondly, unstable modes with higher $\ell$ have higher frequencies.
Thirdly, rotation has a profound impact on the frequency values.
With the increasing rotation rate, modes with the same $\ell$
but different $m$ can have very different frequencies in the observer's frame.
We would like to emphasize that
for some of our prograde sectoral modes with $\ell=4$
rotation can shift
unstable g modes to very high frequencies (e.\,g.\,,
$\nu\approx 10\,\mathrm{d^{-1}}$)
which are typically not associated with the SPB variables.
Moreover, for the higher rotation rates, the  retrograde modes can be reflected, i.\,e., they have frequencies
smaller than rotation frequency and in the observer's frame they have
formally negative values, whereas they are observed as prograde modes.
Examples are the $\left(4,\,-4\right)$ modes at $V_\mathrm{rot}=200~\mathrm{km\,s^{-1}}$.

Furthermore, in the case of the axisymmetric modes, increasing the rotation velocity
acts in the same direction as increasing the mode degree. The frequencies of the $m=0$ modes with the same $\ell$
become higher as $V_\mathrm{rot}$ increases. For these modes we have a pure effect of the Coriolis force.
In the case of the sectoral prograde modes, mainly the Doppler effects is seen because their eigenvalues $\lambda$
change very slightly with rotation \citep[e.\,g.,][]{2015MNRAS.446.1438D}.

In Table\,\ref{ranges_nu_selected_models}, we give the ranges of the radial orders and frequencies of unstable
modes for a few selected models with masses 4 and 9\,$\mathrm M_{\sun}$. In general, for all modes but the sectoral prograde ones,
for higher rotation rates we have more pulsational modes in a given frequency range. However,
their stability conditions
depend on the particular model.

\subsection{Instability domains of mixed gravity--Rossby modes}

\begin{figure*}

	\includegraphics[width=1.8\columnwidth, angle=270]{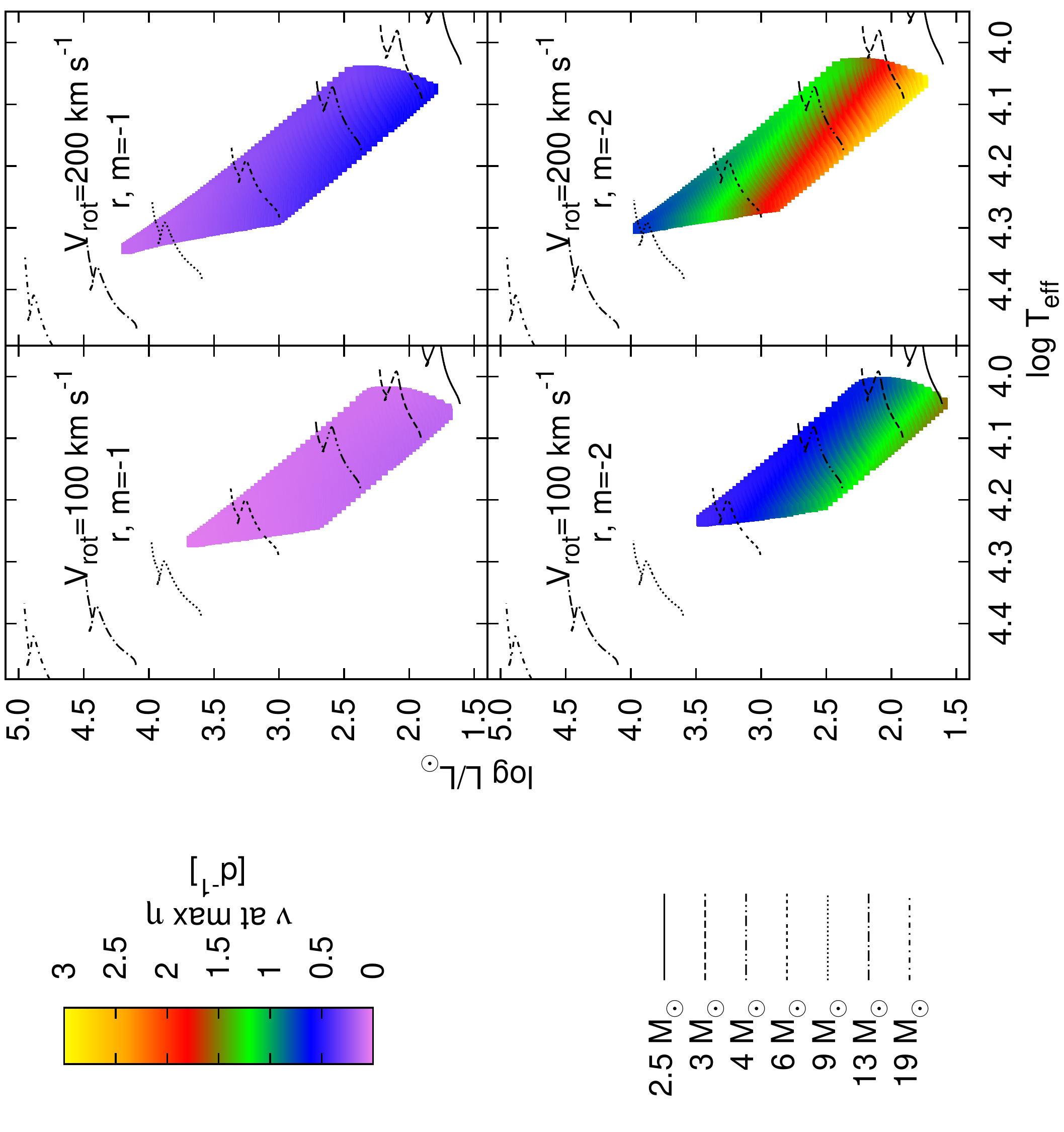}
    \caption{The same as in Fig.\,\ref{l1_OP_X0_7Z0_015_ov0_0} but for the modes with $r$ $m=-1,\,-2$
             and $V_\mathrm{rot}=100$ and 200 $\mathrm{km\,s^{-1}}$.
            }
    \label{l-1-2_OP_X0_7Z0_015_ov0_0}
\end{figure*}

Mixed gravity--Rossby  ($r$) modes
are retrograde ones with $m=-\,\ell$. They may be excited and visible in the light variations if rotation is
fast enough \citep{2005A&A...443..557S, 2005MNRAS.364..573T}.
For these modes, the restoring forces are both buoyancy and Coriolis force as compared with the sole buoyancy force
in the case of gravity modes. The $r$ modes in the considered models are driven by 
the $\kappa$-mechanism operating in the same
Z--bump layer as in the case of gravity modes.

As far as observations are concerned, \citet{2005ApJ...635L..77W} suggested that some frequency peaks
detected in HD\,163868 in the MOST data may be identified as $r$ modes. However,
\citet{WD_JDD_AP2007} showed that the whole oscillation spectrum of the star can
be explained solely by g modes. Subsequently, \citet{WS2015ident}
carried out mode identification for 31 SPB stars with available multicolour ground--based
photometry and found that scarcely two frequencies observed in  two stars may be associated,
but with rather low probability, with $r$ modes.
Nevertheless, we think that in the era of high--precision space photometry,
detection of $r$ modes is only a question of time, hence the motivation to consider them.

We computed the $r$ mode instability strips for $m=-1,\,-2,\,-3$ and $-4$
for the reference models described in the previous subsection. The results are presented
in Fig.\,\ref{l-1-2_OP_X0_7Z0_015_ov0_0} (for $m=-1$ and $-2$) as well as
in Fig.\,\ref{l-3-4_OP_X0_7Z0_015_ov0_0} (for $m=-3$ and $-4$), and are summarised in Table\,\ref{r_tab}.

\begin{table*}
	\caption{The same as in Tab.\,\ref{reference_ranges} but for $r$ modes and $V_\mathrm{rot}=100,~200$ km\,s$^{-1}$.
	}
	\label{r_tab}
	\centering
	\begin{tabular}{llcccccccc} 
		\hline
	\multicolumn{2}{l}{Grid} & $V_\mathrm{rot}$                   & mode                      & $M$      & $\log T_\mathrm{eff}$ & $\log L/\mathrm L_{\sun}$ & $\log g$ & $\nu$ &$\Omega/\Omega_\mathrm{crit}$\\
\multicolumn{2}{l}{}    & $\left[\mathrm{km\,s^{-1}}\right]$ & $\left(r,\,m \right)$ & $\left[\mathrm M_{\sun}\right]$   &                       &                           &  & $\left[\mathrm d^{-1}\right]$ &\\
		\hline
		\multirow{9}{*}{\rotatebox[origin=c]{90}{OP~~~$X_0=0.70$}} & \multirow{9}{*}{\rotatebox[origin=c]{90}{$Z=0.015~~~\alpha_\mathrm{ov}=0.0$}} & \multirow{4}{*}{100} & $\left(r,-1\right)$ & 2.65 -- 7.80 & 4.025 -- 4.268 & 1.71 -- 3.67 & 3.67 -- 4.34 & 0.0258 -- 0.2217 & 0.30 -- 0.46\\
 &    &                     & $\left( r,-2\right)$ & 2.50 -- 6.80 & 4.010 -- 4.235 & 1.61 -- 3.46 & 3.69 -- 4.34 & 0.4390 -- 1.5478 & 0.30 -- 0.46\\
 &    &                     & $\left( r,-3\right)$ & 2.45 -- 6.20 & 3.999 -- 4.210 & 1.58 -- 3.31 & 3.70 -- 4.34 & 0.8538 -- 2.8766 & 0.31 -- 0.46\\
 &    &                     & $\left( r,-4\right)$ & 2.40 -- 5.70 & 3.990 -- 4.189 & 1.54 -- 3.18 & 3.70 -- 4.34 & 1.2868 -- 4.1687 & 0.32 -- 0.47\\
 \cline{3-10}
 &    &\multirow{5}{*}{200} & $\left( r,-1\right)$ & 2.85 -- 11.00 & 4.046 -- 4.333 & 1.82 -- 4.17 & 3.56 -- 4.28 & 0.0990 -- 0.7868 & 0.56 -- 0.86\\
 &    &                     & $\left( r,-2\right)$ & 2.75 -- 9.40 & 4.033 -- 4.301 & 1.76 -- 3.94 & 3.59 -- 4.28 & 0.7133 -- 3.0166 & 0.57 -- 0.87\\
 &    &                     & $\left( r,-3\right)$ & 2.65 -- 8.40 & 4.023 -- 4.277 & 1.70 -- 3.77 & 3.60 -- 4.28 & 1.3785 -- 5.4245 & 0.58 -- 0.87\\
 &    &                     & $\left( r,-4\right)$ & 2.60 -- 7.70 & 4.014 -- 4.258 & 1.67 -- 3.64 & 3.62 -- 4.28 & 2.0822 -- 7.7980 & 0.58 -- 0.88\\
		\hline
\end{tabular}

\end{table*}

The first important finding is that with increasing rotation velocity, from $V_\mathrm{rot}=100\,\mathrm{km\,s^{-1}}$
to $200\,\mathrm{km\,s^{-1}}$, the instability strip of the $r$--modes for a given $m$ expands and is shifted towards higher masses and effective temperatures.
The second conclusion is that if we go towards more negative values of $m$, the instability
strip shrinks and moves towards lower masses and effective temperatures. In the case of
$V_\mathrm{rot}=100\,\mathrm{km\,s^{-1}}$ and the $\left(r,\,-1\right)$ modes, the instability begins at a mass $M=2.65\,\mathrm{M_{\sun}}$,
the effective temperature $\log T_\mathrm{eff}=4.0248$, and extends
over 5.15\,$\mathrm{M_{\sun}}$ in mass and about 7900\,K in effective temperature,
whereas in the case of $\left(r,\,-4\right)$ the instability begins at a mass $M=2.40\,\mathrm{M_{\sun}}$,
the effective temperature $\log T_\mathrm{eff}=3.9915$, and extends
only over 3.40\,$\mathrm{M_{\sun}}$ in mass and about 5800\,K in effective temperature.
We would like to emphasise that the lower boundary of the $\left(r,\,-4\right)$ modes is below 10\,000\,K.
Perhaps this fact partly explains the presence of  a significant number of pulsating stars
between the SPB and $\delta$ Sct instability strip found by \citet{2013A&A...554A.108M} in the young open cluster NGC\,3766.
In this region,  the instability strip of new--old class
of variables (i.e.,\, Maia stars, postulated already in the last century by \cite{Struve1955})
may be located.
The problem with the explanation of Maia stars in terms
of $r$ modes arises from their low visibility. But  we do not know
the intrinsic amplitudes of the modes.
The majority of variable stars discovered by \citet{2013A&A...554A.108M} appear to be fast rotators \citep{2016A&A...595L...1M}.
\citet{2014A&A...569A..18S} tried to explain these stars in terms of prograde sectoral and $r$ modes.
They concluded that prograde sectoral modes in combination with the gravity darkening effect and fast rotation
provide a satisfactory explanation  of these stars.

We recall the fact that the $r$ modes for a given model
are unstable in a very narrow range of frequencies,
which broadens with increasing value of $V_{\rm rot}$ and shrinks with increasing value of $|m|$ \citep[e.\,g.,][]{WD_JDD_AP2007}.
In addition, the range of frequencies of unstable modes in the whole instability domain increases with increasing $|m|$
(see Table\,\ref{r_tab}). This is due to the fact that for higher $|m|$ the frequencies of unstable modes vary more rapidly with the mass and effective
temperature of the model
than in the case of modes with lower $|m|$.



\section{The effects of input parameters on the extent of the SPB instability strip}
\label{par_effects}

\begin{table*}
	\caption{The same as in Table\,\ref{reference_ranges} but for other model grids, i.\,e.,
	         a grid with increased hydrogen abundance, $X_0=0.75$, a grid with decreased
	         metallicity, $Z=0.010$, a grid with overshooting from the convective core,
	         $\alpha_\mathrm{ov}=0.2$ and a  grid with the OPAL tables.
	}
	\label{rest_ranges}
	\centering
	\begin{tabular}{llcccccccc} 
		\hline
	\multicolumn{2}{l}{Grid} & $V_\mathrm{rot}$                   & mode                      & $M$ & $\log T_\mathrm{eff}$ & $\log L/\mathrm L_{\sun}$ & $\log g$ & $\nu$ &$\Omega/\Omega_\mathrm{crit}$\\
\multicolumn{2}{l}{}    & $\left[\mathrm{km\,s^{-1}}\right]$ & $\left( \ell,\,m \right)$ & $\left[\mathrm M_{\sun}\right]$   &                       &                           &  & $\left[\mathrm d^{-1}\right]$ &\\
		\hline
		\multirow{7}{*}{\rotatebox[origin=c]{90}{OP~~~$X_0=0.75$}} & \multirow{7}{*}{\rotatebox[origin=c]{90}{$Z=0.015~~\alpha_\mathrm{ov}=0.0$}} & 0                  &  (1)                        & 3.05 -- 10.50 & 4.037 -- 4.309 & 1.85 -- 4.03 & 3.62 -- 4.36 & 0.2008 -- 1.5589 & 0.00 \\
		                    \cline{3-10}
 &    &\multirow{3}{*}{100} &  $\left(1,\,-1\right)$ & 3.30 -- 20.00$^*$ & 4.070 -- 4.411$^*$ & 1.97 -- 4.89$^*$ & 3.41$^*$ -- 4.34 & 0.0944 -- 1.2553 & 0.27 -- 0.44\\
 &    &                     &  $\left(1,\,+0\right)$ & 3.15 -- 12.10 & 4.051 -- 4.335 & 1.90 -- 4.23 & 3.57 -- 4.34 & 0.2904 -- 1.7049 & 0.28 -- 0.45\\
 &    &                     &  $\left(1,\,+1\right)$ & 2.95 -- 9.40 & 4.025 -- 4.284 & 1.79 -- 3.86 & 3.62 -- 4.34 & 0.4947 -- 2.2073 & 0.29 -- 0.45\\
		                    \cline{3-10}
 &    &\multirow{3}{*}{200}&  $\left(1,\,-1\right)$ & 3.60 -- 20.00$^*$ & 4.093 -- 4.435$^*$ & 2.11 -- 4.89$^*$ & 3.34$^*$ -- 4.29 & (0.0774) -- 1.1671 & 0.52 -- 0.83\\
 &    &                    &  $\left(1,\,+0\right)$ & 3.35 -- 20.00$^*$ & 4.064 -- 4.412$^*$ & 1.99 -- 4.89$^*$ & 3.34$^*$ -- 4.29 & 0.2758 -- 1.9813 & 0.54 -- 0.85\\
 &    &                    &  $\left(1,\,+1\right)$ & 2.95 -- 9.80 & 4.019 -- 4.285  & 1.78 -- 3.92 & 3.56 -- 4.29 & 0.7070 -- 3.0357 & 0.56 -- 0.87\\
		\hline
		\multirow{7}{*}{\rotatebox[origin=c]{90}{OP~~~$X_0=0.70$}} & \multirow{7}{*}{\rotatebox[origin=c]{90}{$Z=0.010~~\alpha_\mathrm{ov}=0.0$}} & 0                  &  (1)                        & 2.80 -- 7.50 & 4.056 -- 4.284 & 1.87 -- 3.65 & 3.75 -- 4.41 & 0.2849 -- 1.6024 & 0.00 \\
		                    \cline{3-10}
 &  &\multirow{3}{*}{100} &  $\left(1,\,-1\right)$ & 3.00 -- 9.90 & 4.086 -- 4.341 & 1.98 -- 4.05 & 3.69 -- 4.39 & 0.1566 -- 1.3162 & 0.28 -- 0.43\\
 &  &                     &  $\left(1,\,+0\right)$ & 2.90 -- 8.40 & 4.067 -- 4.307 & 1.93 -- 3.82 & 3.71 -- 4.39 & 0.3818 -- 1.8232 & 0.29 -- 0.44\\
 &  &                     &  $\left(1,\,+1\right)$ & 2.75 -- 6.80 & 4.044 -- 4.261 & 1.87 -- 3.50 & 3.74 -- 4.39 & 0.6368 -- 2.3479 & 0.30 -- 0.45\\
		                    \cline{3-10}
 &  &\multirow{3}{*}{200}&  $\left(1,\,-1\right)$ & 3.25 -- 20.00$^*$ & 4.108 -- 4.441$^*$ & 2.10 -- 4.96$^*$ & 3.43$^*$ -- 4.34 & (0.0375) -- 1.2402 & 0.53 -- 0.82\\
 &  &                    &  $\left(1,\,+0\right)$ & 3.00 -- 10.60 & 4.079 -- 4.347 & 1.97 -- 4.15 & 3.62 -- 4.34 & 0.4527 -- 2.1763 & 0.55 -- 0.83\\
 &  &                    &  $\left(1,\,+1\right)$ & 2.75 -- 7.10 & 4.038 -- 4.261 & 1.83 -- 3.56 & 3.68 -- 4.34 & 0.9279 -- 3.3072 & 0.58 -- 0.86\\
		\hline
		\multirow{9}{*}{\rotatebox[origin=c]{90}{OP~~~$X_0=0.70$}} & \multirow{9}{*}{\rotatebox[origin=c]{90}{$Z=0.015~~\alpha_\mathrm{ov}=0.2$}} & \multirow{2}{*}{0}                  &  (1)                        & 2.75 -- 10.90 & 4.033 -- 4.323 & 1.77 -- 4.24 & 3.48 -- 4.36 & 0.1746 -- 1.6049 & 0.00 \\
 &  &                     &  (1)                        & 18.50 -- 20.00$^*$ & 4.395 -- 4.404$^*$ & 4.92 -- 5.01$^*$ & 3.29$^*$ -- 3.32 & 0.2129 -- 0.2393 & 0.00 \\
		                    \cline{3-10}
 &  &\multirow{3}{*}{100} &  $\left(1,\,-1\right)$ & 3.00 -- 20.00$^*$ & 4.067 -- 4.428$^*$ & 1.91 -- 5.01$^*$ & 3.25$^*$ -- 4.34 & 0.0762 -- 1.3118 & 0.28 -- 0.47\\
 &  &                     &  $\left(1,\,+0\right)$ & 2.85 -- 20.00$^*$ & 4.047 -- 4.411$^*$ & 1.83 -- 5.01$^*$ & 3.25$^*$ -- 4.34 & 0.2096 -- 1.7686 & 0.29 -- 0.48\\
 &  &                     &  $\left(1,\,+1\right)$ & 2.65 -- 9.30 & 4.021 -- 4.290 & 1.71 -- 4.01 & 3.49 -- 4.34 & 0.4271 -- 2.3164 & 0.29 -- 0.49\\
		                    \cline{3-10}
 &  &\multirow{3}{*}{200}&  $\left(1,\,-1\right)$  & 3.30 -- 20.00$^*$ & 4.091 -- 4.451$^*$ & 2.06 -- 5.00$^*$ & 3.10$^*$ -- 4.28 & (0.0923) -- 1.2089 & 0.53 -- 0.88\\
 &  &                     &  $\left(1,\,+0\right)$ & 3.00 -- 20.00$^*$ & 4.061 -- 4.430$^*$ & 1.90 -- 5.00$^*$ & 3.10$^*$ -- 4.28 & 0.2118 -- 2.0505 & 0.55 -- 0.90\\
 &  &                     &  $\left(1,\,+1\right)$ & 2.65 -- 10.50 & 4.015 -- 4.302 & 1.70 -- 4.18 & 3.39 -- 4.28 & 0.5787 -- 3.1558 & 0.57 -- 0.92\\
 &  &                     &  $\left(1,\,+1\right)$ & 16.00 -- 20.00$^*$ & 4.362 -- 4.394$^*$ & 4.74 -- 5.00$^*$ & 3.10$^*$ -- 3.24 & 0.3463 -- 0.4498 & 0.67 -- 0.72\\
\hline
		\multirow{7}{*}{\rotatebox[origin=c]{90}{OPAL~~~$X_0=0.70$}} & \multirow{7}{*}{\rotatebox[origin=c]{90}{$Z=0.015~~\alpha_\mathrm{ov}=0.0$}} & 0                  &  (1)    & 2.65 -- 7.10 & 4.026 -- 4.246 & 1.72 -- 3.53 & 3.68 -- 4.36 & 0.2291 -- 1.6062 & 0.00 \\
		                    \cline{3-10}
 & &\multirow{3}{*}{100}&  $\left(1,\,-1\right)$ & 2.95 -- 9.40 & 4.063 -- 4.307 & 1.90 -- 3.95 & 3.63 -- 4.33 & 0.1066 -- 1.2949 & 0.29 -- 0.45\\
 & &                    &  $\left(1,\,+0\right)$ & 2.75 -- 8.00 & 4.042 -- 4.270 & 1.78 -- 3.71 & 3.65 -- 4.34 & 0.3310 -- 1.7702 & 0.30 -- 0.46\\
 & &                    &  $\left(1,\,+1\right)$ & 2.55 -- 6.60 & 4.013 -- 4.223 & 1.65 -- 3.42 & 3.67 -- 4.34 & 0.5799 -- 2.3098 & 0.31 -- 0.47\\
		                    \cline{3-10}
 & &\multirow{3}{*}{200}& $\left(1,\,-1\right)$ & 3.20 -- 20.00$^*$ & 4.087 -- 4.422$^*$ & 2.02 -- 4.95$^*$ & 3.33$^*$ -- 4.28 & (0.0866) -- 1.1975 & 0.56 -- 0.84\\
 & &                    & $\left(1,\,+0\right)$ & 2.95 -- 10.10 & 4.057 -- 4.314 & 1.89 -- 4.05 & 3.56 -- 4.28 & 0.4007 -- 2.0681 & 0.57 -- 0.86\\
 & &                    & $\left(1,\,+1\right)$ & 2.55 -- 6.80 & 4.009 -- 4.224 & 1.65 -- 3.46 & 3.60 -- 4.28 & 0.8585 -- 3.1699 & 0.60 -- 0.89\\

		\hline
\end{tabular}

\end{table*}

The location and extent of the SPB instability strip on the H--R
diagram are sensitive to different parameters of evolutionary models.
Here, we examined the influence of the initial hydrogen abundance, $X_0$,
the metallicity, $Z$, overshooting from the convective core, $\alpha_\mathrm{ov}$, and the opacity data.
Since only dipole modes are considered, figures from Appendix \ref{appD}
should be compared with the reference models presented in
Fig.\,\ref{l1_OP_X0_7Z0_015_ov0_0}.

Increasing the initial hydrogen abundance from $X_0 = 0.70$ to $X_0 = 0.75$
(Fig.\,\ref{l1_OP_X0_7Z0_015_ov0_0} $vs.$ Fig.\,\ref{l1_OP_X0_75Z0_015_ov0_0})
shifts the instability strip to higher effective temperatures and masses.
In the non--rotating case, unstable g modes appear at a mass
higher by 0.3\,$\mathrm{M}_{\sun}$ and disappear at a mass higher by 1.4\,$\mathrm{M}_{\sun}$
compared to the reference models.
The corresponding boundaries in effective temperature are shifted only by about 125 and 185 K, respectively.

If we consider lower boundaries of mass and effective temperature
at the rotation velocity, $V_\mathrm{rot}=200~\mathrm{km\,s^{-1}}$,
the g modes become unstable from a mass higher by $0.3\,\mathrm{M_{\sun}}$
and effective temperature higher by $115$\,K for retrograde modes,
$0.35\,\mathrm{M_{\sun}}$ and $130$\,K for axisymmetric modes
and $0.3\,\mathrm{M_{\sun}}$ and $120$\,K for prograde modes,
compared to the reference models.
Upper boundaries of mass and effective temperature, within the parameter space
of our model grid, exist only for prograde modes.
The boundary parameters for the instability domains are summarised in
Table\,\ref{rest_ranges}.
As one can see, the shifts are rather small compared to those induced
by the effects of rotation. A good example is the case of the (1,\,-1) modes at 
$V_\mathrm{rot}=200~\mathrm{km\,s^{-1}}$. In this case the low mass and effective temperature
boundary is shifted by 0.55\,$\mathrm{M_{\sun}}$ and 1500 K relative to the non--rotating models.

It should be noted that the difference between our reference models and those with increased
hydrogen abundance is associated with the change in the amount of the nuclear fuel
and with the change of the free-free opacities in the central layers
of the star. This results in shifting the evolutionary tracks on the H--R
diagram; for a
given effective temperature and luminosity, an increase of the hydrogen abundance
reduces the mass of the corresponding model. However, this has a little impact on pulsation driving by the
$\kappa$-mechanism.

The effect of decreasing the metallicity on the instability domain of dipole modes is shown in Fig.\,\ref{l1_OP_X0_7Z0_010_ov0_0}
where results for $Z=0.010$ are depicted. Since the g--modes are excited by the $\kappa$-mechanism acting
on the Z opacity bump, it is not a surprise that lower metallicity reduces the size of the instability strip.
In the non--rotating case the g modes become unstable for masses
higher by 0.05\,$\mathrm{M_{\sun}}$ and effective temperature higher by
610\,K and are stabilized for masses lower by 1.6\,$\mathrm{M_{\sun}}$
and effective temperature lower by 955\,K compared to our reference grid.
When a star rotates with $V_\mathrm{rot}=200~\mathrm{km\,s^{-1}}$, the
low effective temperature boundaries of the instability strip for all dipole modes are shifted to values higher
by about 550\,K relative to our reference grid.

Moreover, the instability extension to high masses observed for
$Z=0.015$ for axisymmetric modes disappears.
In the case of retrograde modes, the instability appears in the whole range
masses above $M=3.25\,\mathrm{M_{\sun}}$. Similarly,
only the lower boundary of effective temperature, $\log T_\mathrm{eff}=4.108$, can be defined
for the parameters of our grid of models (see also Table\,\ref{rest_ranges}).
Finally, in the case of prograde modes, the high mass and effective temperature instability border is shifted to lower
values by 1.5\,$\mathrm{M_{\sun}}$ and 905\,K, respectively.

The effect of overshooting from the convective core is shown in Fig.\,\ref{l1_OP_X0_70Z0_015_ov0_2}.
We applied the overshooting law  proposed by \citet{2008MNRAS.385.2061D},
which is more smooth than usual step overshooting.
The calculations were performed with overshooting parameter, $\alpha_\mathrm{ov} = 0.2$.
The convective core overshooting considerably prolongs the main sequence evolution 
causing wider instability domains. It means that for a given effective temperature, the
instability strip reaches higher luminosities compared to the one found for the
reference models. Moreover, overshooting expands the pulsational instability to higher luminosities,
effective temperatures and masses.
In the  non--rotating models with $\alpha_\mathrm{ov} = 0.2$, unstable dipole modes
disappear at 10.9\,$\mathrm{M_{\sun}}$ and appear again
at  about 18.5\,$\mathrm{M_{\sun}}$ in contrast to our reference grid.
As one can conclude from Fig.\,D3, the lower limit of the instability in the effective temperature is almost unchanged.

The instability domains for the dipole g modes obtained with the OPAL opacity
tables are shown in Fig.\,\ref{l1_OPAL_X0_7Z0_015_ov0_0} (see also Table\,\ref{rest_ranges}).
In comparison with our reference models they are smaller.
The upper and lower boundaries of masses and effective temperatures
are shifted towards smaller values.
This is a well--known fact resulting from the deeper location of the Z--bump in the OP data compared
to the OPAL data
\citep[e.\,g.,][]{Pamyatnykh1999}.

For the zero--rotation models, the instability
is shifted at the low mass border by 0.1\,$\mathrm{M_{\sun}}$
and 150\,K and at the high mass border by 2\,$\mathrm{M_{\sun}}$
and 2565\,K.

In the models with $V_\mathrm{rot}=200~\mathrm{km\,s^{-1}}$,
we found an extension of the instability strip to higher masses for retrograde modes but
not for axisymmetric modes.
In comparison with our reference grid, the low temperature instability boundary is shifted towards lower values
by 0.1\,$\mathrm{M_{\sun}}$ and 55\,K for retrograde modes,
by 0.05\,$\mathrm{M_{\sun}}$ and 55\,K for axisymmetric modes and
by 0.1\,$\mathrm{M_{\sun}}$ and 120\,K for prograde modes.

Let us now discuss effects of different parameters
on the values of unstable mode frequencies. For the purpose of clarity and
brevity we focus only on dipole modes in non--rotating models. Let us recall the that
dipole unstable modes in our reference models have frequencies
in the range $\nu=0.2162 - 1.6026\,\mathrm{d^{-1}}$. Decreasing metallicity from
$Z=0.015$ to $Z=0.010$ shifts the minimum frequency to slightly higher values by 0.07$\,\mathrm{d^{-1}}$,
whereas the maximum frequency remains almost unchanged.
A higher initial hydrogen abundance, $X_0=0.75$,
shifts the minimum and maximum frequencies towards lower values, by 0.02 and 0.04$\,\mathrm{d^{-1}}$,
respectively. Assuming $\alpha_\mathrm{ov}=0.2$ lowers the minimum frequency (by 0.04$\,\mathrm{d^{-1}}$) but does not change the highest frequency.
Finally, using the OPAL tables results in increasing
the minimum frequency by 0.01$\,\mathrm{d^{-1}}$
whereas the value of the maximum frequency remains almost unchanged.
As one can see, all these effects are noticeable but far smaller than the effects of rotation.

\section{Instability domains of the dipole modes for fixed value of $\Omega/\Omega_\mathrm{crit}$}
\label{fixed_omega_section}
\begin{figure*}
	\includegraphics[width=2.7\columnwidth, angle=270]{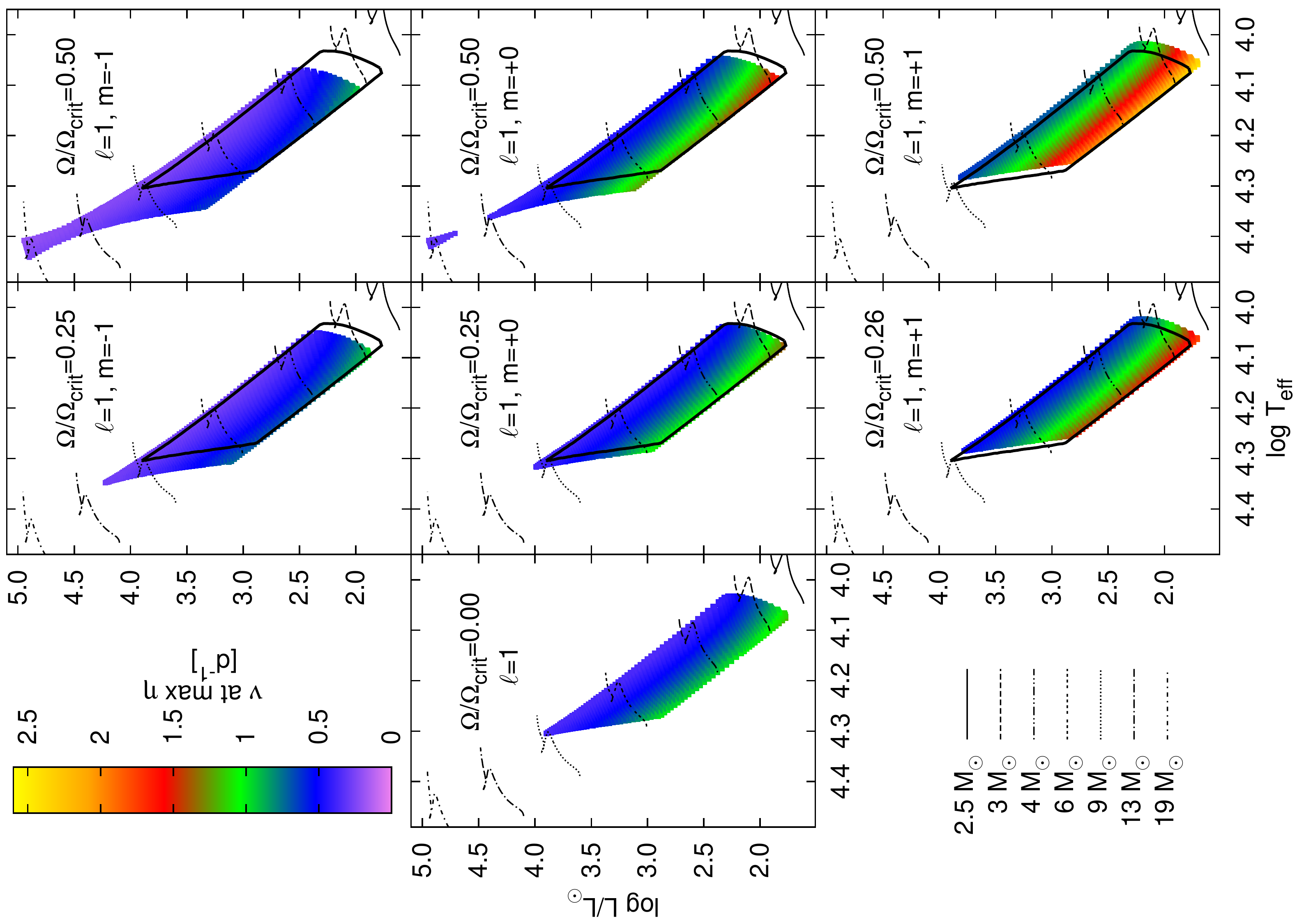}
    \caption{The same as in Fig.\,\ref{l1_OP_X0_7Z0_015_ov0_0} but for the fixed ratio of the angular rotation velocity to its
    critical value, $\Omega/\Omega_\mathrm{crit}=0$, 0.25 and 0.5.
            }
    \label{l1_OP_X0_7Z0_015_ov0_0_Om}
\end{figure*}

In order to compare our results with those of \citet{Townsend2005}
and  to show how non--constant $\Omega/\Omega_\mathrm{crit}$ affects instability strips, we computed
a grid of models with the values of $\Omega/\Omega_\mathrm{crit}$ fixed at 0, 0.25 and 0.50.
The remaining parameters were the same as for the reference models (see Section\,\ref{sub_sec_g}).
The computations were limited to the dipole modes.
The results are shown in Fig.\,\ref{l1_OP_X0_7Z0_015_ov0_0_Om} and are summarised in Table\,\ref{fixed_omega_ranges}.

\begin{table*}
	\caption{The same as in Tab.\,\ref{reference_ranges} but for fixed ratios of angular rotation rates to their critical values, $\Omega/\Omega_\mathrm{crit}=0.00$,
	0.25, 0.50, and for dipole modes. In the last column there are given the ranges of the
	equatorial velocity.
	}
	\label{fixed_omega_ranges}
	\centering
	\begin{tabular}{llcccccccc} 
		\hline
		\multicolumn{2}{l}{Grid} & $\Omega/\Omega_\mathrm{crit}$& mode                      & $M$    & $\log T_\mathrm{eff}$ & $\log L/\mathrm L_{\sun}$ & $\log g$ & $\nu$ &$V_\mathrm{rot}$ \\
		    & & & $\left( \ell,\,m \right)$  & $\left[\mathrm M_{\sun}\right]$   &                       &                           &  & $\left[\mathrm d^{-1}\right]$ & [km\,s$^{-1}$]\\
		\hline
		\multirow{8}{*}{\rotatebox[origin=c]{90}{OP~~~$X_0=0.70$}} & \multirow{8}{*}{\rotatebox[origin=c]{90}{$Z=0.015~~~\alpha_\mathrm{ov}=0.0$}} & \multirow{1}{*}{0.00}  & (1)  & 2.75 -- 9.10 & 4.032 -- 4.305 & 1.77 -- 3.90 & 3.67 -- 4.36 & 0.2162 -- 1.6026 &  0\\
		                    \cline{3-10}
& &\multirow{3}{*}{0.25}&  $\left(1,\,-1\right)$ & 2.95 -- 11.40 & 4.051 -- 4.348 & 1.89 -- 4.22 & 3.61 -- 4.35 & 0.1454 -- 1.3268 & 55 -- 88\\
& &                     &  $\left(1,\,+0\right)$ & 2.80 -- 9.70  & 4.038 -- 4.317 & 1.80 -- 3.99 & 3.64 -- 4.35 & 0.2678 -- 1.7010 & 54 -- 86\\
& &                     &  $\left(1,\,+1\right)$ & 2.65 -- 8.40  & 4.023 -- 4.285 & 1.71 -- 3.78 & 3.66 -- 4.35 & 0.4124 -- 2.1157 & 54 -- 84 \\
		                    \cline{3-10}
& &\multirow{4}{*}{0.50}&  $\left(1,\,-1\right)$ & 3.15 -- 20.00$^*$  & 4.069 -- 4.441$^*$ & 1.99 -- 4.94$^*$ & 3.31$^*$ -- 4.32 & 0.0623 -- 1.2546 & 112 -- 183 \\
& &                     &  $\left(1,\,+0\right)$ & 2.90 -- 13.00      & 4.047 -- 4.362     & 1.85 -- 4.40     & 3.51 -- 4.32     & 0.3279 -- 1.9390 & 110 -- 177 \\
& &                     &  $\left(1,\,+0\right)$ & 16.50 -- 20.00$^*$ & 4.394 -- 4.420$^*$ & 4.71 -- 4.94$^*$ & 3.31$^*$ -- 3.42 & 0.2641 -- 0.3406 & 142 -- 148 \\
& &                     &  $\left(1,\,+1\right)$ & 2.65 -- 8.60       & 4.018 -- 4.284     & 1.70 -- 3.81     & 3.61 -- 4.32     & 0.6023 -- 2.7352 & 108 -- 169 \\
 		
		\hline
\end{tabular}
\end{table*}

As one can see, the instability strips are slightly smaller than those computed for the fixed values of the equatorial velocity.
This is due to the fact that the values of $V_\mathrm{rot}$  corresponding to the $\Omega/\Omega_\mathrm{crit}$ values
used in this section are smaller
than those used in preceding sections (see the last column of Table\,\ref{fixed_omega_ranges}).
Thus, the impact of rotation is smaller.

Our instability strips are much larger than those obtained by \citet{Townsend2005}.
In particular, we got much larger extension towards higher masses and higher effective temperatures on the
H--R diagram.
This is mainly due to the difference in the opacity data; \citet{Townsend2005} used the OPAL tables while we, the OP data.
It has been already shown by \citet{Pamyatnykh1999} for non--rotating models that such an extension of the instability strip exists
if the OP opacities are used.
Some differences can result also from adopting various chemical mixtures: GN93 in Townsend's computations $vs.$ AGSS09 in ours.

Besides, we computed and compared instability strips for the non--rotating and rotating evolutionary models.
It turns out that even such simple incorporation of rotation in the equilibrium models
as we have done (see Section\,\ref{sub_sec_g}) affects the instability strips.
Comparing the Townsend's and our instability strips, the reader has to bear in mind these differences.

\section{Influence of rotation on period spacing of high radial--order g modes}
\label{DP}

For the present paper, the period spacing of high--order g modes is a side issue but
recent discoveries of regular period patterns in B--type stars from space data
\citep{Degroote2010N, Papics2012,Papics2014,Ppics2015} make it highly important.
Moreover, the period spacing of g modes has been never discussed for
massive stars such as $\beta$ Cephei variables ($M=8 -16\,\mathrm M_\odot$).
Here, we would like to emphasize that one should be careful while interpreting
dense oscillation spectra obtained from space photometry. A good example
is HD\,50230: \citet{Degroote2010N,2012A&A...542A..88D} claimed that they found regular period spacing in the CoRoT data,
which is a manifestation of asymptotic behaviour,
but our studies \citep{2014IAUS..301..109S} suggested rather that this regularity is spurious,
i.e., the oscillation spectrum is composed of modes with various $(\ell, m)$, and can not be
interpreted according to the asymptotic theory.

For high radial--order g modes asymptotic theory predicts that period spacing,
$\Delta P$, defined as difference between periods of modes with the same spherical
harmonic degree, $\ell$, and consecutive radial orders, $n$,
is constant \citep{1980ApJS...43..469T},
\begin{equation}
\Delta P_\ell =P_{n+1} - P_n = \frac{2\pi^2}{\sqrt{\ell\left(\ell+1\right)} \int_{r_1}^{r_2} N\,d\ln r} = \mathrm{const},
\end{equation}
where $r_1$ and $r_2$ are the inner
and outer turning points of the mode propagation cavity and $N$ is
the Brunt-V\"ais\"al\"a frequency which depends only on the equilibrium
model. However, this result was obtained under zero--rotation and
uniform chemical composition assumptions. When rotation is taken into
account, the eigenvalue $\ell\left(\ell+1\right)$ has to be replaced by
$\lambda$ and  $\Delta P$ in the co--rotating frame is given by \citep[e.\,g.,\,][]{2013MNRAS.429.2500B}
\begin{equation}
\Delta P_{\ell,\,m} = \frac{2\pi^2}{\sqrt{\lambda_{\ell,\,m,\,s\left(n+1\right)}} \int_{r_1}^{r_2} N\,d\ln r
~\left( 1+\frac{1}{2} \frac{d\,\ln \lambda_{\ell,\,m,\,s\left(n\right)}}{d\,\ln s} \right)},
\end{equation}
where subscripts in $\lambda$ were added to emphasize the dependence on the angular indices
and the spin parameter, which is itself a function of rotation and mode frequency.
Therefore, in the case of rotating models $\Delta P$, is no longer constant.
Naturally, rotation changes also the equilibrium model and hence $N$.

\begin{figure*}
	\includegraphics[width=2.7\columnwidth, angle=270]{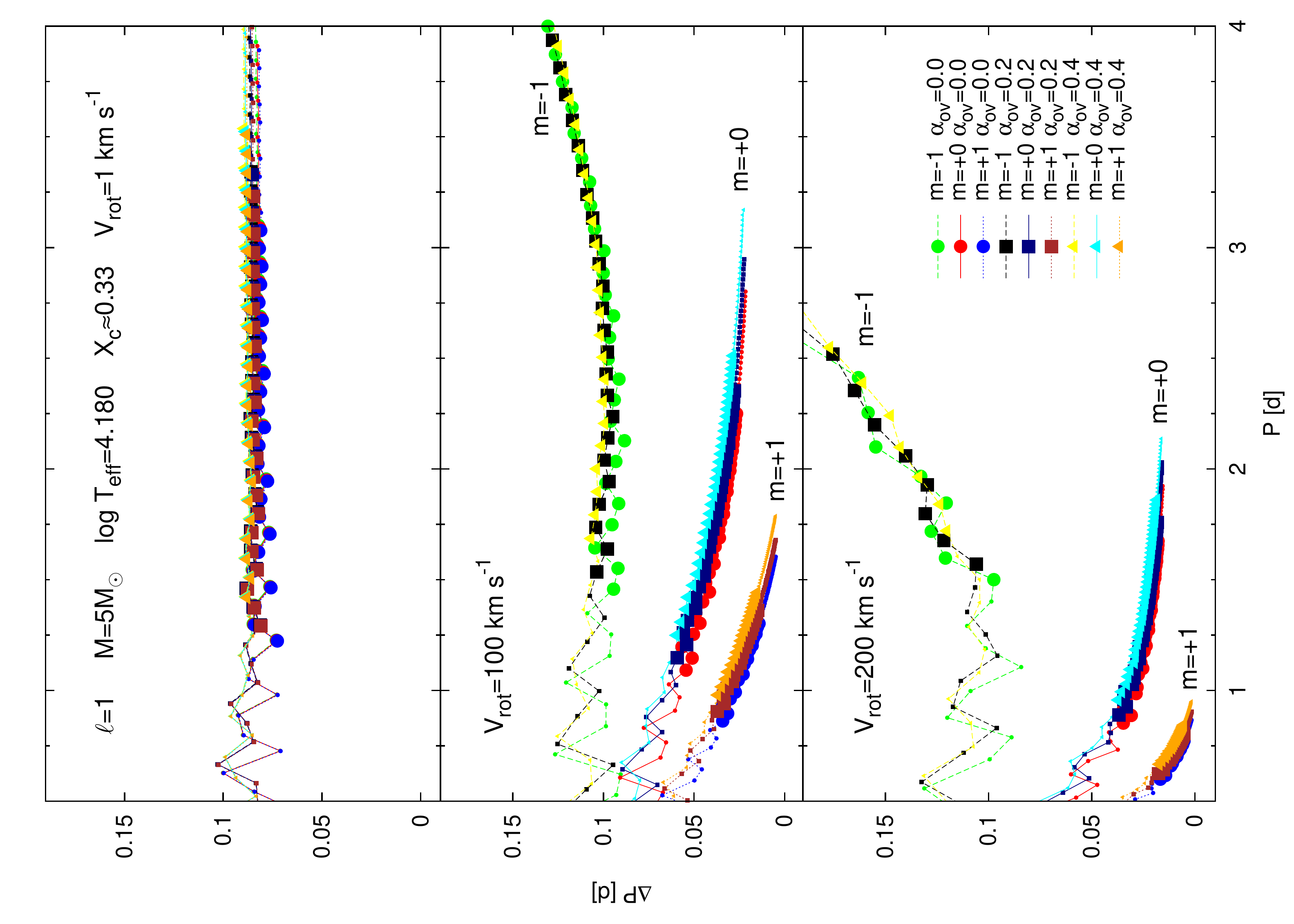}
    \caption{The period spacing for the dipole modes with $m=-1,\,0$ and $+1$ in stellar model with
             $M=5\mathrm{M_{\sun}}$, $\log T_\mathrm{eff}=4.180$,
             $Z=0.017$, $X_c\approx 0.33$, computed with the OP opacities and the AGSS09
             chemical mixture. The three values of the rotation velocity, $V_{\rm rot}=1,\,100,\,200$ km\,s$^{-1}$,
             and three values of the core overshooting parameter, $\alpha_\mathrm{ov}=0.0,\,0.2,\,0.4$, were examined.
             Pulsationally unstable modes are marked by larger symbols.
            }
    \label{PDP_5M}
\end{figure*}

\begin{figure*}
	\includegraphics[width=2.7\columnwidth, angle=270]{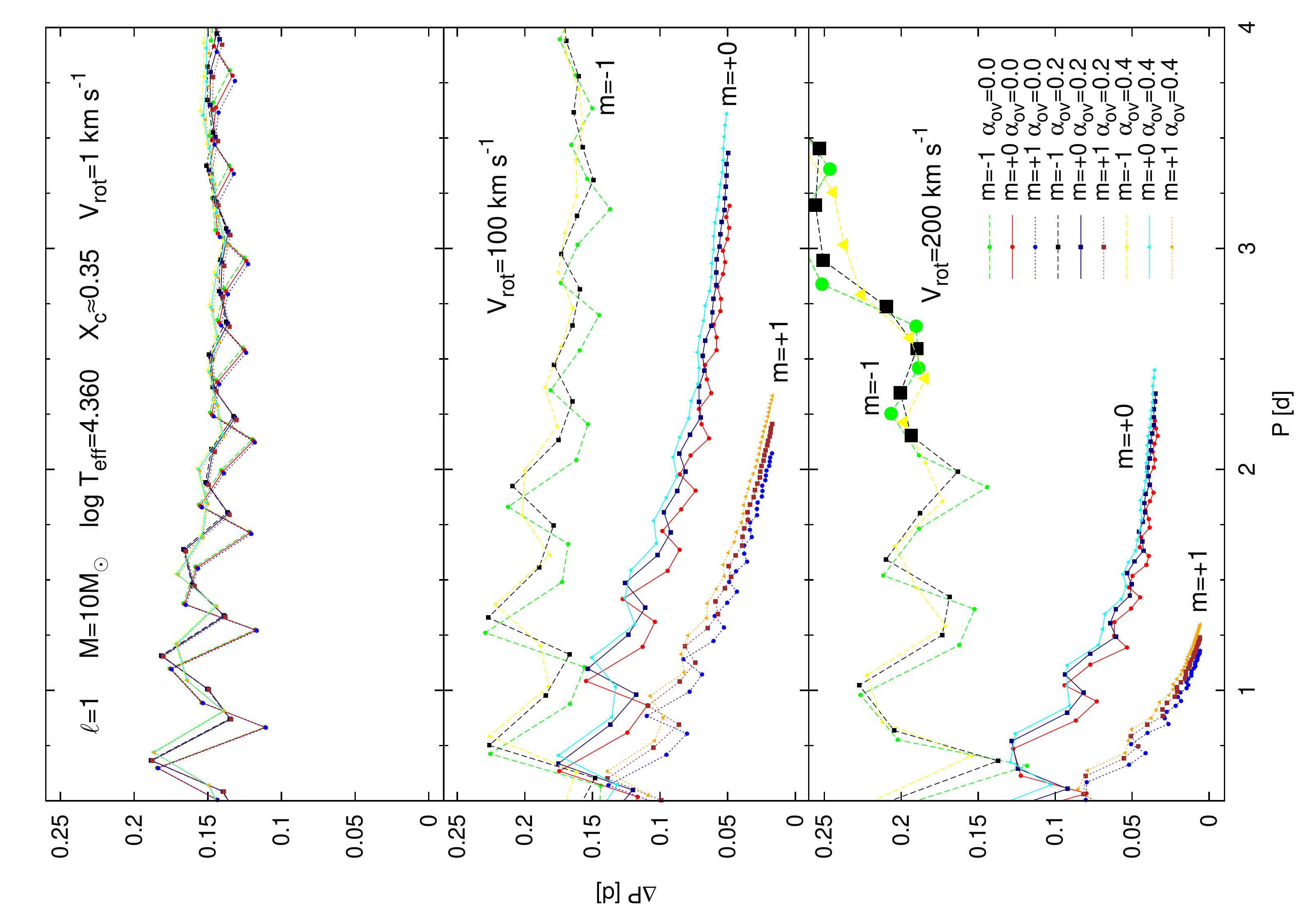}
    \caption{The same as in Fig.\,\ref{PDP_5M} but for $10\,\mathrm M_{\sun}$ model
             with $\log T_\mathrm{eff}=4.360$ and $X_c\approx 0.35$.
            }
    \label{PDP_10M}
\end{figure*}

An appropriate determination of $\Delta P$ is important because it is used
in mode identification as well as an indicator of chemical abundance
gradient on the boundary  of the convective core,
e.\,g.,\, in the case of white dwarfs \citep{1991ApJ...378..326W}
and recently in SPB stars \citep{Degroote2010N,Papics2014,Ppics2015}.
The behaviour of $\Delta P$ for the SPB--like models was studied also
by \citet{WD_PM_AP1993} for non--rotating models and by \citet{2012ASPC..462..103A}
for rotating models but with rather low rotation velocity.
Moreover, \citet{2008MNRAS.386.1487M} derived an analytical approximation for the high--order g--mode periods that
takes into account the effect of the chemical composition gradient near the core, ignoring all effects of rotation.
For the zero-rotation case, their results are compatible with ours but since the authors used different masses
we cannot make an adequate comparison.

In the present paper, we study the effects of rotation on the values of $\Delta P$ in the framework
of the traditional approximation. In Fig.\,\ref{PDP_5M}, we show $\Delta P$ as a function of period for dipole modes
in the model with  $M=5\,\mathrm{M_{\sun}}$ and $\log T_\mathrm{eff}=4.180$.
The calculations were performed for three values of the equatorial rotation: $V_\mathrm{rot}=1,\,100$ and 200
km\,s$^{-1}$. We considered also three values of core overshooting: $\alpha_\mathrm{ov}=0.0,\,0.2$ and 0.4.
As can be seen from Fig.\,\ref{PDP_5M},
the highest value of $\Delta P$ is obtained for retrograde sectoral modes $\left(m=-\ell\right)$
and the lowest for prograde sectoral modes $\left(m=\ell\right)$.

With increasing rotation, the deviation from the constant value of $\Delta P$ becomes more pronounced.
An impact of convective core overshooting is rather minor and it is quantitatively comparable to the
effect of slow rotation of the order of 1\,km\,s$^{-1}$.
Moreover, for intermediate radial orders one can clearly see the oscillatory behaviour of $\Delta P$,
a result already predicted by \citet{WD_PM_AP1993}.
The amplitude of this oscillation depends on the evolutionary stage, and more
precisely, on the chemical composition gradient above the convective core.
Furthermore, the amplitude of these oscillations decreases if core overshooting
(as described by \citet{2008MNRAS.385.2061D}) is included,
this is a natural consequence of adding any partial mixing which blurs the chemical composition gradients.

In general, $\Delta P$ is a function of many variables; the most important are
$\left(\ell,\,m\right)$, $V_\mathrm{rot}$, $M$, $T_\mathrm{eff}$, $Z$, $X_0$,
evolutionary stage or core overshooting.
However, except for very slow rotation, the impact of $V_\mathrm{rot}$ overwhelms the effects of other parameters.

In a similar way as for the $5\,\mathrm{M_{\sun}}$ model,
we tested the effect of rotation and core overshooting on $\Delta P$
in the more massive model with $M=10\,\mathrm{M_{\sun}}$ and $\log T_\mathrm{eff}=4.360$
(see Fig.\,\ref{PDP_10M}). As one can see from the figure, the qualitative properties are quite similar
to those seen in Fig.\,\ref{PDP_5M}
but the values of the mean period spacing are higher for higher--mass models.

The period spacings were modeled for, eg., KIC\,10526294 \citep{2015A&A...580A..27M} or KIC\,7760680 \citep{2016ApJ...823..130M}
without and with the effects of rotation taken into account, respectively.
The main problem in these studies was related to the instability conditions for some observed frequencies.
As has been shown recently by \citet{2017arXiv170101256S}, this problem can be solved by an appropriate enhancement of the opacities.
In \citet{2017arXiv170101256S}, we successfully reproduced both the value of the period spacing and the frequency range of unstable modes
in the rotating SPB star KIC\,7760680.
Earlier, \citet{2013A&A...559A..25S} did not succeed in reproducing the period spacing in HD\,43317.
Let us repeat that it is easy to confuse period spacing resulting from asymptotic behaviour with
accidental distribution of frequencies in dense oscillation spectrum \citep[see][]{2014IAUS..301..109S}.

\section{Summary}
\label{conclusions}

We presented results of extensive computations of gravity and mixed gravity--Rossby
modes instability domains
for stellar models with masses 2--20\,$\mathrm M_{\sun}$. The effects of rotation on pulsations were included
using the traditional approximation which can be safely applied for slow--to--moderate rotators.
We considered high--order g modes with  $\ell$ up to 4 and mixed gravity--Rossby modes
with $|m|$ up to 4. The latter modes become propagative only in the presence of rotation.

We relied on the equilibrium models computed with the Warsaw--New Jersey code which takes into account the mean effects
of the centrifugal force whereas all effects of rotationally-induced mixing are ignored.
Our results give a good qualitative picture of pulsational instability of slow modes in rotating B--type main sequence models.
For a more detailed seismic analysis of individual objects, a more advanced evolutionary code, e.g., MESA, should be used.

We limited our computations to the rotation rate $\Omega \lesssim 0.7\Omega_\mathrm{crit}$. This value was exceeded for a small number of evolved models
with lowest masses. In these cases the frequencies and the instability parameter can be inaccurate.

In comparison with the results of \citet{Townsend2005}, we obtained the g--mode instability domains
on the H--R diagram much more extended towards
higher masses and higher effective temperatures.
This is mainly due to the difference in the opacity data (OPAL $vs.$ OP).
We found that in the case of rotating models, the extension occurs also
for the  OPAL opacity data (cf. Fig.\,D4 for $V_\mathrm{rot}=200~\mathrm{km\,s^{-1}}$).
Thus, another reason for the difference between our and Townsend's results is the higher rotation rate we used.
Some differences may have  arisen from adopting different chemical mixtures (GN93  $vs.$ AGSS09).
Finally, including  mean effects of the centrifugal force in our equilibrium models
also had a not negligible effect.

We would like to emphasize that in the rotating models, the unstable prograde
high radial--order g modes may have quite high frequencies, typically not associated with SPB--like pulsation.
This fact is especially important in the era of high precision space photometry
where modes with higher degree, $\ell \ge 3$, can be detected.
The important result is also that for the rotating models,
we obtained a wider instability strip for a given $\ell$ than in the non--rotating case.
Moreover, the shift of the lower boundary of the effective temperature to the lower values
for  prograde sectoral modes ($\ell=m$), combined with their high frequencies caused by
rotation, can possibly explain the existence of some Maia stars \citep[e.\,g.,][]{2015MNRAS.451.1445B}.
Moreover, $r$ modes can also, at least partially, fill the gap between the SPB and $\delta$ Sct instability
domains, exactly where \citet{2013A&A...554A.108M} found a new class of pulsating stars.
The role of the fast rotation in the phenomenon of pulsating stars located
between the SPB and $\delta$ Scuti instability strip was already discussed in the literature
\citep[e.g.][]{2013A&A...554A.108M,2014A&A...569A..18S,2016A&A...595L...1M, 2017arXiv170100937D,2017arXiv170202306S}.

Variable stars similar  to those of \citet{2013A&A...554A.108M} were also observed in NGC\,457
\citep{2014AcA....64...89M} and NGC\,884 \citep{2010A&A...515A..16S, 2013AJ....146..102S}.
Some of them exhibit much too high frequencies for the SPB stars or seem to lie below the classical
SPB instability strip. Using parameters given by \citet{2013AJ....146..102S}, we plotted variables observed in
NGC\,884 on the H--R diagram and compared with our instability strips. It turned out that the location 
and the high frequencies of all but one Maia-like star can be explained by prograde sectoral modes
excited in fast rotating models of SPB stars (see also Appendix \ref{appE}).

Recently, \citet{WS2015ident} compiled the parameters of SPB stars.
The observed parameters of the SPB stars in  their list are consistent
in terms of both the position on the H--R diagram and the excited frequencies
with the calculations presented in the present paper.

There is also well known problem with the so called macroturbulence, i.e.
the line--profile bradening cause by other factors than rotation \citep[e.g.][]{2014A&A...562A.135S}.
Recently, \citet{2009A&A...508..409A} showed that macroturbulence could be a signature of collective effect
of pulsations. However, \citet{2017A&A...597A..22S, 2017A&A...597A..23G} argued that alone heat driven pulsations
can not explain the occurrence of  macroturbulence. Not all stars with macroturbulent
broadening fall into instability strips. In Section \ref{sec_l_1_4}
we obtained wider instability strips than \citet{2017A&A...597A..22S}
and \citet{2017A&A...597A..23G}. In particular, we showed
that the instability of high--order g modes begins at earlier evolutionary stage for massive stars
than found by
\citet{2017A&A...597A..22S} and \citet{2017A&A...597A..23G}.
These differences are caused mainly by the circumstance that they used an older chemical mixture (GN93) and ignored the effects of rotation.
Taking into account modes with $\ell$ up to 20 as was done by \citet{2017A&A...597A..23G}
could give us even more extend instability domains than those presented in the present paper.
Unfortunately, we did not study instability condition beyond the main sequence where
many stars with significant macroturbulent broadening are located.
This task is planned for the near future.

Finally, we showed that the initial hydrogen abundance, metallicity, overshooting from the convective
core and a source of the opacity data have a minor influence on the extent of the SPB instability
domains in comparison with the effects of rotation.

\section*{Acknowledgements}
We thank Miko{\l}aj Jerzykiewicz for careful reading of the manuscript and language corrections.
This work was financially supported by the Polish National Science Centre grant 2015/17/B/ST9/02082.
Calculations have been carried out using
resources provided by Wroc{\l}aw Centre for Networking and Supercomputing (http://wcss.pl), grant no. 265.






\bibliographystyle{mnras}
\bibliography{szewczuk}

\begin{thebibliography}{}
\makeatletter
\relax
\def\mn@urlcharsother{\let\do\@makeother \do\$\do\&\do\#\do\^\do\_\do\%\do\~}
\def\mn@doi{\begingroup\mn@urlcharsother \@ifnextchar [ {\mn@doi@}
  {\mn@doi@[]}}
\def\mn@doi@[#1]#2{\def\@tempa{#1}\ifx\@tempa\@empty \href
  {http://dx.doi.org/#2} {doi:#2}\else \href {http://dx.doi.org/#2} {#1}\fi
  \endgroup}
\def\mn@eprint#1#2{\mn@eprint@#1:#2::\@nil}
\def\mn@eprint@arXiv#1{\href {http://arxiv.org/abs/#1} {{\tt arXiv:#1}}}
\def\mn@eprint@dblp#1{\href {http://dblp.uni-trier.de/rec/bibtex/#1.xml}
  {dblp:#1}}
\def\mn@eprint@#1:#2:#3:#4\@nil{\def\@tempa {#1}\def\@tempb {#2}\def\@tempc
  {#3}\ifx \@tempc \@empty \let \@tempc \@tempb \let \@tempb \@tempa \fi \ifx
  \@tempb \@empty \def\@tempb {arXiv}\fi \@ifundefined
  {mn@eprint@\@tempb}{\@tempb:\@tempc}{\expandafter \expandafter \csname
  mn@eprint@\@tempb\endcsname \expandafter{\@tempc}}}

\bibitem[\protect\citeauthoryear{{Aerts} \& {Dupret}}{{Aerts} \&
  {Dupret}}{2012}]{2012ASPC..462..103A}
{Aerts} C.,  {Dupret} M.-A.,  2012, in {Shibahashi} H.,  {Takata} M.,
  {Lynas-Gray} A.~E.,  eds,  Astronomical Society of the Pacific Conference
  Series Vol. 462, Progress in Solar/Stellar Physics with Helio- and
  Asteroseismology. p.~103 (\mn@eprint {arXiv} {1108.6248})

\bibitem[\protect\citeauthoryear{{Aerts}, {Puls}, {Godart}  \&
  {Dupret}}{{Aerts} et~al.}{2009}]{2009A&A...508..409A}
{Aerts} C.,  {Puls} J.,  {Godart} M.,   {Dupret} M.-A.,  2009, \mn@doi [\aap]
  {10.1051/0004-6361/200810471}, \href
  {http://cdsads.u-strasbg.fr/abs/2009A%26A...508..409A} {508, 409}

\bibitem[\protect\citeauthoryear{{Aprilia}, {Lee}  \& {Saio}}{{Aprilia}
  et~al.}{2011}]{2011MNRAS.412.2265A}
{Aprilia} {Lee} U.,   {Saio} H.,  2011, \mn@doi [\mnras]
  {10.1111/j.1365-2966.2010.18048.x}, \href
  {http://cdsads.u-strasbg.fr/abs/2011MNRAS.412.2265A} {412, 2265}

\bibitem[\protect\citeauthoryear{{Asplund}, {Grevesse}  \& {Sauval}}{{Asplund}
  et~al.}{2005}]{2005ASPC..336...25A}
{Asplund} M.,  {Grevesse} N.,   {Sauval} A.~J.,  2005, in {Barnes} III T.~G.,
  {Bash} F.~N.,  eds,  Astronomical Society of the Pacific Conference Series
  Vol. 336, Cosmic Abundances as Records of Stellar Evolution and
  Nucleosynthesis. p.~25

\bibitem[\protect\citeauthoryear{{Asplund}, {Grevesse}, {Sauval}  \&
  {Scott}}{{Asplund} et~al.}{2009}]{Asplund2009}
{Asplund} M.,  {Grevesse} N.,  {Sauval} A.~J.,   {Scott} P.,  2009, \mn@doi
  [\araa] {10.1146/annurev.astro.46.060407.145222}, \href
  {http://cdsads.u-strasbg.fr/abs/2009ARA%26A..47..481A} {47, 481}

\bibitem[\protect\citeauthoryear{{Bahcall}, {Pinsonneault}  \&
  {Wasserburg}}{{Bahcall} et~al.}{1995}]{1995RvMP...67..781B}
{Bahcall} J.~N.,  {Pinsonneault} M.~H.,   {Wasserburg} G.~J.,  1995, \mn@doi
  [Reviews of Modern Physics] {10.1103/RevModPhys.67.781}, \href
  {http://cdsads.u-strasbg.fr/abs/1995RvMP...67..781B} {67, 781}

\bibitem[\protect\citeauthoryear{{Bailey} et~al.,}{{Bailey}
  et~al.}{2015}]{2015Natur.517...56B}
{Bailey} J.~E.,  et~al., 2015, \mn@doi [\nat] {10.1038/nature14048}, \href
  {http://cdsads.u-strasbg.fr/abs/2015Natur.517...56B} {517, 56}

\bibitem[\protect\citeauthoryear{{Ballot}, {Ligni{\`e}res}, {Prat}, {Reese}  \&
  {Rieutord}}{{Ballot} et~al.}{2012}]{Ballot2012}
{Ballot} J.,  {Ligni{\`e}res} F.,  {Prat} V.,  {Reese} D.~R.,   {Rieutord} M.,
  2012, in {Shibahashi} H.,  {Takata} M.,   {Lynas-Gray} A.~E.,  eds,
  Astronomical Society of the Pacific Conference Series Vol. 462, Progress in
  Solar/Stellar Physics with Helio- and Asteroseismology. p.~389

\bibitem[\protect\citeauthoryear{{Ballot}, {Ligni{\`e}res}  \&
  {Reese}}{{Ballot} et~al.}{2013}]{2013LNP...865...91B}
{Ballot} J.,  {Ligni{\`e}res} F.,   {Reese} D.~R.,  2013, in {Goupil} M.,
  {Belkacem} K.,  {Neiner} C.,  {Ligni{\`e}res} F.,   {Green} J.~J.,  eds,
  Lecture Notes in Physics, Berlin Springer Verlag Vol. 865, Lecture Notes in
  Physics, Berlin Springer Verlag. p.~91, \mn@doi{10.1007/978-3-642-33380-4_5}

\bibitem[\protect\citeauthoryear{{Balona} et~al.,}{{Balona}
  et~al.}{2011}]{Balona2011}
{Balona} L.~A.,  et~al., 2011, \mn@doi [\mnras]
  {10.1111/j.1365-2966.2011.18311.x}, \href
  {http://cdsads.u-strasbg.fr/abs/2011MNRAS.413.2403B} {413, 2403}

\bibitem[\protect\citeauthoryear{{Balona}, {Baran}, {Daszy{\'n}ska-Daszkiewicz}
   \& {De Cat}}{{Balona} et~al.}{2015}]{2015MNRAS.451.1445B}
{Balona} L.~A.,  {Baran} A.~S.,  {Daszy{\'n}ska-Daszkiewicz} J.,   {De Cat} P.,
   2015, \mn@doi [\mnras] {10.1093/mnras/stv1017}, \href
  {http://cdsads.u-strasbg.fr/abs/2015MNRAS.451.1445B} {451, 1445}

\bibitem[\protect\citeauthoryear{{Bouabid}, {Dupret}, {Salmon},
  {Montalb{\'a}n}, {Miglio}  \& {Noels}}{{Bouabid}
  et~al.}{2013}]{2013MNRAS.429.2500B}
{Bouabid} M.-P.,  {Dupret} M.-A.,  {Salmon} S.,  {Montalb{\'a}n} J.,  {Miglio}
  A.,   {Noels} A.,  2013, \mn@doi [\mnras] {10.1093/mnras/sts517}, \href
  {http://cdsads.u-strasbg.fr/abs/2013MNRAS.429.2500B} {429, 2500}

\bibitem[\protect\citeauthoryear{{Castelli} \& {Kurucz}}{{Castelli} \&
  {Kurucz}}{2003}]{2003IAUS..210P.A20C}
{Castelli} F.,  {Kurucz} R.~L.,  2003, in {Piskunov} N.,  {Weiss} W.~W.,
  {Gray} D.~F.,  eds,  IAU Symposium Vol. 210, Modelling of Stellar
  Atmospheres. p.~A20

\bibitem[\protect\citeauthoryear{{Chapellier}, {Le Contel}, {Le Contel},
  {Mathias}  \& {Valtier}}{{Chapellier} et~al.}{2006}]{2006A&A...448..697C}
{Chapellier} E.,  {Le Contel} D.,  {Le Contel} J.~M.,  {Mathias} P.,
  {Valtier} J.-C.,  2006, \mn@doi [\aap] {10.1051/0004-6361:20053815}, \href
  {http://cdsads.u-strasbg.fr/abs/2006A%26A...448..697C} {448, 697}

\bibitem[\protect\citeauthoryear{{Colgan} et~al.,}{{Colgan}
  et~al.}{2013}]{2013HEDP....9..369C}
{Colgan} J.,  et~al., 2013, \mn@doi [High Energy Density Physics]
  {10.1016/j.hedp.2013.03.001}, \href
  {http://cdsads.u-strasbg.fr/abs/2013HEDP....9..369C} {9, 369}

\bibitem[\protect\citeauthoryear{{Colgan}, {Kilcrease}, {Magee}, {Abdallah},
  {Sherrill}, {Fontes}, {Hakel}  \& {Zhang}}{{Colgan}
  et~al.}{2015}]{2015HEDP...14...33C}
{Colgan} J.,  {Kilcrease} D.~P.,  {Magee} N.~H.,  {Abdallah} J.,  {Sherrill}
  M.~E.,  {Fontes} C.~J.,  {Hakel} P.,   {Zhang} H.~L.,  2015, \mn@doi [High
  Energy Density Physics] {10.1016/j.hedp.2015.02.006}, \href
  {http://cdsads.u-strasbg.fr/abs/2015HEDP...14...33C} {14, 33}

\bibitem[\protect\citeauthoryear{{Cox}, {Morgan}, {Rogers}  \&
  {Iglesias}}{{Cox} et~al.}{1992}]{1992ApJ...393..272C}
{Cox} A.~N.,  {Morgan} S.~M.,  {Rogers} F.~J.,   {Iglesias} C.~A.,  1992,
  \mn@doi [\apj] {10.1086/171504}, \href
  {http://cdsads.u-strasbg.fr/abs/1992ApJ...393..272C} {393, 272}

\bibitem[\protect\citeauthoryear{{Cugier}}{{Cugier}}{2014}]{2014A&A...565A..76C}
{Cugier} H.,  2014, \mn@doi [\aap] {10.1051/0004-6361/201220507}, \href
  {http://cdsads.u-strasbg.fr/abs/2014A%26A...565A..76C} {565, A76}

\bibitem[\protect\citeauthoryear{{Daszynska-Daszkiewicz}, {Dziembowski}  \&
  {Pamyatnykh}}{{Daszynska-Daszkiewicz} et~al.}{2007}]{JDD_WD_AP2007}
{Daszynska-Daszkiewicz} J.,  {Dziembowski} W.~A.,   {Pamyatnykh} A.~A.,  2007,
  \actaa, \href {http://cdsads.u-strasbg.fr/abs/2007AcA....57...11D} {57, 11}

\bibitem[\protect\citeauthoryear{{Daszy{\'n}ska-Daszkiewicz}, {Dziembowski}  \&
  {Pamyatnykh}}{{Daszy{\'n}ska-Daszkiewicz} et~al.}{2008}]{JDD_WD_AP2008}
{Daszy{\'n}ska-Daszkiewicz} J.,  {Dziembowski} W.~A.,   {Pamyatnykh} A.~A.,
  2008, \mn@doi [Journal of Physics Conference Series]
  {10.1088/1742-6596/118/1/012024}, \href
  {http://cdsads.u-strasbg.fr/abs/2008JPhCS.118a2024D} {118, 012024}

\bibitem[\protect\citeauthoryear{{Daszy{\'n}ska-Daszkiewicz}, {Dziembowski},
  {Jerzykiewicz}  \& {Handler}}{{Daszy{\'n}ska-Daszkiewicz}
  et~al.}{2015}]{2015MNRAS.446.1438D}
{Daszy{\'n}ska-Daszkiewicz} J.,  {Dziembowski} W.~A.,  {Jerzykiewicz} M.,
  {Handler} G.,  2015, \mn@doi [\mnras] {10.1093/mnras/stu2216}, \href
  {http://cdsads.u-strasbg.fr/abs/2015MNRAS.446.1438D} {446, 1438}

\bibitem[\protect\citeauthoryear{{Daszynska-Daszkiewicz}, {Walczak}  \&
  {Pamyatnykh}}{{Daszynska-Daszkiewicz} et~al.}{2017}]{2017arXiv170100937D}
{Daszynska-Daszkiewicz} J.,  {Walczak} P.,   {Pamyatnykh} A.,  2017, preprint,
  \href {http://cdsads.u-strasbg.fr/abs/2017arXiv170100937D} {} (\mn@eprint
  {arXiv} {1701.00937})

\bibitem[\protect\citeauthoryear{{Degroote} et~al.,}{{Degroote}
  et~al.}{2010}]{Degroote2010N}
{Degroote} P.,  et~al., 2010, \mn@doi [\nat] {10.1038/nature08864}, \href
  {http://cdsads.u-strasbg.fr/abs/2010Natur.464..259D} {464, 259}

\bibitem[\protect\citeauthoryear{{Degroote} et~al.,}{{Degroote}
  et~al.}{2012}]{2012A&A...542A..88D}
{Degroote} P.,  et~al., 2012, \mn@doi [\aap] {10.1051/0004-6361/201118548},
  \href {http://cdsads.u-strasbg.fr/abs/2012A%26A...542A..88D} {542, A88}

\bibitem[\protect\citeauthoryear{{Dziembowski} \& {Pamyatnykh}}{{Dziembowski}
  \& {Pamyatnykh}}{2008}]{2008MNRAS.385.2061D}
{Dziembowski} W.~A.,  {Pamyatnykh} A.~A.,  2008, \mn@doi [\mnras]
  {10.1111/j.1365-2966.2008.12964.x}, \href
  {http://cdsads.u-strasbg.fr/abs/2008MNRAS.385.2061D} {385, 2061}

\bibitem[\protect\citeauthoryear{{Dziembowski}, {Moskalik}  \&
  {Pamyatnykh}}{{Dziembowski} et~al.}{1993}]{WD_PM_AP1993}
{Dziembowski} W.~A.,  {Moskalik} P.,   {Pamyatnykh} A.~A.,  1993, \mn@doi
  [\mnras] {10.1093/mnras/265.3.588}, \href
  {http://cdsads.u-strasbg.fr/abs/1993MNRAS.265..588D} {265, 588}

\bibitem[\protect\citeauthoryear{{Dziembowski}, {Daszy{\'n}ska-Daszkiewicz}  \&
  {Pamyatnykh}}{{Dziembowski} et~al.}{2007}]{WD_JDD_AP2007}
{Dziembowski} W.~A.,  {Daszy{\'n}ska-Daszkiewicz} J.,   {Pamyatnykh} A.~A.,
  2007, \mn@doi [\mnras] {10.1111/j.1365-2966.2006.11139.x}, \href
  {http://cdsads.u-strasbg.fr/abs/2007MNRAS.374..248D} {374, 248}

\bibitem[\protect\citeauthoryear{{Ferguson}, {Alexander}, {Allard}, {Barman},
  {Bodnarik}, {Hauschildt}, {Heffner-Wong}  \& {Tamanai}}{{Ferguson}
  et~al.}{2005}]{FergusonAlexander2005}
{Ferguson} J.~W.,  {Alexander} D.~R.,  {Allard} F.,  {Barman} T.,  {Bodnarik}
  J.~G.,  {Hauschildt} P.~H.,  {Heffner-Wong} A.,   {Tamanai} A.,  2005,
  \mn@doi [\apj] {10.1086/428642}, \href
  {http://cdsads.u-strasbg.fr/abs/2005ApJ...623..585F} {623, 585}

\bibitem[\protect\citeauthoryear{{Frost}}{{Frost}}{1902}]{1902ApJ....15..340F}
{Frost} E.~B.,  1902, \mn@doi [\apj] {10.1086/140929}, \href
  {http://cdsads.u-strasbg.fr/abs/1902ApJ....15..340F} {15}

\bibitem[\protect\citeauthoryear{{Gautschy} \& {Saio}}{{Gautschy} \&
  {Saio}}{1993}]{Gautschy1993}
{Gautschy} A.,  {Saio} H.,  1993, \mn@doi [\mnras] {10.1093/mnras/262.1.213},
  \href {http://cdsads.u-strasbg.fr/abs/1993MNRAS.262..213G} {262, 213}

\bibitem[\protect\citeauthoryear{{Godart}, {Sim{\'o}n-D{\'{\i}}az}, {Herrero},
  {Dupret}, {Gr{\"o}tsch-Noels}, {Salmon}  \& {Ventura}}{{Godart}
  et~al.}{2017}]{2017A&A...597A..23G}
{Godart} M.,  {Sim{\'o}n-D{\'{\i}}az} S.,  {Herrero} A.,  {Dupret} M.~A.,
  {Gr{\"o}tsch-Noels} A.,  {Salmon} S.~J.~A.~J.,   {Ventura} P.,  2017, \mn@doi
  [\aap] {10.1051/0004-6361/201628856}, \href
  {http://cdsads.u-strasbg.fr/abs/2017A%26A...597A..23G} {597, A23}

\bibitem[\protect\citeauthoryear{{Grevesse} \& {Noels}}{{Grevesse} \&
  {Noels}}{1993}]{1993oee..conf...15G}
{Grevesse} N.,  {Noels} A.,  1993, in {Prantzos} N.,  {Vangioni-Flam} E.,
  {Casse} M.,  eds, Origin and Evolution of the Elements. pp 15--25

\bibitem[\protect\citeauthoryear{{Handler} et~al.,}{{Handler}
  et~al.}{2006}]{2006MNRAS.365..327H}
{Handler} G.,  et~al., 2006, \mn@doi [\mnras]
  {10.1111/j.1365-2966.2005.09728.x}, \href
  {http://cdsads.u-strasbg.fr/abs/2006MNRAS.365..327H} {365, 327}

\bibitem[\protect\citeauthoryear{{Huang} \& {Gies}}{{Huang} \&
  {Gies}}{2006}]{2006ApJ...648..580H}
{Huang} W.,  {Gies} D.~R.,  2006, \mn@doi [\apj] {10.1086/505782}, \href
  {http://cdsads.u-strasbg.fr/abs/2006ApJ...648..580H} {648, 580}

\bibitem[\protect\citeauthoryear{{Iglesias} \& {Rogers}}{{Iglesias} \&
  {Rogers}}{1996}]{Iglesias1996}
{Iglesias} C.~A.,  {Rogers} F.~J.,  1996, \mn@doi [\apj] {10.1086/177381},
  \href {http://cdsads.u-strasbg.fr/abs/1996ApJ...464..943I} {464, 943}

\bibitem[\protect\citeauthoryear{{Jerzykiewicz}, {Handler}, {Shobbrook},
  {Pigulski}, {Medupe}, {Mokgwetsi}, {Tlhagwane}  \&
  {Rodr{\'{\i}}guez}}{{Jerzykiewicz} et~al.}{2005}]{2005MNRAS.360..619J}
{Jerzykiewicz} M.,  {Handler} G.,  {Shobbrook} R.~R.,  {Pigulski} A.,  {Medupe}
  R.,  {Mokgwetsi} T.,  {Tlhagwane} P.,   {Rodr{\'{\i}}guez} E.,  2005, \mn@doi
  [\mnras] {10.1111/j.1365-2966.2005.09088.x}, \href
  {http://cdsads.u-strasbg.fr/abs/2005MNRAS.360..619J} {360, 619}

\bibitem[\protect\citeauthoryear{{Lee}}{{Lee}}{2001}]{Lee2001}
{Lee} U.,  2001, \mn@doi [\apj] {10.1086/321554}, \href
  {http://cdsads.u-strasbg.fr/abs/2001ApJ...557..311L} {557, 311}

\bibitem[\protect\citeauthoryear{{Lee}}{{Lee}}{2006}]{2006MNRAS.365..677L}
{Lee} U.,  2006, \mn@doi [\mnras] {10.1111/j.1365-2966.2005.09751.x}, \href
  {http://cdsads.u-strasbg.fr/abs/2006MNRAS.365..677L} {365, 677}

\bibitem[\protect\citeauthoryear{{Lee}}{{Lee}}{2008}]{2008CoAst.157..203L}
{Lee} U.,  2008, Communications in Asteroseismology, \href
  {http://cdsads.u-strasbg.fr/abs/2008CoAst.157..203L} {157, 203}

\bibitem[\protect\citeauthoryear{{Lee} \& {Saio}}{{Lee} \&
  {Saio}}{1989}]{Lee1989}
{Lee} U.,  {Saio} H.,  1989, \mn@doi [\mnras] {10.1093/mnras/237.4.875}, \href
  {http://cdsads.u-strasbg.fr/abs/1989MNRAS.237..875L} {237, 875}

\bibitem[\protect\citeauthoryear{{Lee} \& {Saio}}{{Lee} \&
  {Saio}}{1997}]{Lee_Saio1997}
{Lee} U.,  {Saio} H.,  1997, \apj, \href
  {http://cdsads.u-strasbg.fr/abs/1997ApJ...491..839L} {491, 839}

\bibitem[\protect\citeauthoryear{{Marsh Boyer}, {McSwain}, {Aragona}  \&
  {Ou-Yang}}{{Marsh Boyer} et~al.}{2012}]{2012AJ....144..158M}
{Marsh Boyer} A.~N.,  {McSwain} M.~V.,  {Aragona} C.,   {Ou-Yang} B.,  2012,
  \mn@doi [\aj] {10.1088/0004-6256/144/6/158}, \href
  {http://cdsads.u-strasbg.fr/abs/2012AJ....144..158M} {144, 158}

\bibitem[\protect\citeauthoryear{{McNamara}, {Jackiewicz}  \&
  {McKeever}}{{McNamara} et~al.}{2012}]{McNamara2012}
{McNamara} B.~J.,  {Jackiewicz} J.,   {McKeever} J.,  2012, \mn@doi [\aj]
  {10.1088/0004-6256/143/4/101}, \href
  {http://cdsads.u-strasbg.fr/abs/2012AJ....143..101M} {143, 101}

\bibitem[\protect\citeauthoryear{{Miglio}, {Montalb{\'a}n}  \&
  {Dupret}}{{Miglio} et~al.}{2007a}]{2007CoAst.151...48M}
{Miglio} A.,  {Montalb{\'a}n} J.,   {Dupret} M.-A.,  2007a, \mn@doi
  [Communications in Asteroseismology] {10.1553/cia151s48}, \href
  {http://cdsads.u-strasbg.fr/abs/2007CoAst.151...48M} {151, 48}

\bibitem[\protect\citeauthoryear{{Miglio}, {Montalb{\'a}n}  \&
  {Dupret}}{{Miglio} et~al.}{2007b}]{Miglio2007}
{Miglio} A.,  {Montalb{\'a}n} J.,   {Dupret} M.-A.,  2007b, \mn@doi [\mnras]
  {10.1111/j.1745-3933.2006.00267.x}, \href
  {http://cdsads.u-strasbg.fr/abs/2007MNRAS.375L..21M} {375, L21}

\bibitem[\protect\citeauthoryear{{Miglio}, {Montalb{\'a}n}, {Noels}  \&
  {Eggenberger}}{{Miglio} et~al.}{2008}]{2008MNRAS.386.1487M}
{Miglio} A.,  {Montalb{\'a}n} J.,  {Noels} A.,   {Eggenberger} P.,  2008,
  \mn@doi [\mnras] {10.1111/j.1365-2966.2008.13112.x}, \href
  {http://cdsads.u-strasbg.fr/abs/2008MNRAS.386.1487M} {386, 1487}

\bibitem[\protect\citeauthoryear{{Moravveji}}{{Moravveji}}{2016}]{Moravveji2016}
{Moravveji} E.,  2016, \mn@doi [\mnras] {10.1093/mnrasl/slv142}, \href
  {http://cdsads.u-strasbg.fr/abs/2016MNRAS.455L..67M} {455, L67}

\bibitem[\protect\citeauthoryear{{Moravveji}, {Aerts}, {P{\'a}pics}, {Triana}
  \& {Vandoren}}{{Moravveji} et~al.}{2015}]{2015A&A...580A..27M}
{Moravveji} E.,  {Aerts} C.,  {P{\'a}pics} P.~I.,  {Triana} S.~A.,   {Vandoren}
  B.,  2015, \mn@doi [\aap] {10.1051/0004-6361/201425290}, \href
  {http://cdsads.u-strasbg.fr/abs/2015A%26A...580A..27M} {580, A27}

\bibitem[\protect\citeauthoryear{{Moravveji}, {Townsend}, {Aerts}  \&
  {Mathis}}{{Moravveji} et~al.}{2016}]{2016ApJ...823..130M}
{Moravveji} E.,  {Townsend} R.~H.~D.,  {Aerts} C.,   {Mathis} S.,  2016,
  \mn@doi [\apj] {10.3847/0004-637X/823/2/130}, \href
  {http://cdsads.u-strasbg.fr/abs/2016ApJ...823..130M} {823, 130}

\bibitem[\protect\citeauthoryear{{Moskalik} \& {Dziembowski}}{{Moskalik} \&
  {Dziembowski}}{1992}]{1992A&A...256L...5M}
{Moskalik} P.,  {Dziembowski} W.~A.,  1992, \aap, \href
  {http://cdsads.u-strasbg.fr/abs/1992A%26A...256L...5M} {256, L5}

\bibitem[\protect\citeauthoryear{{Mowlavi}, {Barblan}, {Saesen}  \&
  {Eyer}}{{Mowlavi} et~al.}{2013}]{2013A&A...554A.108M}
{Mowlavi} N.,  {Barblan} F.,  {Saesen} S.,   {Eyer} L.,  2013, \mn@doi [\aap]
  {10.1051/0004-6361/201321065}, \href
  {http://cdsads.u-strasbg.fr/abs/2013A%26A...554A.108M} {554, A108}

\bibitem[\protect\citeauthoryear{{Mowlavi}, {Saesen}, {Semaan}, {Eggenberger},
  {Barblan}, {Eyer}, {Ekstr{\"o}m}  \& {Georgy}}{{Mowlavi}
  et~al.}{2016}]{2016A&A...595L...1M}
{Mowlavi} N.,  {Saesen} S.,  {Semaan} T.,  {Eggenberger} P.,  {Barblan} F.,
  {Eyer} L.,  {Ekstr{\"o}m} S.,   {Georgy} C.,  2016, \mn@doi [\aap]
  {10.1051/0004-6361/201629175}, \href
  {http://cdsads.u-strasbg.fr/abs/2016A%26A...595L...1M} {595, L1}

\bibitem[\protect\citeauthoryear{{Mo{\'z}dzierski}, {Pigulski}, {Kopacki},
  {Ko{\l}aczkowski}  \& {St{\c e}{\'s}licki}}{{Mo{\'z}dzierski}
  et~al.}{2014}]{2014AcA....64...89M}
{Mo{\'z}dzierski} D.,  {Pigulski} A.,  {Kopacki} G.,  {Ko{\l}aczkowski} Z.,
  {St{\c e}{\'s}licki} M.,  2014, \actaa, \href
  {http://cdsads.u-strasbg.fr/abs/2014AcA....64...89M} {64, 89}

\bibitem[\protect\citeauthoryear{{Pamyatnykh}}{{Pamyatnykh}}{1999}]{Pamyatnykh1999}
{Pamyatnykh} A.~A.,  1999, \actaa, \href
  {http://cdsads.u-strasbg.fr/abs/1999AcA....49..119P} {49, 119}

\bibitem[\protect\citeauthoryear{{Pamyatnykh}, {Dziembowski}, {Handler}  \&
  {Pikall}}{{Pamyatnykh} et~al.}{1998}]{Pamyatnykh1998}
{Pamyatnykh} A.~A.,  {Dziembowski} W.~A.,  {Handler} G.,   {Pikall} H.,  1998,
  \aap, \href {http://cdsads.u-strasbg.fr/abs/1998A%26A...333..141P} {333, 141}

\bibitem[\protect\citeauthoryear{{P{\'a}pics} et~al.,}{{P{\'a}pics}
  et~al.}{2012}]{Papics2012}
{P{\'a}pics} P.~I.,  et~al., 2012, \mn@doi [\aap]
  {10.1051/0004-6361/201218809}, \href
  {http://cdsads.u-strasbg.fr/abs/2012A%26A...542A..55P} {542, A55}

\bibitem[\protect\citeauthoryear{{P{\'a}pics}, {Moravveji}, {Aerts},
  {Tkachenko}, {Triana}, {Bloemen}  \& {Southworth}}{{P{\'a}pics}
  et~al.}{2014}]{Papics2014}
{P{\'a}pics} P.~I.,  {Moravveji} E.,  {Aerts} C.,  {Tkachenko} A.,  {Triana}
  S.~A.,  {Bloemen} S.,   {Southworth} J.,  2014, \mn@doi [\aap]
  {10.1051/0004-6361/201424094}, \href
  {http://cdsads.u-strasbg.fr/abs/2014A%26A...570A...8P} {570, A8}

\bibitem[\protect\citeauthoryear{{P{\'a}pics}, {Tkachenko}, {Aerts}, {Van
  Reeth}, {De Smedt}, {Hillen}, {{\O}stensen}  \& {Moravveji}}{{P{\'a}pics}
  et~al.}{2015}]{Ppics2015}
{P{\'a}pics} P.~I.,  {Tkachenko} A.,  {Aerts} C.,  {Van Reeth} T.,  {De Smedt}
  K.,  {Hillen} M.,  {{\O}stensen} R.,   {Moravveji} E.,  2015, \mn@doi [\apjl]
  {10.1088/2041-8205/803/2/L25}, \href
  {http://cdsads.u-strasbg.fr/abs/2015ApJ...803L..25P} {803, L25}

\bibitem[\protect\citeauthoryear{{Pigulski} et~al.,}{{Pigulski}
  et~al.}{2016}]{2016A&A...588A..55P}
{Pigulski} A.,  et~al., 2016, \mn@doi [\aap] {10.1051/0004-6361/201527872},
  \href {http://cdsads.u-strasbg.fr/abs/2016A%26A...588A..55P} {588, A55}

\bibitem[\protect\citeauthoryear{{Rogers} \& {Nayfonov}}{{Rogers} \&
  {Nayfonov}}{2002}]{2002ApJ...576.1064R}
{Rogers} F.~J.,  {Nayfonov} A.,  2002, \mn@doi [\apj] {10.1086/341894}, \href
  {http://cdsads.u-strasbg.fr/abs/2002ApJ...576.1064R} {576, 1064}

\bibitem[\protect\citeauthoryear{{Saesen} et~al.,}{{Saesen}
  et~al.}{2010}]{2010A&A...515A..16S}
{Saesen} S.,  et~al., 2010, \mn@doi [\aap] {10.1051/0004-6361/200913236}, \href
  {http://cdsads.u-strasbg.fr/abs/2010A%26A...515A..16S} {515, A16}

\bibitem[\protect\citeauthoryear{{Saesen}, {Briquet}, {Aerts}, {Miglio}  \&
  {Carrier}}{{Saesen} et~al.}{2013}]{2013AJ....146..102S}
{Saesen} S.,  {Briquet} M.,  {Aerts} C.,  {Miglio} A.,   {Carrier} F.,  2013,
  \mn@doi [\aj] {10.1088/0004-6256/146/4/102}, \href
  {http://cdsads.u-strasbg.fr/abs/2013AJ....146..102S} {146, 102}

\bibitem[\protect\citeauthoryear{{Saio}, {Ekstr{\"o}m}, {Mowlavi}, {Georgy},
  {Saesen}, {Eggenberger}, {Semaan}  \& {Salmon}}{{Saio}
  et~al.}{2017}]{2017arXiv170202306S}
{Saio} H.,  {Ekstr{\"o}m} S.,  {Mowlavi} N.,  {Georgy} C.,  {Saesen} S.,
  {Eggenberger} P.,  {Semaan} T.,   {Salmon} S.~J.~A.~J.,  2017, preprint,
  \href {http://cdsads.u-strasbg.fr/abs/2017arXiv170202306S} {} (\mn@eprint
  {arXiv} {1702.02306})

\bibitem[\protect\citeauthoryear{{Salmon}, {Montalb{\'a}n}, {Morel}, {Miglio},
  {Dupret}  \& {Noels}}{{Salmon} et~al.}{2012}]{2012MNRAS.422.3460S}
{Salmon} S.,  {Montalb{\'a}n} J.,  {Morel} T.,  {Miglio} A.,  {Dupret} M.-A.,
  {Noels} A.,  2012, \mn@doi [\mnras] {10.1111/j.1365-2966.2012.20857.x}, \href
  {http://cdsads.u-strasbg.fr/abs/2012MNRAS.422.3460S} {422, 3460}

\bibitem[\protect\citeauthoryear{{Salmon}, {Montalb{\'a}n}, {Reese}, {Dupret}
  \& {Eggenberger}}{{Salmon} et~al.}{2014}]{2014A&A...569A..18S}
{Salmon} S.~J.~A.~J.,  {Montalb{\'a}n} J.,  {Reese} D.~R.,  {Dupret} M.-A.,
  {Eggenberger} P.,  2014, \mn@doi [\aap] {10.1051/0004-6361/201323259}, \href
  {http://cdsads.u-strasbg.fr/abs/2014A%26A...569A..18S} {569, A18}

\bibitem[\protect\citeauthoryear{{Savonije}}{{Savonije}}{2005}]{2005A&A...443..557S}
{Savonije} G.~J.,  2005, \mn@doi [\aap] {10.1051/0004-6361:20053328}, \href
  {http://cdsads.u-strasbg.fr/abs/2005A%26A...443..557S} {443, 557}

\bibitem[\protect\citeauthoryear{{Savonije}}{{Savonije}}{2013}]{2013A&A...559A..25S}
{Savonije} G.~J.,  2013, \mn@doi [\aap] {10.1051/0004-6361/201321961}, \href
  {http://cdsads.u-strasbg.fr/abs/2013A%26A...559A..25S} {559, A25}

\bibitem[\protect\citeauthoryear{{Seaton}}{{Seaton}}{1996}]{Seaton1996}
{Seaton} M.~J.,  1996, \mn@doi [\mnras] {10.1093/mnras/279.1.95}, \href
  {http://cdsads.u-strasbg.fr/abs/1996MNRAS.279...95S} {279, 95}

\bibitem[\protect\citeauthoryear{{Seaton}}{{Seaton}}{2005}]{Seaton2005}
{Seaton} M.~J.,  2005, \mn@doi [\mnras] {10.1111/j.1365-2966.2005.00019.x},
  \href {http://cdsads.u-strasbg.fr/abs/2005MNRAS.362L...1S} {362, L1}

\bibitem[\protect\citeauthoryear{{Sim{\'o}n-D{\'{\i}}az} \&
  {Herrero}}{{Sim{\'o}n-D{\'{\i}}az} \& {Herrero}}{2014}]{2014A&A...562A.135S}
{Sim{\'o}n-D{\'{\i}}az} S.,  {Herrero} A.,  2014, \mn@doi [\aap]
  {10.1051/0004-6361/201322758}, \href
  {http://cdsads.u-strasbg.fr/abs/2014A%26A...562A.135S} {562, A135}

\bibitem[\protect\citeauthoryear{{Sim{\'o}n-D{\'{\i}}az}, {Godart}, {Castro},
  {Herrero}, {Aerts}, {Puls}, {Telting}  \&
  {Grassitelli}}{{Sim{\'o}n-D{\'{\i}}az} et~al.}{2017}]{2017A&A...597A..22S}
{Sim{\'o}n-D{\'{\i}}az} S.,  {Godart} M.,  {Castro} N.,  {Herrero} A.,  {Aerts}
  C.,  {Puls} J.,  {Telting} J.,   {Grassitelli} L.,  2017, \mn@doi [\aap]
  {10.1051/0004-6361/201628541}, \href
  {http://cdsads.u-strasbg.fr/abs/2017A%26A...597A..22S} {597, A22}

\bibitem[\protect\citeauthoryear{{Stankov} \& {Handler}}{{Stankov} \&
  {Handler}}{2005}]{2005ApJS..158..193S}
{Stankov} A.,  {Handler} G.,  2005, \mn@doi [\apjs] {10.1086/429408}, \href
  {http://cdsads.u-strasbg.fr/abs/2005ApJS..158..193S} {158, 193}

\bibitem[\protect\citeauthoryear{{Strom}, {Wolff}  \& {Dror}}{{Strom}
  et~al.}{2005}]{2005AJ....129..809S}
{Strom} S.~E.,  {Wolff} S.~C.,   {Dror} D.~H.~A.,  2005, \mn@doi [\aj]
  {10.1086/426748}, \href {http://cdsads.u-strasbg.fr/abs/2005AJ....129..809S}
  {129, 809}

\bibitem[\protect\citeauthoryear{{Struve}}{{Struve}}{1955}]{Struve1955}
{Struve} O.,  1955, \skytel, \href
  {http://cdsads.u-strasbg.fr/abs/1955S%26T....14..461S} {14}

\bibitem[\protect\citeauthoryear{{Szewczuk} \&
  {Daszy{\'n}ska-Daszkiewicz}}{{Szewczuk} \&
  {Daszy{\'n}ska-Daszkiewicz}}{2015}]{WS2015ident}
{Szewczuk} W.,  {Daszy{\'n}ska-Daszkiewicz} J.,  2015, \mn@doi [\mnras]
  {10.1093/mnras/stv715}, \href
  {http://cdsads.u-strasbg.fr/abs/2015MNRAS.450.1585S} {450, 1585}

\bibitem[\protect\citeauthoryear{{Szewczuk}, {Daszy{\'n}ska-Daszkiewicz}  \&
  {Dziembowski}}{{Szewczuk} et~al.}{2014}]{2014IAUS..301..109S}
{Szewczuk} W.,  {Daszy{\'n}ska-Daszkiewicz} J.,   {Dziembowski} W.,  2014, in
  {Guzik} J.~A.,  {Chaplin} W.~J.,  {Handler} G.,   {Pigulski} A.,  eds,  IAU
  Symposium Vol. 301, Precision Asteroseismology. pp 109--112 (\mn@eprint
  {arXiv} {1311.2818}), \mn@doi{10.1017/S1743921313014178}

\bibitem[\protect\citeauthoryear{{Szewczuk}, {Daszy{\'n}ska-Daszkiewicz}  \&
  {Walczak}}{{Szewczuk} et~al.}{2017}]{2017arXiv170101256S}
{Szewczuk} W.,  {Daszy{\'n}ska-Daszkiewicz} J.,   {Walczak} P.,  2017,
  preprint, \href {http://cdsads.u-strasbg.fr/abs/2017arXiv170101256S} {}
  (\mn@eprint {arXiv} {1701.01256})

\bibitem[\protect\citeauthoryear{{Tassoul}}{{Tassoul}}{1980}]{1980ApJS...43..469T}
{Tassoul} M.,  1980, \mn@doi [\apjs] {10.1086/190678}, \href
  {http://cdsads.u-strasbg.fr/abs/1980ApJS...43..469T} {43, 469}

\bibitem[\protect\citeauthoryear{{Townsend}}{{Townsend}}{2003a}]{Townsend2003b}
{Townsend} R.~H.~D.,  2003a, \mn@doi [\mnras]
  {10.1046/j.1365-8711.2003.06379.x}, \href
  {http://cdsads.u-strasbg.fr/abs/2003MNRAS.340.1020T} {340, 1020}

\bibitem[\protect\citeauthoryear{{Townsend}}{{Townsend}}{2003b}]{Townsend2003a}
{Townsend} R.~H.~D.,  2003b, \mn@doi [\mnras]
  {10.1046/j.1365-8711.2003.06640.x}, \href
  {http://cdsads.u-strasbg.fr/abs/2003MNRAS.343..125T} {343, 125}

\bibitem[\protect\citeauthoryear{{Townsend}}{{Townsend}}{2005a}]{Townsend2005}
{Townsend} R.~H.~D.,  2005a, \mn@doi [\mnras]
  {10.1111/j.1365-2966.2005.09002.x}, \href
  {http://cdsads.u-strasbg.fr/abs/2005MNRAS.360..465T} {360, 465}

\bibitem[\protect\citeauthoryear{{Townsend}}{{Townsend}}{2005b}]{2005MNRAS.364..573T}
{Townsend} R.~H.~D.,  2005b, \mn@doi [\mnras]
  {10.1111/j.1365-2966.2005.09585.x}, \href
  {http://cdsads.u-strasbg.fr/abs/2005MNRAS.364..573T} {364, 573}

\bibitem[\protect\citeauthoryear{{Turck-Chi{\`e}ze} et~al.,}{{Turck-Chi{\`e}ze}
  et~al.}{2013}]{2013HEDP....9..473T}
{Turck-Chi{\`e}ze} S.,  et~al., 2013, \mn@doi [High Energy Density Physics]
  {10.1016/j.hedp.2013.04.004}, \href
  {http://cdsads.u-strasbg.fr/abs/2013HEDP....9..473T} {9, 473}

\bibitem[\protect\citeauthoryear{{Waelkens}}{{Waelkens}}{1991}]{Waelkens1991}
{Waelkens} C.,  1991, \aap, \href
  {http://cdsads.u-strasbg.fr/abs/1991A%26A...246..453W} {246, 453}

\bibitem[\protect\citeauthoryear{{Walczak}, {Fontes}, {Colgan}, {Kilcrease}  \&
  {Guzik}}{{Walczak} et~al.}{2015}]{PW2015}
{Walczak} P.,  {Fontes} C.~J.,  {Colgan} J.,  {Kilcrease} D.~P.,   {Guzik}
  J.~A.,  2015, \mn@doi [\aap] {10.1051/0004-6361/201526824}, \href
  {http://cdsads.u-strasbg.fr/abs/2015A%26A...580L...9W} {580, L9}

\bibitem[\protect\citeauthoryear{{Walker} et~al.,}{{Walker}
  et~al.}{2005}]{2005ApJ...635L..77W}
{Walker} G.~A.~H.,  et~al., 2005, \mn@doi [\apjl] {10.1086/499362}, \href
  {http://cdsads.u-strasbg.fr/abs/2005ApJ...635L..77W} {635, L77}

\bibitem[\protect\citeauthoryear{{Winget} et~al.,}{{Winget}
  et~al.}{1991}]{1991ApJ...378..326W}
{Winget} D.~E.,  et~al., 1991, \mn@doi [\apj] {10.1086/170434}, \href
  {http://cdsads.u-strasbg.fr/abs/1991ApJ...378..326W} {378, 326}

\bibitem[\protect\citeauthoryear{{Zdravkov} \& {Pamyatnykh}}{{Zdravkov} \&
  {Pamyatnykh}}{2008}]{Zdravkov2008}
{Zdravkov} T.,  {Pamyatnykh} A.~A.,  2008, \mn@doi [Journal of Physics
  Conference Series] {10.1088/1742-6596/118/1/012079}, \href
  {http://cdsads.u-strasbg.fr/abs/2008JPhCS.118a2079Z} {118, 012079}

\makeatother
\end{thebibliography}


\appendix

\section{The distribution of $\Omega/\Omega_\mathrm{crit}$ across the considered mass range}
\label{appA}

\begin{figure*}
	\includegraphics[width=2.7\columnwidth, angle=270]{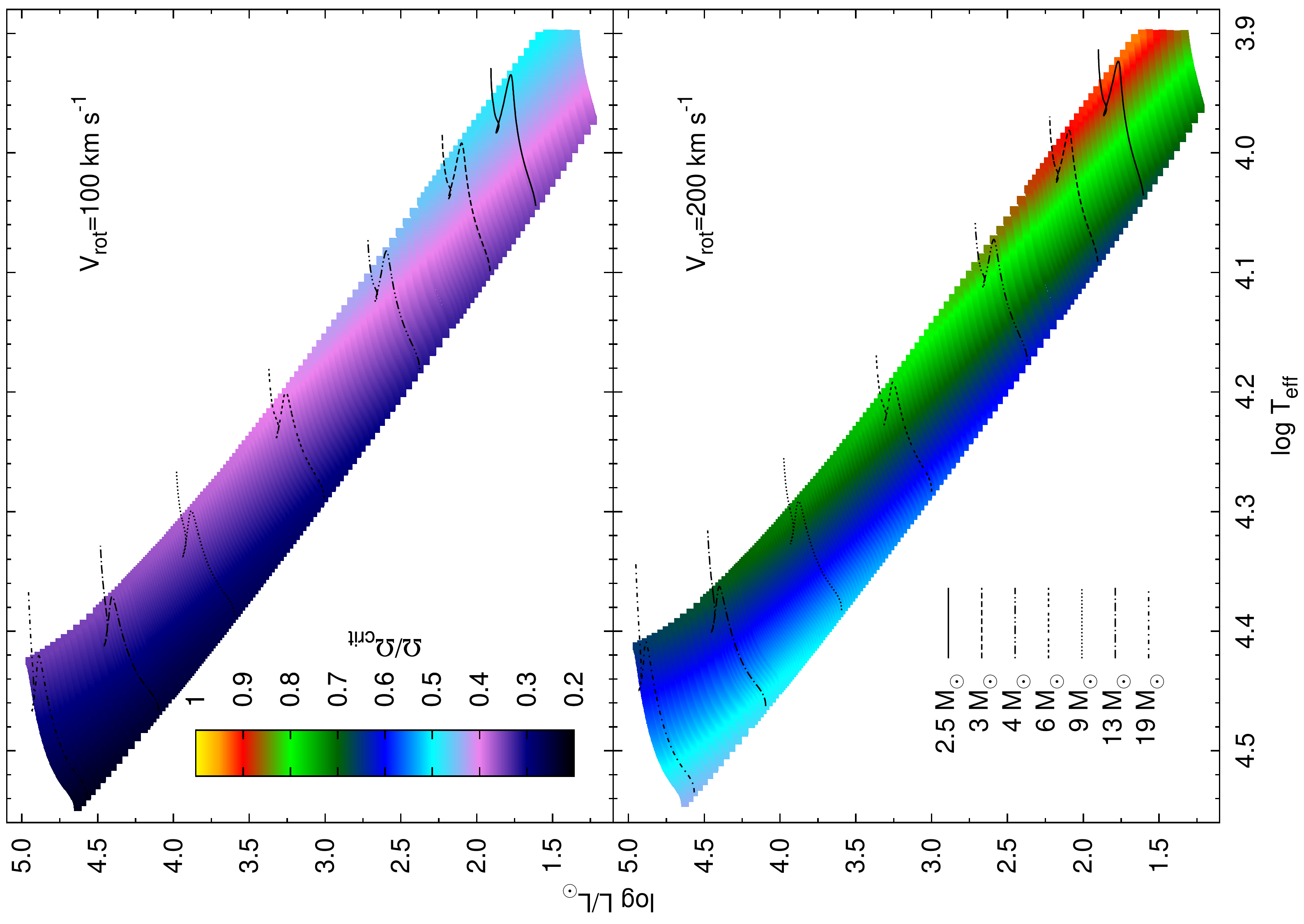}
    \caption{The distribution of the value of  $\Omega/\Omega_\mathrm{crit}$
corresponding to the rotation velocity, $V_{\rm rot}=100$ (the upper panel) and 200\,km\,s$^{-1}$
(the lower panel) across the H--R diagram.
Models from our reference grid with the masses $M=2-20\,\mathrm{M_{\sun}}$ are considered.
            }
    \label{figA1}
\end{figure*}

In Fig.\,\ref{figA1}, we depicted the distribution of the values of $\Omega/\Omega_\mathrm{crit}$
corresponding to the rotation velocity, $V_{\rm rot}=100$ and 200\,km\,s$^{-1}$.
The reference models, as described in the text, were considered, i.e., the models
with masses $2-20\,\mathrm{M_{\sun}}$ computed for 
$X_0=0.7$, $Z=0.015$, OP data and the AGSS09 chemical mixture.

\section{Instability domains for the g modes with $\ell=3,~4$
and mixed gravity--Rossby modes with $m=-3,~ -4$}
\label{appB}

\begin{figure*}
	\includegraphics[width=2.7\columnwidth, angle=270]{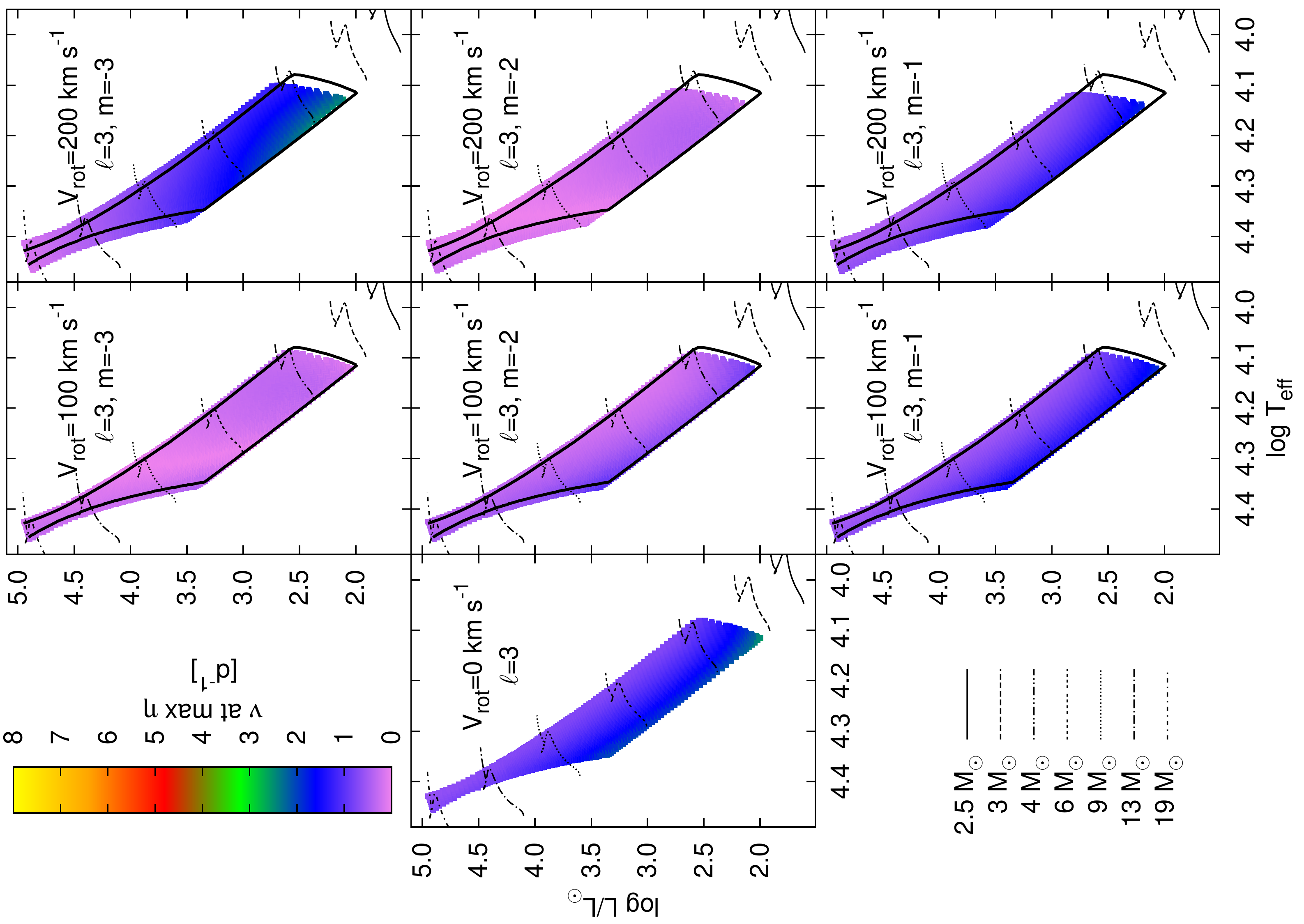}
    \caption{The same as in Fig.\,2 but for modes with $\ell=3$
             and $m\le-1$.
            }
    \label{l3a_OP_X0_7Z0_015_ov0_0}
\end{figure*}

\begin{figure*}
	\includegraphics[width=2.7\columnwidth, angle=270]{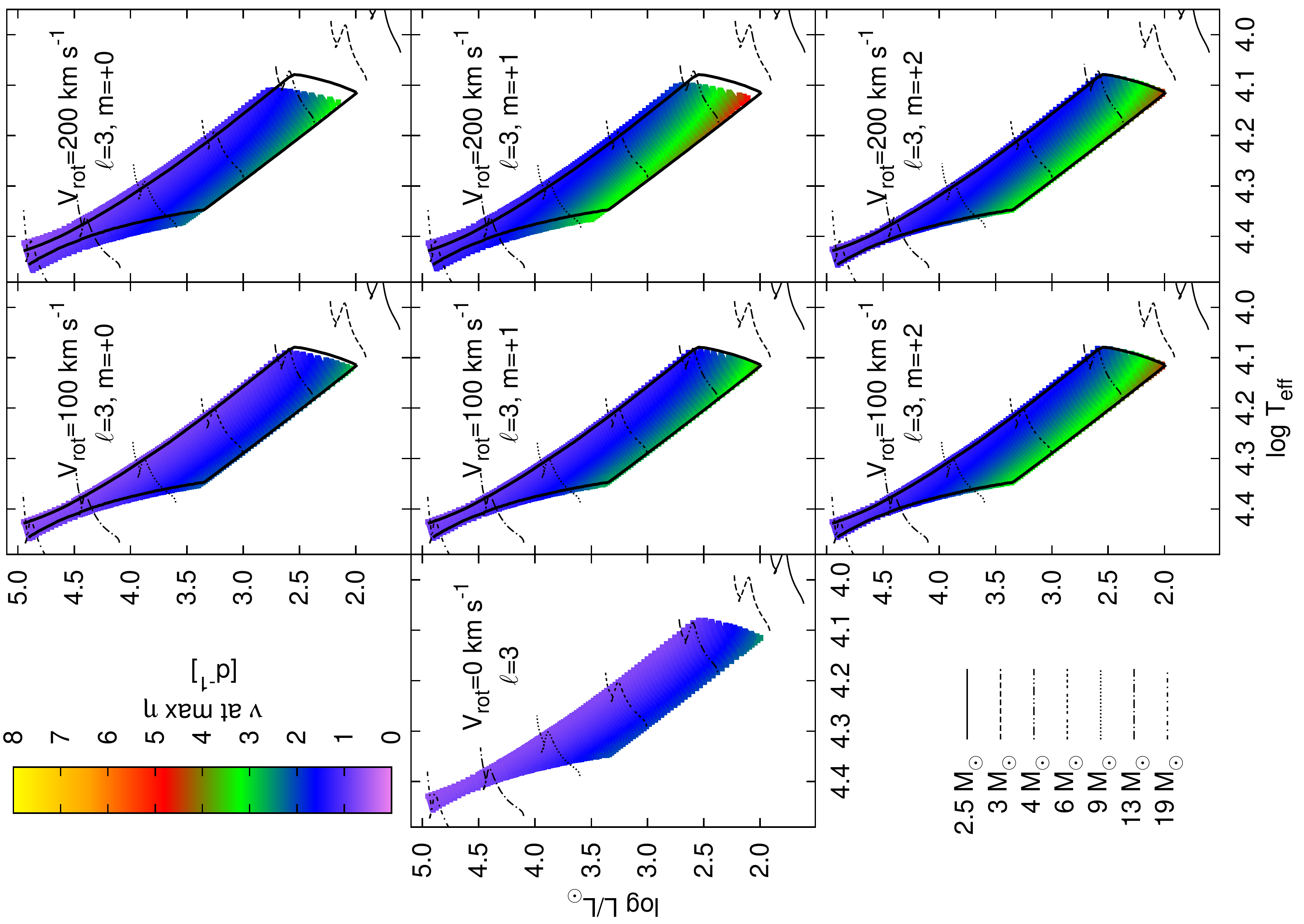}
    \caption{The same as in Fig.\,2 but for modes with $\ell=3$
             and $0\le m\le+2$.
            }
    \label{l3b_OP_X0_7Z0_015_ov0_0}
\end{figure*}

\begin{figure*}
	\includegraphics[width=2.7\columnwidth, angle=270]{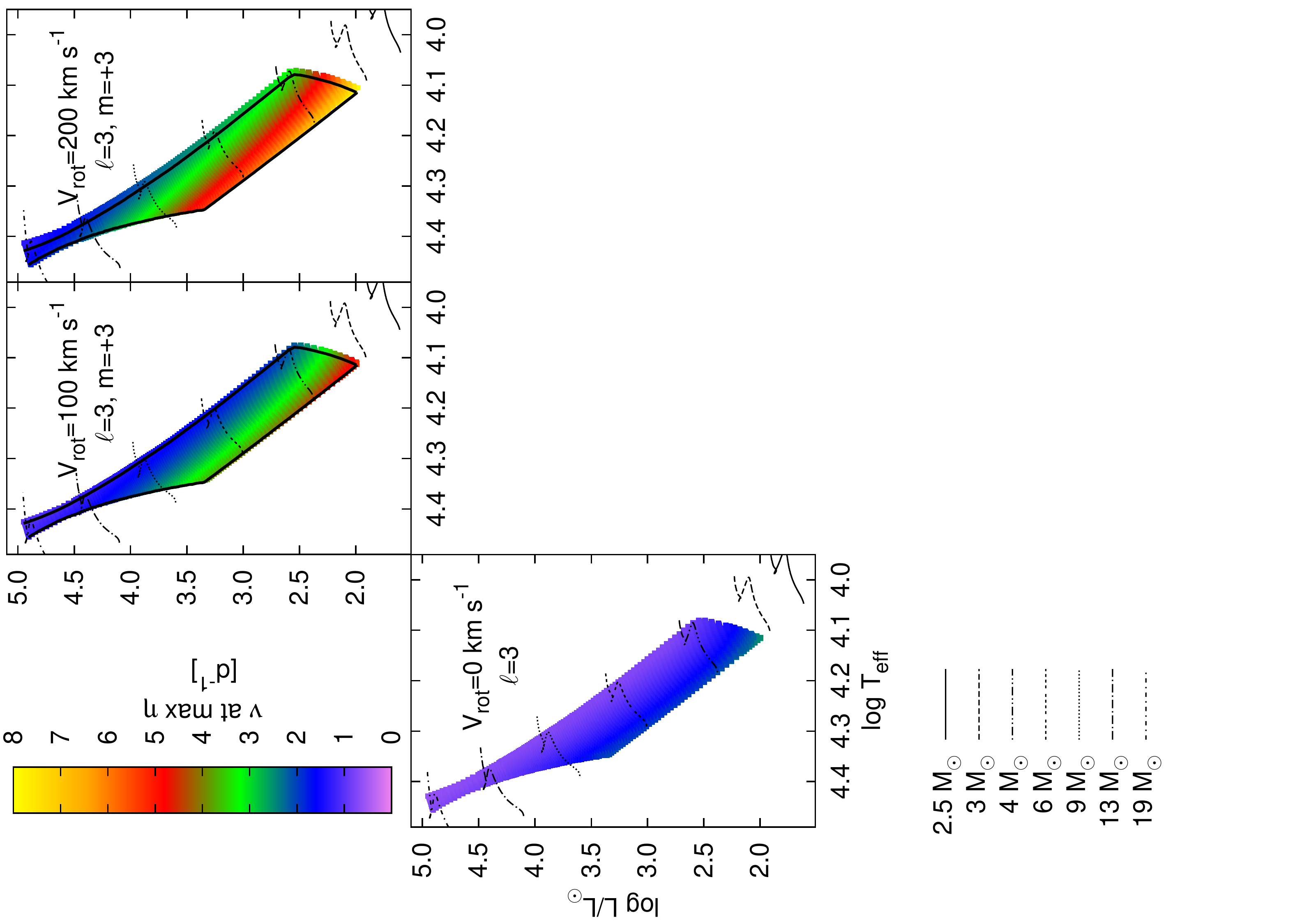}
    \caption{The same as in Fig.\,2 but for modes with $\ell=3$
             and $m=3$.
            }
    \label{l3c_OP_X0_7Z0_015_ov0_0}
\end{figure*}

\begin{figure*}

	\includegraphics[width=2.7\columnwidth, angle=270]{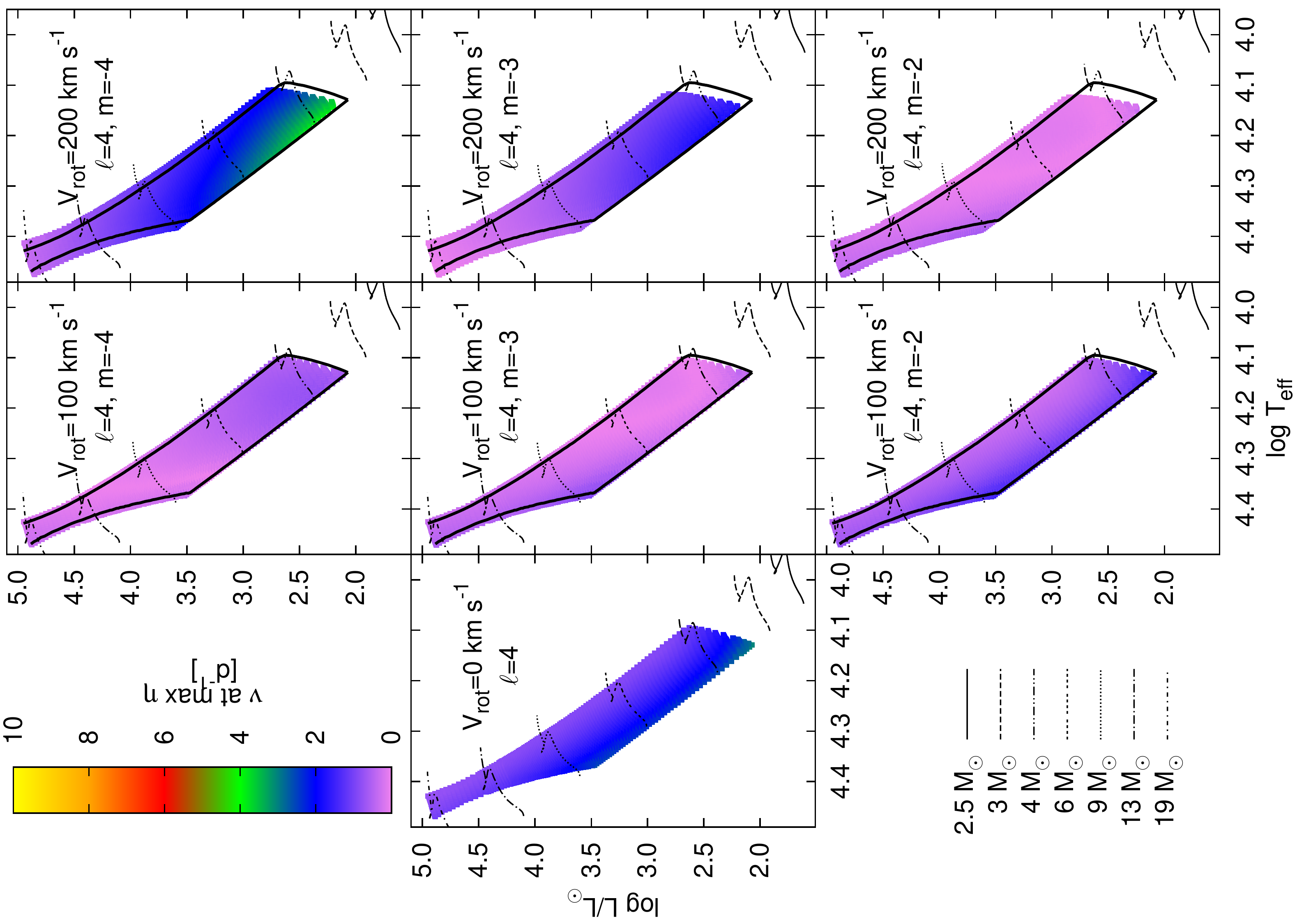}
    \caption{The same as in Fig.\,2 but for modes with $\ell=4$
             and $m\le-2$.
            }
    \label{l4a_OP_X0_7Z0_015_ov0_0}
\end{figure*}

\begin{figure*}

	\includegraphics[width=2.7\columnwidth, angle=270]{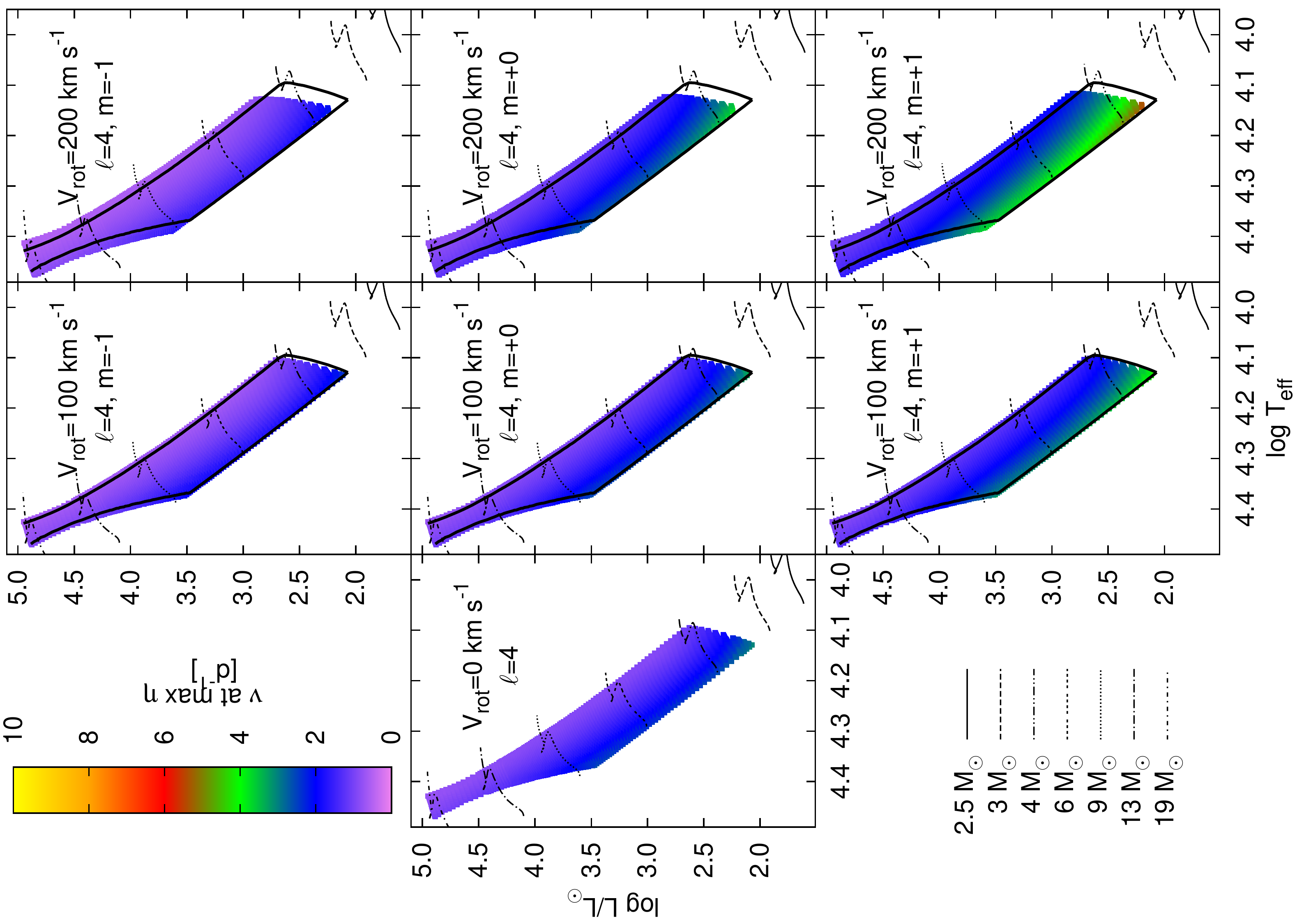}
    \caption{The same as in Fig.\,2 but for modes with $\ell=4$
             and $-1\le m\le+1$.
            }
    \label{l4b_OP_X0_7Z0_015_ov0_0}
\end{figure*}

\begin{figure*}

	\includegraphics[width=2.7\columnwidth, angle=270]{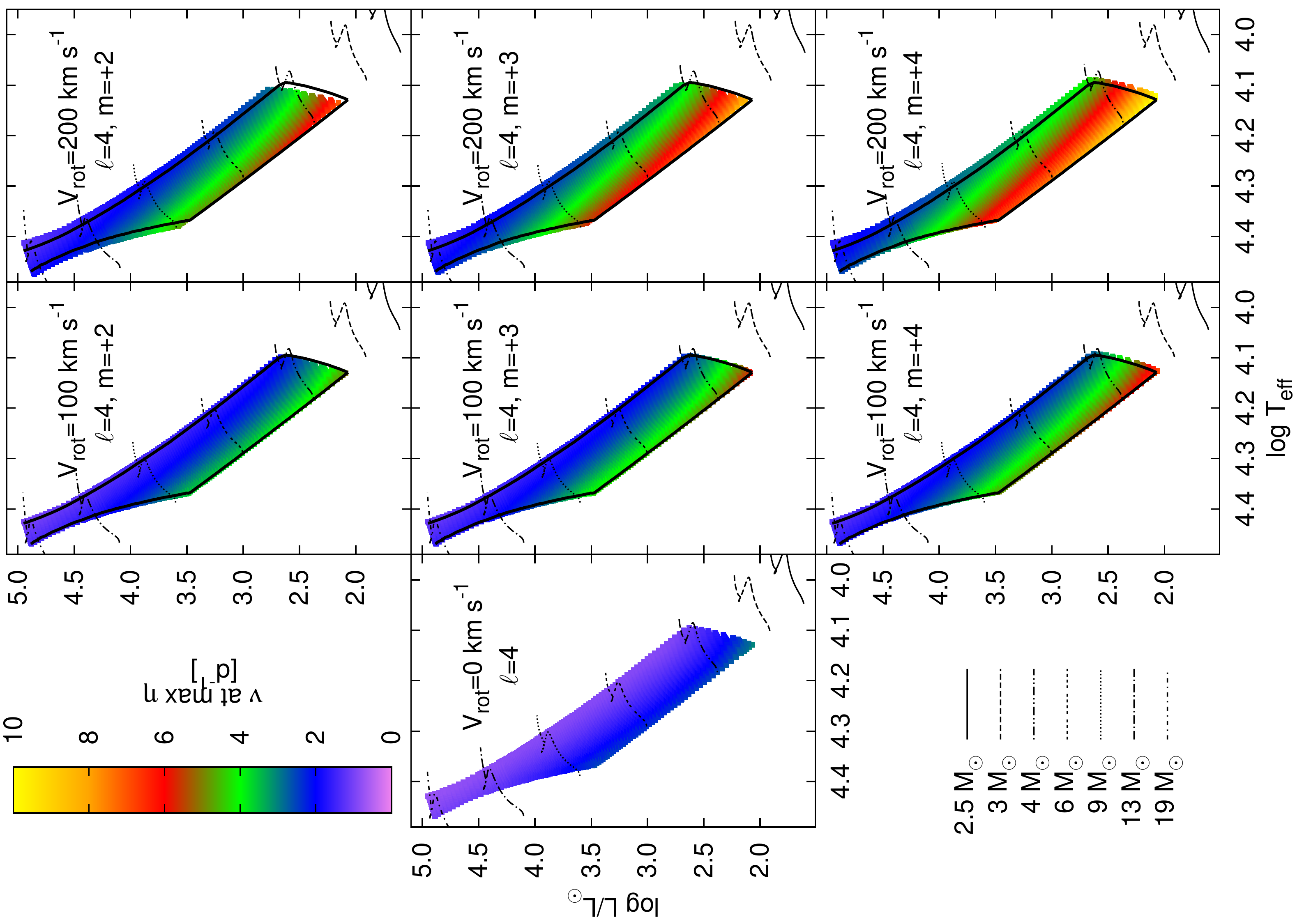}
    \caption{The same as in Fig.\,2 but for modes with $\ell=4$
             and $m\ge+2$.
            }
    \label{l4c_OP_X0_7Z0_015_ov0_0}
\end{figure*}

\begin{figure*}

	\includegraphics[width=1.8\columnwidth, angle=270]{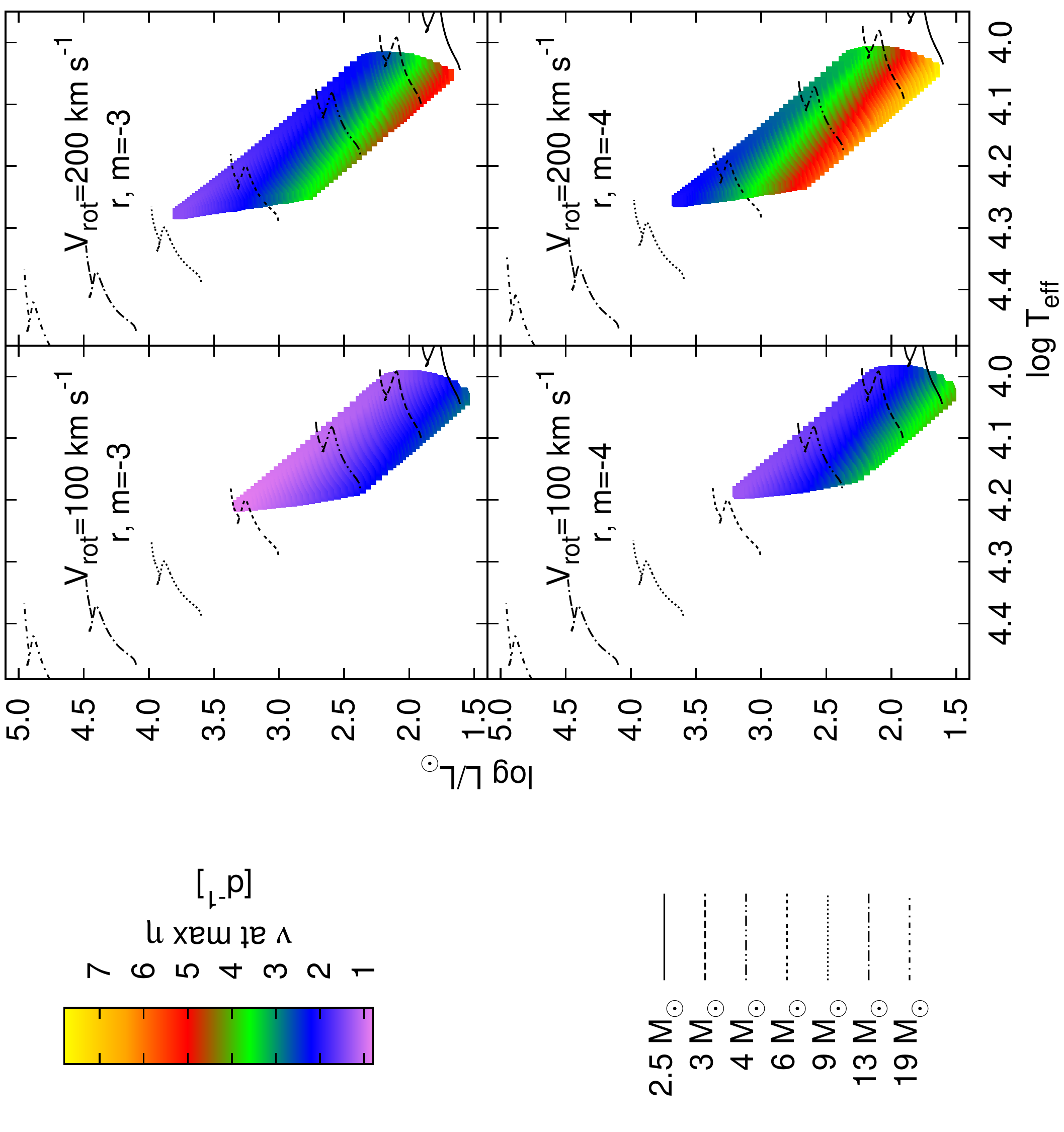}
    \caption{The same as in Fig.\,2 but for $r$ $m=-3,\,-4$ modes
             and $V_\mathrm{rot}=100$ and 200 $\mathrm{km\,s^{-1}}$.
            }
    \label{l-3-4_OP_X0_7Z0_015_ov0_0}
\end{figure*}

Here, we show the pulsation instability strips for the g modes
with $\ell=3$ and 4
(Fig\,\ref{l3a_OP_X0_7Z0_015_ov0_0}--\ref{l4c_OP_X0_7Z0_015_ov0_0})
and $r$ modes with $m=-3$ and $-4$ (Fig\,\ref{l-3-4_OP_X0_7Z0_015_ov0_0})
excited in the rotating reference models described in Section\,2.
Again the three values of the rotation velocity were considered, $V_{\rm rot}=0,~100,~200$ km\,s$^{-1}$.

\section{Frequency evolution of unstable g modes in selected models}
\label{appC}

\begin{figure*}
	\includegraphics[width=2.7\columnwidth, angle=270]{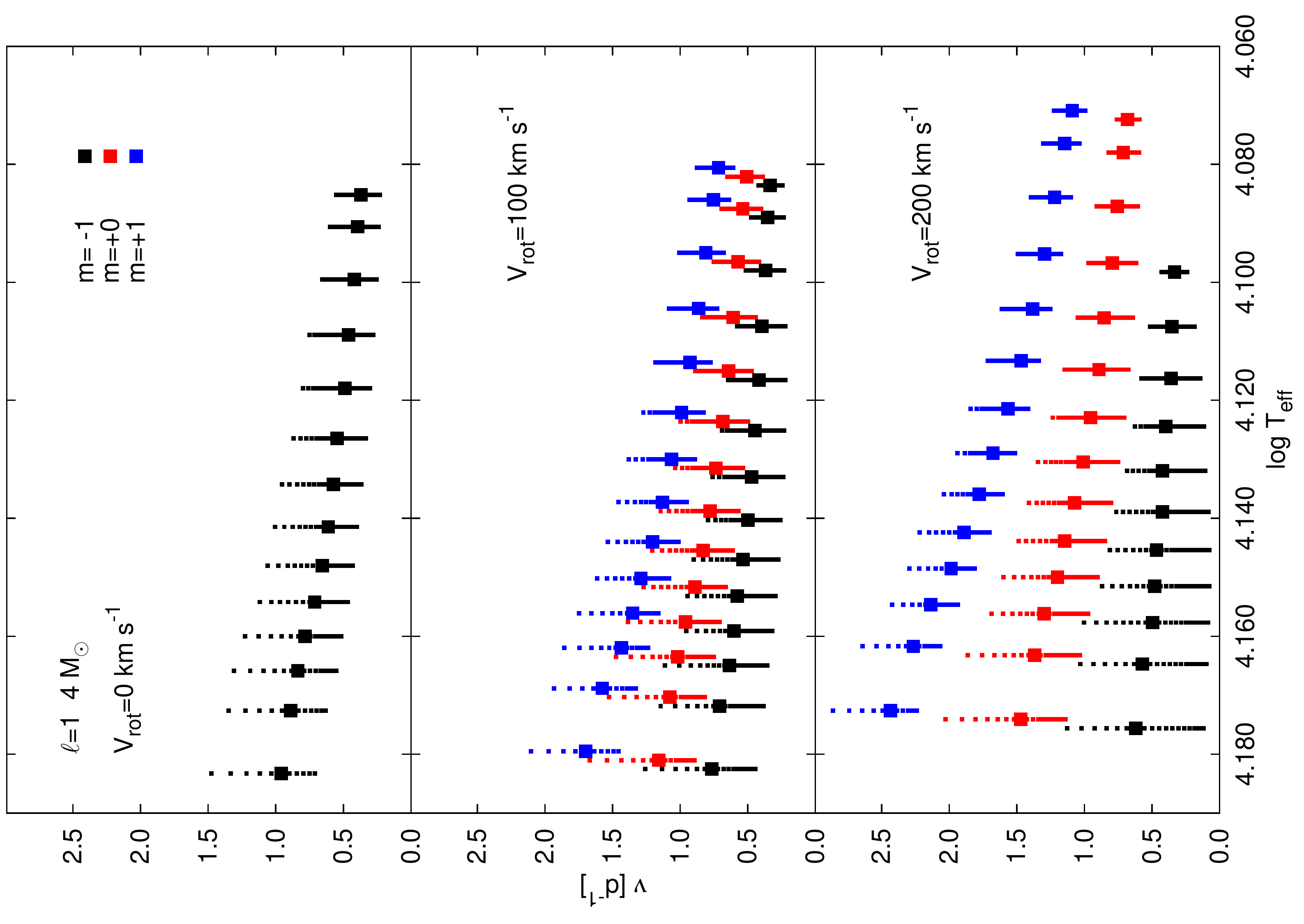}
    \caption{Evolution of the frequencies of unstable dipole modes in a $4\,\mathrm{M_{\sun}}$ model
    for three values of rotation velocity. Frequencies corresponding to modes with the maximum values
    of the instability parameter $\eta$ are indicated by large symbols.}
    \label{f_ranges_4}
\end{figure*}

\begin{figure*}

	\includegraphics[width=2.7\columnwidth, angle=270]{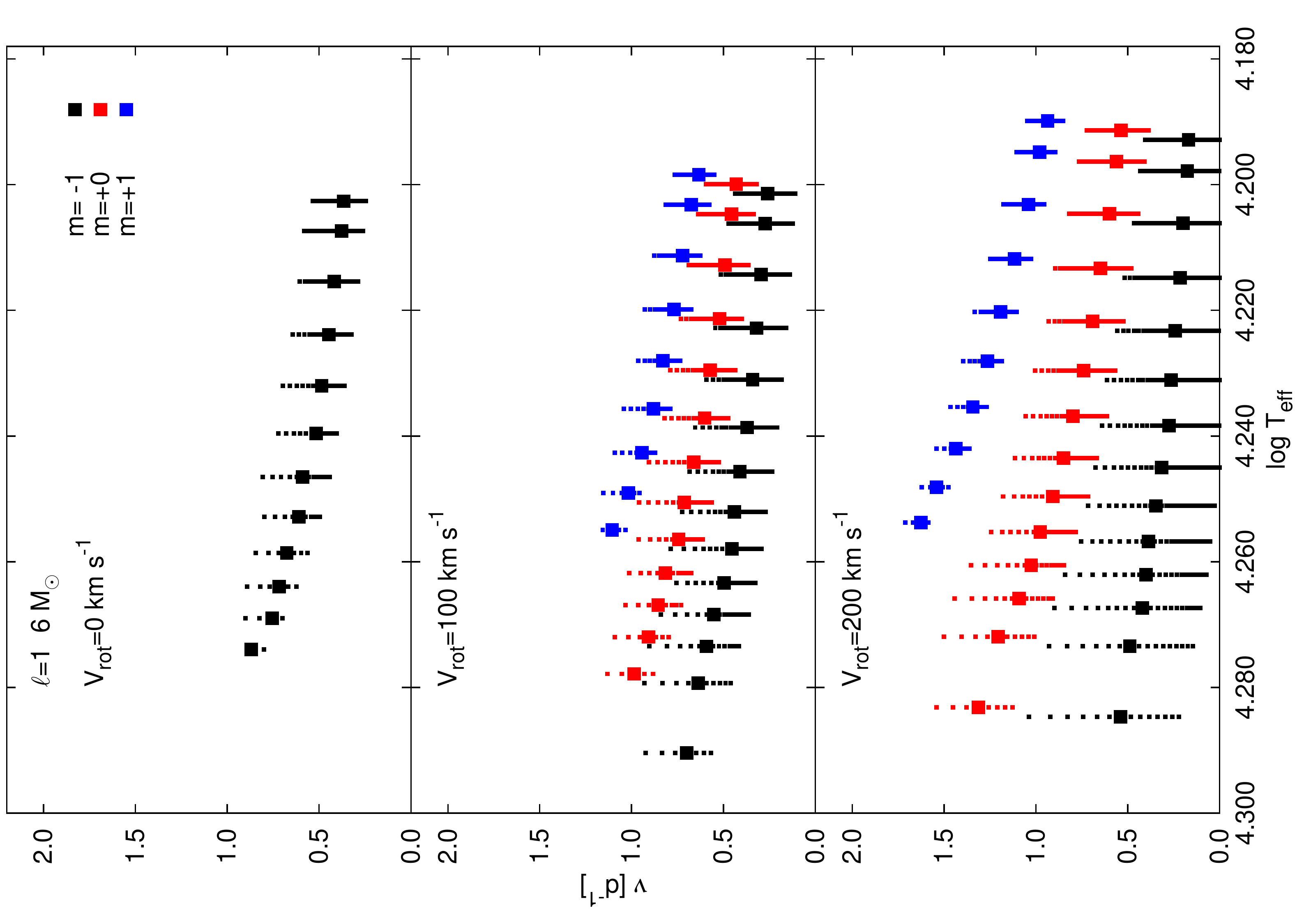}
    \caption{The same as in Fig.\,\ref{f_ranges_4} but for $6\,\mathrm{M_{\sun}}$ model.}
    \label{f_ranges_6}
\end{figure*}

\begin{figure*}

	\includegraphics[width=2.7\columnwidth, angle=270]{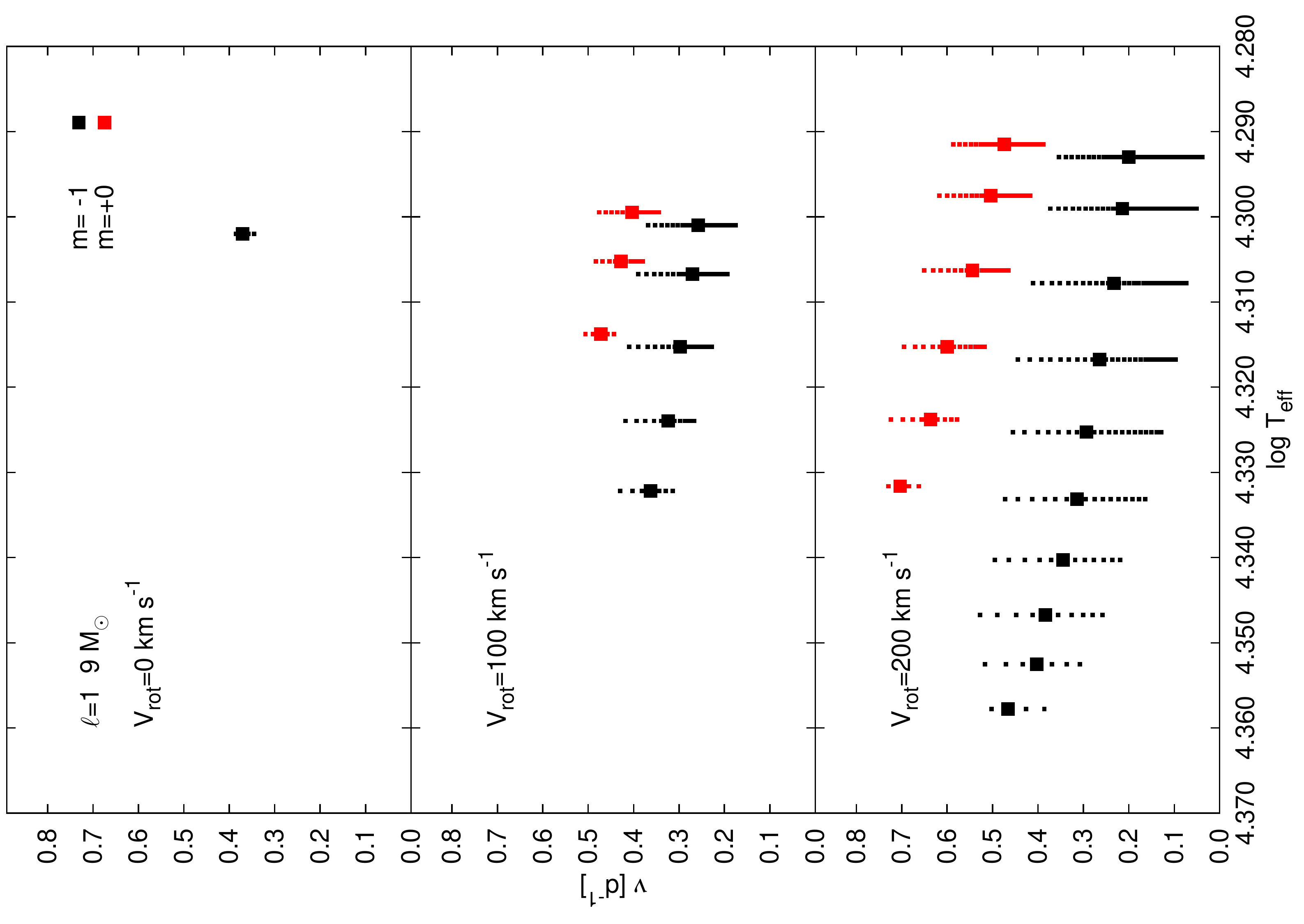}
    \caption{The same as in Fig.\,\ref{f_ranges_4} but for $9\,\mathrm{M_{\sun}}$ model.}
    \label{f_ranges_9}
\end{figure*}

\begin{figure*}

	\includegraphics[width=2.7\columnwidth, angle=270]{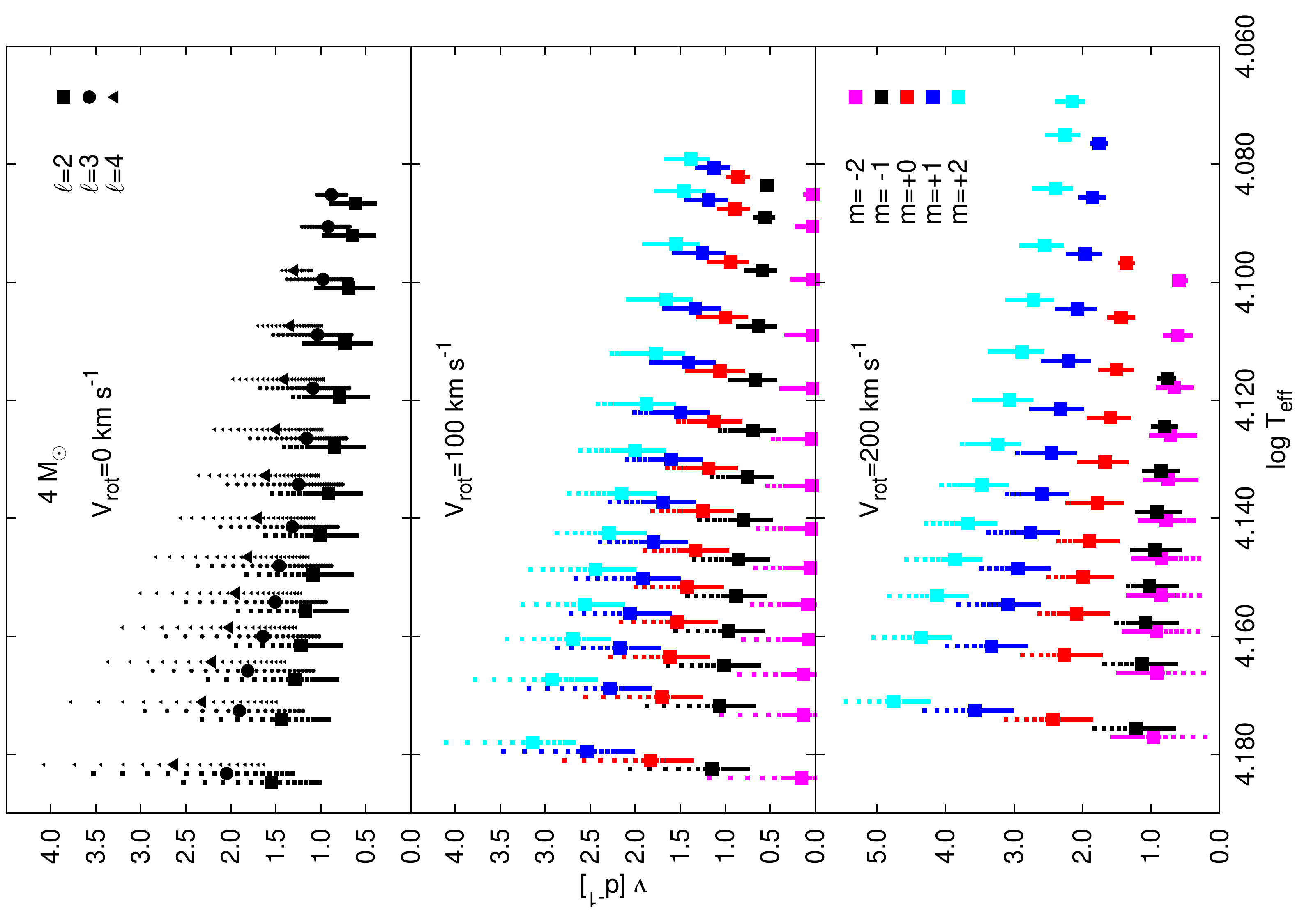}
    \caption{The same as in Fig.\,\ref{f_ranges_4} but for $\ell=2-4$ in the non--rotating models
    (the upper panel)
    and for $\ell=2$ in the rotating models with the rotation velocity $V_\mathrm{rot}=100$\,km\,s$^{-1}$
    (the middle panel) and $V_\mathrm{rot}=200$\,km\,s$^{-1}$ (the bottom panel).}
    \label{f_ranges_l2-4}
\end{figure*}

Besides calculations of the instability strips, an interesting issue is the change of the pulsational frequencies
and of the range of unstable mode frequencies with evolution. We present such results for the modes excited in our reference models
with masses $M=4,~6,~9\,\mathrm M_{\sun}$.
As before,  three values of the rotation velocity were chosen, $V_{\rm rot}=0,~100,~200$ km\,s$^{-1}$.
Figs.\,\ref{f_ranges_4}--\ref{f_ranges_9}
are for the dipole modes whereas Fig.\,\ref{f_ranges_l2-4} is for modes with $\ell=2,~3, ~4$ for a mass $M=4\,\mathrm M_{\sun}$.

\section{Effects of the input parameters on the extent of the SPB instability strip}
\label{appD}

Here, we show figures presenting  effects of the input parameters on the extent of the instability strip
of the dipole g--modes. There are shown the effects of the initial hydrogen abundance, $X_0$,
metallicity, $Z$, overshooting from the convective core, $\alpha_\mathrm{ov}$, and the opacity data.
To see the influence of these parameters, all figures have to be compared
with Fig.\,2 of Section\,2.1.

\begin{figure*}
	\includegraphics[width=2.7\columnwidth, angle=270]{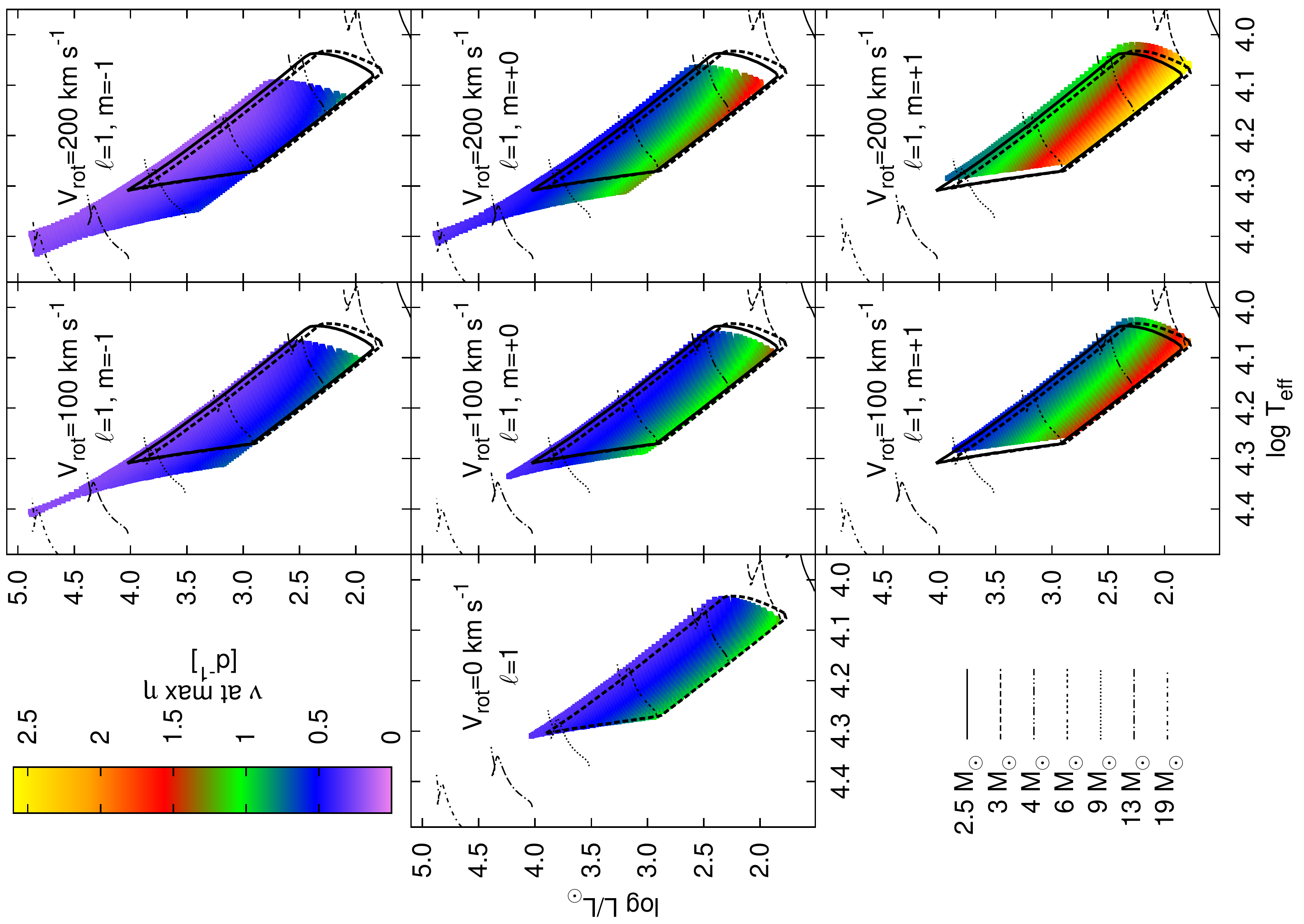}
    \caption{The same as in Fig.\,2 but for $X_0=0.75$.
    In addition, we marked the instability strips computed for non--rotating models with $X_0=0.70$
    (thick dashed line)
    and $X_0=0.75$ (thick solid line).
    }
    \label{l1_OP_X0_75Z0_015_ov0_0}
\end{figure*}

\begin{figure*}

	\includegraphics[width=2.7\columnwidth, angle=270]{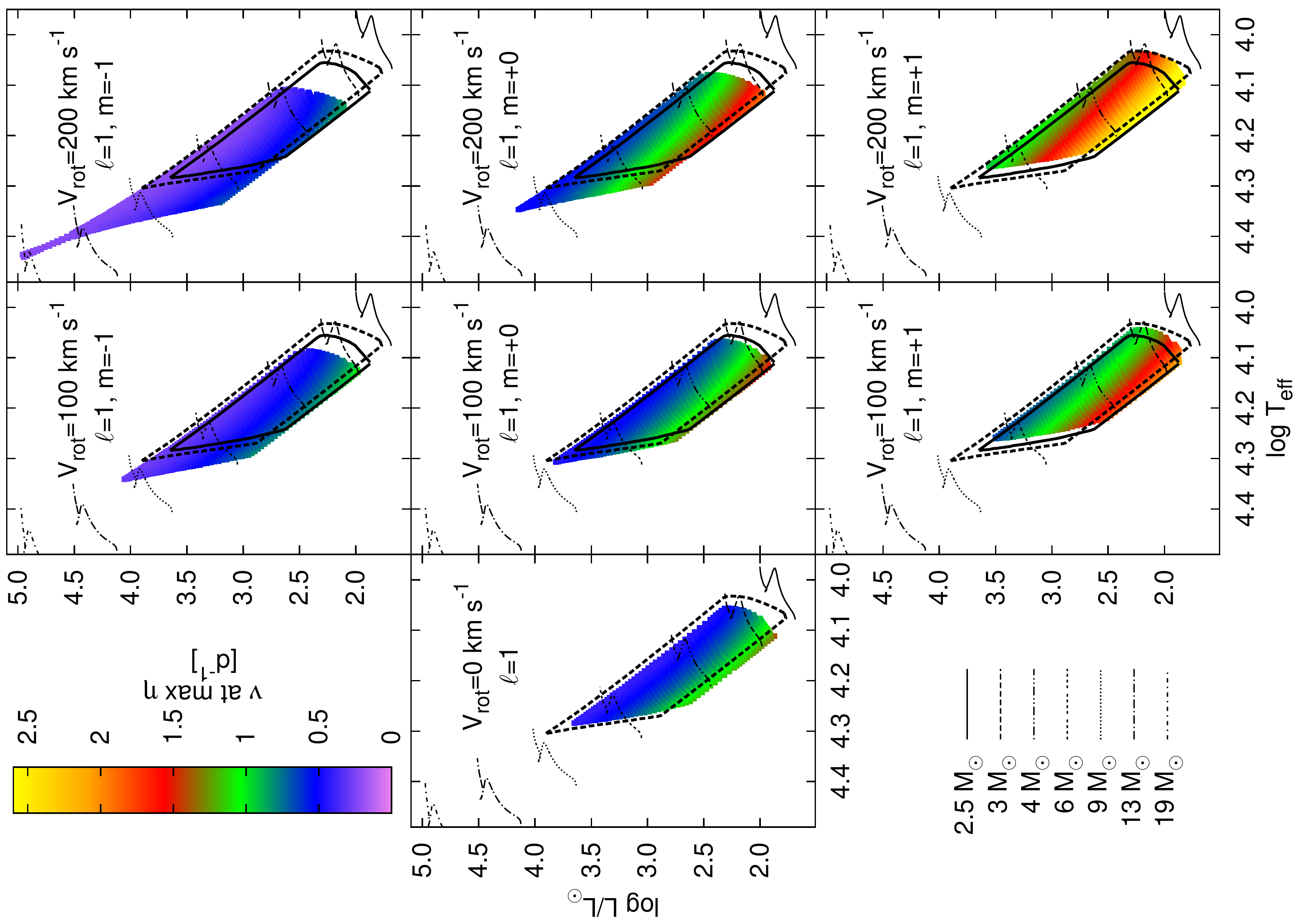}
    \caption{The same as in Fig.\,D1 but for $Z=0.010$.
    In addition, we marked the instability strips computed for non--rotating models with $Z=0.015$
    (thick dashed line)
    and $Z=0.010$ (thick solid line).}
    \label{l1_OP_X0_7Z0_010_ov0_0}
\end{figure*}

\begin{figure*}
	\includegraphics[width=2.7\columnwidth, angle=270]{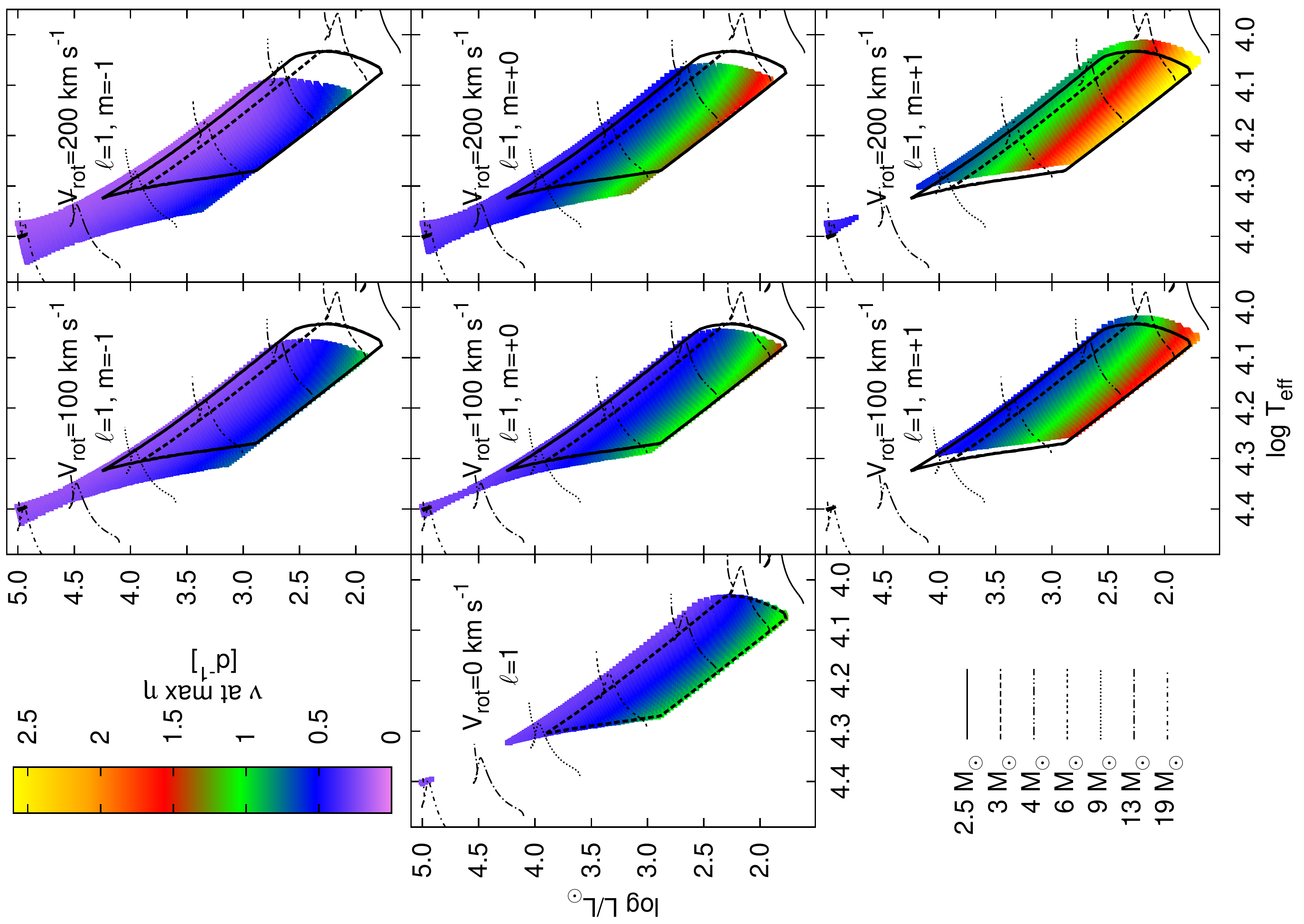}
    \caption{The same as in Fig.\,D1 but for $\alpha_\mathrm{ov}=0.2$.
    In addition, we marked the instability strips computed for non--rotating models with $\alpha_\mathrm{ov}=0.0$
    (thick dashed line)
    and $\alpha_\mathrm{ov}=0.2$ (thick solid line).}
    \label{l1_OP_X0_70Z0_015_ov0_2}
\end{figure*}

\begin{figure*}

	\includegraphics[width=2.7\columnwidth, angle=270]{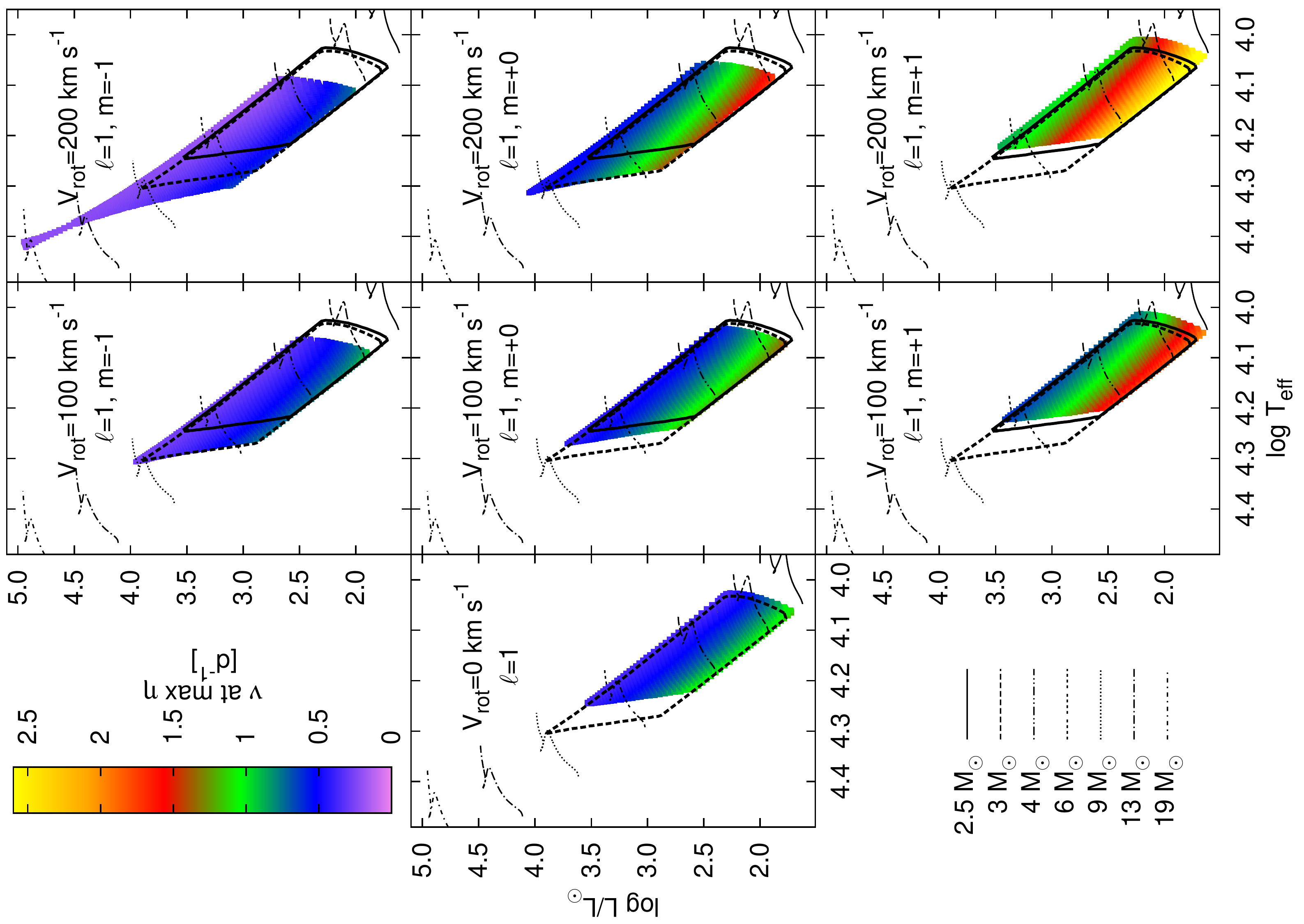}
    \caption{The same as in Fig.\,D1 but for the OPAL opacity tables.
    In addition, we marked the instability strips computed for non--rotating models with the OP tables
    (thick dashed line)
    and the OPAL tables (thick solid line).}
    \label{l1_OPAL_X0_7Z0_015_ov0_0}
\end{figure*}

\section{Maia or SPB stars in NGC\,884}
\label{appE}

NGC\,884 is a young open cluster, $\log(t/yr)=7.12-7.28$, containing many SPB stars with unusually high frequencies
\citep{2010A&A...515A..16S, 2013AJ....146..102S}. For some of them, there are known the values of the projected rotation velocity;
they are all fast rotators with $V_\mathrm{rot} \sin i$ well above 100 km\,s$^{-1}$ in most cases
\citep{2005AJ....129..809S, 2006ApJ...648..580H, 2012AJ....144..158M}.

\begin{figure*}
	\includegraphics[width=1.4\columnwidth, angle=270]{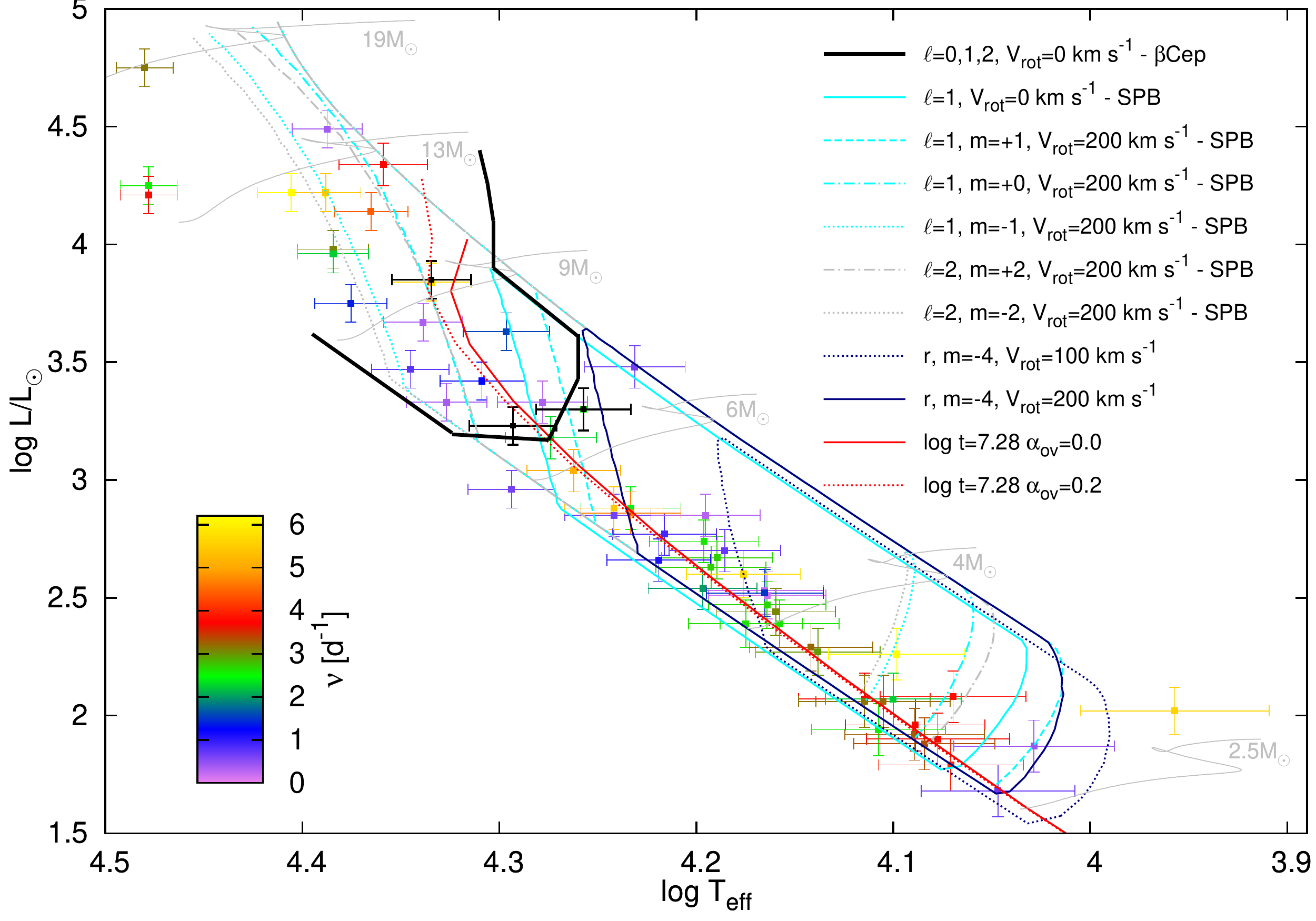}
    \caption{Variable stars observed in NGC\,884 (Saesen et al. 2010, 3013). Colours code the frequency of the dominant mode.
            For clarity, stars with frequencies above 6.3 d$^{-1}$ are marked in black.
            High--order $g$ modes instability strips for dipole modes
            excited in non--rotating and in rotating models are enclosed by 
            cyan lines,
            high--order $g$ modes instability strips for some quadrupole  modes excited in rotating models are enclosed by grey lines
            and instability strips for the $r,\,m=-4$ modes are enclosed by navy-blue lines.
            The cool border of the $\beta$ Cephei instability strip for $\ell=0$, 1, 2 and non--rotating
            models is marked by the black continuous line.
            In addition, there are shown the isochrones corresponding to the maximum age of the cluster,
            $\log(t/yr)=7.28$ (Saesen et al. 2013) for rotating models, $V_\mathrm{rot}=200$ km\,s$^{-1}$,
            without overshooting from convective core (red continues line) and with overshooting from convective core
            (red dotted line). Some evolutionary tracks calculated with  $V_\mathrm{rot}=200$ km\,s$^{-1}$
            and $\alpha_\mathrm{ov}=0.0$ are shown as well.
            }
    \label{NGC884}
\end{figure*}

We plotted all variables from the cluster on
the H--R diagram (Fig.\,\ref{NGC884}) and marked dominant frequency
and the instability strips presented in the main paper. The stars, especially in the lower part of the diagram,
are located near the ZAMS in accordance with the young age of the cluster. Moreover,
as one can see from the figure, in the region where the SPB stars should be located there are
many variables with frequencies in the range from about 2 to 6\,d$^{-1}$.
Non--rotating models cannot explain these frequencies, but unstable prograde sectoral dipole and quadrupole
modes in rotating models can explain easily the range of the observed frequencies.

The only exception is the star Oo\,2151. It falls close to the $r,\,m=-4$ mode instability domain but the frequency
seems to be too high. However, the star's luminosity greatly exceeds that of other stars with similar $T_\mathrm{eff}$ in the cluster.
Moreover, the determination of the effective temperature of the star is uncertain because
it lacks Geneva colour indices from which \citet{2013AJ....146..102S}
have derived $T_\mathrm{eff}$ of most SPB variables of the cluster.

In the case of multiperiodic variables we also compared further frequencies with our models.
On first sight, it seems that only two cases, $\nu_2=12.0$\,d$^{-1}$ in Oo\,2323 and $\nu_3=12.6$\,d$^{-1}$ in Oo\,2566
cannot be explained by our models. But then there is no certainty that they are of pulsational
origin.

\bsp	
\label{lastpage}
\end{document}